\documentclass[traditabstract]{aa} 
\usepackage[varg]{txfonts}
\usepackage{graphicx,rotating,amssymb}
\usepackage{booktabs}
\usepackage{subfig}
\usepackage{float,capt-of}
\usepackage{natbib}
\usepackage{multirow}
\usepackage[english]{babel}
\usepackage{morefloats}
\usepackage{amsmath}

\begin{document}

   \title{Outflows and complex stellar kinematics in \\ SDSS star forming galaxies}

   \author{C. Cicone, 
	\inst{1,2,3} 
        	\and
      	R. Maiolino  \inst{2,3}
	\and
	A. Marconi \inst{4, 5}
	}
	\institute{Institute for Astronomy, Department of Physics, ETH Zurich, Wolfgang-Pauli-Strasse 27, CH-8093 Zurich, Switzerland \\ 
	\email{claudia.cicone@phys.ethz.ch}
	\and Cavendish Laboratory, University of Cambridge 19 J. J. Thomson Avenue, Cambridge CB3 0HE, UK\\ 
	 \and Kavli Institute for Cosmology, University of Cambridge, Madingley Road, Cambridge CB3 0HA, UK\\ 
	\and  Dipartimento di Fisica e Astronomia, Universit\`a degli studi di Firenze, Via G. Sansone 1, 50019, Sesto Fiorentino (Firenze), Italy \\
	\and INAF - Osservatorio Astrofisico di Arcetri, Largo E. Fermi 5, 50125, Firenze, Italy \\
	}

 \date{Received xx xx xx / Accepted: xx xx xx}

\abstract{We investigate the properties of star formation-driven outflows 
by using a large spectroscopic sample of $\sim$160,000 local ``normal''
star forming galaxies, drawn from the Sloan digital sky survey (SDSS), spanning a wide range of star formation rates (SFRs) and
stellar masses (M$_*$). The galaxy sample is divided into a fine grid of bins in the ${\rm M_*-SFR}$
parameter space, for each of which we produce a composite spectrum by stacking together the SDSS spectra
of the galaxies contained in that bin. We exploit the high signal-to-noise of the stacked spectra to study 
the emergence of faint features of optical emission lines that may trace galactic outflows and would otherwise be too faint
to detect in individual galaxy spectra. We adopt a novel approach that relies on the comparison
between the line-of-sight velocity distribution (LoSVD) of the ionised gas (as traced by the [OIII]$\lambda$5007 and
H$\alpha$+[NII]$\lambda\lambda$6548,6583 emission lines) and the LoSVD of the stars, which are used as a reference tracing virial motions.
Significant deviations of the gas kinematics from the stellar kinematics in the high velocity tail of the LoSVDs are interpreted
as a signature of outflows.
Our results suggest that the incidence of ionised outflows increases with SFR and specific SFR. 
The outflow velocity ($v_{\rm out}$) is found to correlate tightly with the SFR for 
$\rm SFR>1~M_{\odot}~yr^{-1}$, whereas at lower SFRs the 
dependence of $v_{\rm out}$ on SFR is nearly flat. The outflow velocity, although with a much
larger scatter, appears to increase also with
the stellar velocity dispersion ($\sigma_*$), and we infer velocities as high as
$v_{\rm out} \sim (6-8) \sigma_*$.
Strikingly, we detect the signature of ionised outflows only in galaxies located above the main sequence (MS) of
star forming galaxies in the M$_*$--SFR diagram, and the incidence of such outflows
increases sharply with the offset from the MS.
This result suggests that star formation-driven outflows may be responsible for shaping the upper envelope of the MS
by providing a self-regulating mechanism for star formation.
Finally, our complementary analysis of the stellar kinematics reveals the presence of blue asymmetries of a few 10 km s$^{-1}$ 
in the stellar LoSVDs. The origin of such asymmetries is not clear, but a possibility is that these trace the presence, in local galaxies of a large
number of high velocity runaway stars and hypervelocity stars in radial trajectories.}

 
 \maketitle

\section{Introduction}

Baryons in galaxies are engaged in a complex intertwining of mechanisms
whose combined effects drive galaxy evolution. Key players of this game
are gas, stars and active galactic nuclei (AGN). Gas is the primary ingredient of
both star formation and black hole accretion, and its properties are therefore crucial
for shaping galaxies. Star formation and AGN activity can in turn heavily influence 
the physical, chemical and kinematical conditions of gas within galaxies, thereby exerting
a ``feedback'' on galaxy evolution.
In a simplistic way, our current knowledge of the processes occurring in galaxies and
involving gas and stars 
can be summarised as follows. 
The presence of gas within galaxies is a key prerequisite for star formation. 
In local galaxies, all known star formation
takes place in dense molecular clouds, mostly in giant molecular clouds (GMCs, see review by \citealt{Scoville13}).
The star formation process can be either smooth and quiescent, probably fostered by cosmological inflows of cold gas into galaxies
(e.g. \citealt{Dekel+09, Bouche+10}), or bursty, triggered by dynamical processes such as 
galaxy interactions \citep{Scoville13}. 
Recent studies suggested that
the latter mode of star formation does not contribute significantly to the cosmic star formation rate (SFR) density, accounting  
for only 10\% of it even at $z\sim 2$, i.e. the peak of the cosmic SFR density
\citep{Rodighiero+11}. Although we still lack conclusive observational evidence for 
the presence of cold accretion flows feeding galaxy halos (but see promising observations by \citealt{Cantalupo+14} and \citealt{Hennawi+15} at
$z\sim2$),
the hypothesis that the building up of baryonic mass is a secular process
has been suggested by the observation of a tight correlation between
 stellar mass (${\rm M_{*}}$) and star formation rate in star forming galaxies \citep{Bouche+10,Lilly+13}, 
 dubbed star formation ``main sequence'' (MS). The MS is a
 well-established property of the star forming galaxy population at $z\sim0$ \citep{Noeske+07, Peng+10} and 
is believed to persist at least up to $z\sim4$ \citep{Schreiber+15}, 
with an intrinsic scatter of only $\sim0.3$~dex.

Once star formation is activated, supernovae enrich the interstellar medium 
(ISM) with metals and, along with young and massive stars, inject energy and 
 momentum into the ISM, thereby altering the physical conditions of the gas surrounding the sites
 of star formation and at the same time driving multi-phase 
 galactic-scale outflows that propel part of the gas out of these regions, and in some
 cases can even expel it from the galaxy.
 The investigation of stellar feedback has triggered a wealth of both theoretical 
 and observational works. For theoretical work we refer 
 to the early review by \cite{Chevalier77} for energy-driven outflows, to
 \cite{Murray+05} for momentum-driven outflows, and to
 \cite{Hopkins+14} for cosmological zoom-in simulations exploring the combined effect of 
 multiple stellar feedback mechanisms on galaxy evolution.
 For observational work we refer to the reviews by \cite{Veilleux+05} and \cite{Erb15},
 the latter one focussing on high redshift.
 
Gas heating and removal resulting from intense episodes of star formation
 is believed to play a key role in galaxy evolution. For example, stellar feedback may be the 
 main responsible for:
 (i) the low efficiency of star formation observed in both low and high mass
galaxies \citep{Behroozi+10,Papastergis+12}, 
 although for more massive galaxies an additional feedback from AGNs 
 is often invoked (e.g. reviews by \citealt{Cattaneo+09} and \citealt{Kormendy+Ho13}); (ii) the lower gas-phase metal content
 of low-mass galaxies with respect to more massive galaxies 
 \citep{Dave+11b}, and (iii) 
 the consequent enrichment of the circumgalactic medium (CGM) with metals \citep{MacLow+Ferrara99}. Moreover,
 it has been suggested that galactic outflows driven by star formation 
 may have aided the reionization of the early Universe, by creating
gas-free paths around young star clusters thus facilitating the leakage of ionising photons 
from the first galaxies \citep{Heckman+11,Erb15}.
 
Although blueshifted absorption lines at optical and UV wavelengths are probably the most direct 
and accessible tool to identify outflows of ionised (and atomic) gas, especially in distant galaxies, 
nebular emission lines such as [OIII]$\lambda$5007, H$\alpha$ and 
[NII]$\lambda\lambda$6548,6584 are also widely employed to study galactic outflows,
both in local star forming galaxies (e.g. \citealt{Heckman+90,Veilleux+95,Lehnert+Heckman96,Soto+12,Westmoquette+12,
Rupke+Veilleux13, Rodriguez-Zaurin+13,Bellocchi+13,Cazzoli+14,Arribas+14})
and at higher redshifts (e.g. \citealt{Shapiro+09,Newman+12,Harrison+12,Cano-Diaz+12,Genzel+14,Forster-Schreiber+14,Carniani+15}).
However, despite the increasing number of observational studies targeting galactic outflows
and the remarkable advances in the field,
a {\it systematic and unbiased} investigation of star formation-driven outflows 
in a large sample of galaxies is still missing. Indeed, 
since the outflow signature can be very faint and difficult 
to detect in individual normal galaxies, previous studies focused on rather small samples, mostly characterised
by ``extreme'' properties (e.g. powerful and massive starbursts, 
ultra-luminous infrared galaxies (ULIRGs)), hence not representative of the general galaxy population,
although significant improvement has been recently made in this regard (e.g.
\citealt{Chen+10,Martin+12,Rubin+14,Chisholm+14}).

In this paper we study star formation-driven outflows of ionised gas as traced by faint broad
wings of optical emission lines ([OIII]$\lambda$5007, H$\alpha$ and 
[NII]$\lambda\lambda$6548,6584) by using the largest and most unbiased spectroscopic sample
of local galaxies currently available, the Sloan digital sky survey (SDSS).
The SDSS, in conjunction with the catalogues provided by the MPA-JHU analysis, 
constitutes a huge database, whose noteworthy potential in terms of studying the
general properties of galaxies can be further exploited by stacking multiple spectra together
to obtain very high signal-to-noise composite spectra
(e.g. \citealt{Chen+10,Andrews+Martini13}).
The stacking technique has numerous advantages, and it is particularly useful when searching for
faint spectral features, such as those impressed by galactic outflows, which can be missed
in the noise of single galaxy spectra. 

We divide the ${\rm M_{*}-SFR}$
parameter space into a fine grid of bins, each of which includes
from several tens to thousands of sources. The spectra within each bin
are then stacked together to produce a single, very high signal-to-noise
composite spectrum that is representative of the spectrum of a ``typical'' galaxy in that bin,
in which we can study the emergence of faint broad wings of optical emission lines
that may trace ionised outflows.
Moreover, since stellar mass and star formation rate are intrinsically correlated
in most star-forming galaxies (because most of them follow the MS), 
by stacking galaxies in bins of SFR and M$_*$ 
we can investigate the trends between the outflow properties and these two parameters 
independently, thus breaking their degeneracy resulting from the MS.
The SDSS database and the use of the stacking technique 
allow us to explore the incidence and properties of ionised outflows over a wide dynamical range
of galaxy properties, i.e. ${\rm M_{*} \in [2 \times 10^7, 6 \times 10^{11}]~M_{\odot}}$ and
${\rm SFR \in [2 \times 10^{-3}, 2 \times 10^2 ]~M_{\odot}~yr^{-1}}$, which is unprecedented
in outflow studies.

A rather controversial issue in studies of galactic outflows through {\it emission} line profiles is
the definition of ``broad component'', which is supposed to trace
gas in outflow, and of
``narrow component'', which instead traces virial motions in the galaxy. 
Such an issue has not been addressed in a systematic way by previous studies: most 
of them just rely on the $\chi^2$ improvement
to establish the necessity of a broad component to fit the emission line profile
(e.g. \citealt{Harrison+12,Westmoquette+12,Bellocchi+13,Arribas+14}).
Moreover, even if a broad component is present, 
its association with galactic flows is not straightforward: a broad component in emission
may trace the high velocity tail of virialised motions within the galaxy, or galaxy interactions.
In this study we attempt to overcome this problem by adopting a novel strategy, consisting in directly
comparing the line-of-sight velocity distribution (hereafter, LoSVD) of the 
ionised gas with the LoSVD of the stars, which is taken as a reference tracing virial motions. 
Accordingly, gaseous outflows are identified by deviations of the gas kinematics from the stellar
kinematics, and in particular by the presence of an excess of gas velocity with respect to the stars in the
high velocity tail of the LoSVD.

Summarising, the main goal of this work is to investigate
the incidence and properties of galactic outflows of ionised gas as a function of a wide 
range of galaxy properties, breaking the degeneracy between SFR and ${\rm M_{*}}$. 
Moreover, for the first time, we push outflow studies to stellar masses and SFRs as low
as ${\rm 2 \times 10^7~M_{\odot}}$ and ${\rm 2\times 10^{-3} M_{\odot}~yr^{-1}}$, respectively, a regime
that has not been probed before in this field.

A ${\rm H_0}$ = 70 km s$^{-1}$ Mpc$^{-1}$, ${\rm \Omega_{M}}$=0.27, 
${\rm \Omega_{\Lambda}}$=0.73 cosmology is adopted throughout this paper.

\begin{figure*}[tbp]
\centering
    \includegraphics[clip=true, trim=-1.5cm 0cm 0.cm -1.5cm,angle=90,scale=0.5]{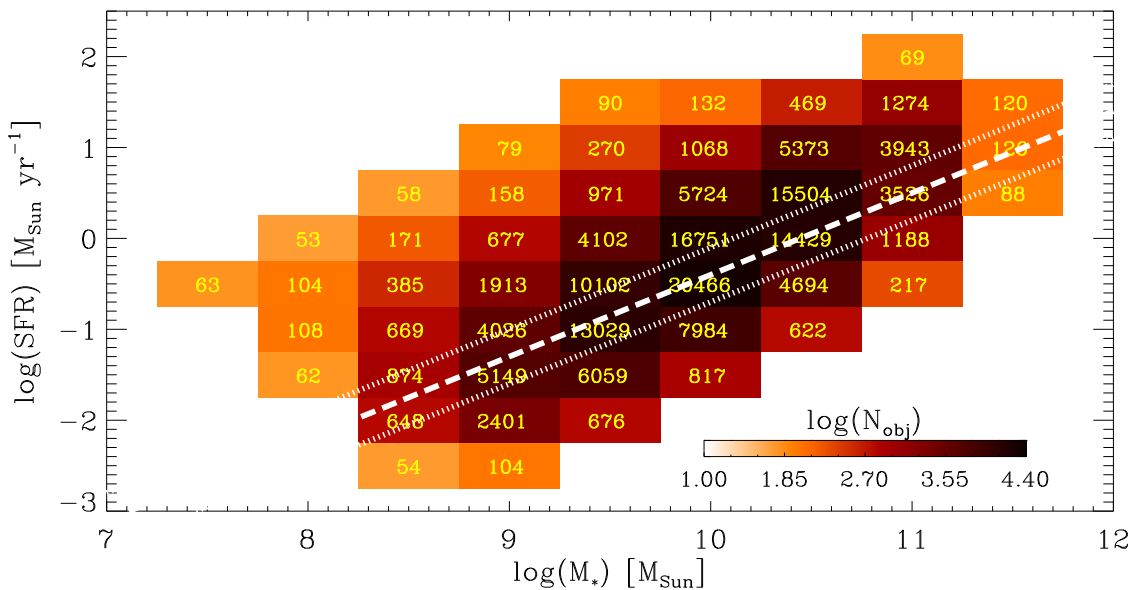}\\
     \caption{The SDSS sample of star-forming, non active galaxies selected for this study, plotted 
     in the M$_*$-SFR parameter space. The grid of 50 bins used for the stacking is shown, 
     with the bins colour-coded according to the number of spectra enclosed. The exact number of sources within
     each bin is indicated by the yellow text. The white dashed and dotted lines 
     represent the MS relation of local star-forming galaxies with its $\pm1\sigma$ scatter ($=0.3$~dex).
     The MS relation has been taken from \cite{Peng+10} and adapted to this dataset as explained in $\S$~\ref{sec:stacking}.}
   \label{fig:grid_sf}
\end{figure*}

\section{Methods}

In this section we present the SDSS sample of star forming, non active galaxies
selected for this study ($\S$~\ref{sec:sample_sel}). We also describe in detail the stacking procedure ($\S$~\ref{sec:stacking}) and the
fitting techniques applied to the stellar continuum and to the [OIII]$\lambda$5007 and
H$\alpha$+[NII]$\lambda\lambda$6548,6583 nebular lines to obtain the line-of-sight 
velocity distributions of stars and ionised gas from the galaxy composite spectra ($\S$~\ref{sec:stellar_sub}-
\ref{sec:losvd_fit}).

\subsection{Sample selection}\label{sec:sample_sel}

Our galaxy sample is entirely drawn from the spectroscopic sample 
of the Sloan digital sky survey, data release 7 (SDSS DR7, \citealt{Abazajian+09}).  
We operate a first selection by including only those galaxies enclosed in the MPA-JHU 
catalogue \footnote{Available at \texttt{http://www.mpa-garching.mpg.de/SDSS/}}, which provides 
intrinsic and derived spectral parameters for each source,
such as redshift, emission line fluxes, stellar mass, and star formation rate \citep{Tremonti+04, Kauffmann+03c,
Brinchmann+04, Salim+07}. The MPA-JHU analysis encompasses only SDSS DR7 sources at
$z<0.7$.

Among the MPA-JHU galaxies, we selected only sources classified as ``star-forming''  (or ``HII'') according
to their location on the log([OIII]/H$\beta$) vs log([NII]/H$\alpha$) diagram 
(i.e. the ``BPT'' diagram, \citealt{BPT81}), by adopting the same division criteria as \citet{Kauffmann+03c}. 
Such selection is needed in order to avoid contamination
from AGNs, which is of primary importance to constrain the star 
formation-driven feedback mechanisms in action in normal local galaxies.

We note that the BPT selection may introduce a potential bias in our sample of star forming galaxies.
This is due to the fact that purely star forming galaxies may still exhibit extended areas where the optical line 
ratios are typical of low-ionisation emission line regions (LINERs). As a consequence, an ``HII'' galaxy 
may be erroneously classified as ``Composite'' or ``LINER'' when its optical emission 
is averaged over a large aperture (such as the SDSS spectroscopic fibre).
Extended LINER-like gas excitation in purely star forming galaxies can have different origins.
In the vast majority of local isolated (inactive) galaxies, the most plausible source of extended
LINER-type excitation are evolved (post-AGB) stars 
\citep{CidFernandes+11,Yan+Blanton12, Singh+13,Belfiore+15}.
This is in general not the case for strongly star forming galaxies such as (U)LIRGs, 
where there is evidence for extended LINER excitation associated with 
widespread shocks. Indeed, both theoretical models \citep{Allen+08} 
and spatially resolved integral field spectroscopy (IFS) 
observations \citep{Monreal-Ibero+06, Monreal-Ibero+10, Soto+Martin12, Rich+14, Rich+15} have
shown that galaxy regions dominated by shock excitation exhibit optical line ratios typical of LINERs.

Since extended shocks can be associated with strong outflows produced by vigorous star formation 
(e.g. \citealt{Sharp+Bland-Hawthorn10}), our sample selection, by leaving out LINERs and 
Composite galaxies, may be biased against (U)LIRGs experiencing strong feedback from star formation. 
Although this is certainly a caveat to keep in mind, the nature of the widespread 
shocks that may dominate the ionised gas excitation over extended regions in (U)LIRGs 
is most likely linked to the merger process.
This argument is based on the lack of observational evidence 
for shock-like excitation significantly affecting the optical line fluxes 
in {\it isolated (U)LIRGs} \citep{Monreal-Ibero+10, Rich+15}.
Indeed, the point made by resolved IFS studies \citep{Soto+Martin12, Rich+14, Rich+15} about a possible 
misclassification of star forming galaxies hosting widespread shocks, when aperture-integrated spectra are used,
concerns {\it only (U)LIRGs in a merger stage}. 
In contrast, for isolated (U)LIRGs, the contribution from shocks to extended optical line emission 
is negligible \citep{Monreal-Ibero+10, Rich+15}. 
Therefore, the types of shocks that can affect integrated optical line 
fluxes to the point that an ``HII'' galaxy is classified as ``Composite'' or ``LINER'' are galaxy-wide shocks produced by a complex
interplay of merger-driven processes, and not just by star-formation-driven outflows. For their sample of local ULIRGs, 
\cite{Monreal-Ibero+06} explicitly 
concluded that the LINER-type optical line ratios observed in the extranuclear regions,
several kiloparsecs away from the circumnuclear starburst, are due to shocks associated with tidally-induced large-scale flows
produced during the merger and not to star formation-driven ``superwinds'' (which instead tend to dominate the gas kinematics in the 
inner regions of these ULIRGs that are characterised mostly by HII-type excitation). A similar conclusion was reached by \cite{Monreal-Ibero+10}
who investigated the excitation source of ionised gas in a sample of 32 local LIRGs using IFS data, and 
found that the presence and relevance of shock-type excitation is more strongly correlated with the interaction class of the system 
than with the star formation activity (with isolated galaxies being largely dominated by HII-type excitation).

In summary, based on the observational evidence available from IFS studies, 
our BPT sample selection could lead to two possible biases:
(i) a bias against ageing isolated galaxies dominated by old stellar populations and/or (ii) a bias against merger-driven shocks. 
The first bias would affect massive ``retired'' galaxies 
with stellar populations older than 1~Gyr and little ongoing star formation (e.g. \citealt{CidFernandes+11, Singh+13}),
hence with very little or no star formation feedback at work. The second bias would affect mergers, which are only a minor fraction 
of the local galaxy population \citep{Patton+Atfield08} and so unimportant in our sample\footnote{Our sample is largely
representative of the local star forming galaxy population, in contrast to previous outflow studies focussing mostly on the local (U)LIRG
population (for which the merger fraction is much higher).}. 
Moreover, even if hypothesising that our analysis under-represents
merging (U)LIRGs experiencing strong feedback, our conclusions (e.g. $\S$~\ref{sec:discussion_MS}) 
would not be affected but would actually be even strengthened: in this case
we would be under-representing galaxies with high SFRs/sSFRs (i.e.
galaxies above the Main Sequence in the M$_*$-SFR diagram, e.g. \citealt{Combes+13, KilerciEser+14}) 
with strong outflows, hence our investigation and 
the relevant results could be regarded as conservative.

In addition to the BPT selection of ``HII'' galaxies, we require 
a signal-to-noise greater than five in all four emission lines used in the BPT diagram (i.e. [OIII], H$\beta$, [NII], H$\alpha$).
As a consequence, the most passive galaxies, which do not show prominent
nebular lines, are under-represented in our sample.
We note that, by imposing a threshold SNR on the [OIII] and [NII] line fluxes, we may introduce a bias with
metallicity. However, since local galaxies define a tight surface in the $\rm M_*-\mathcal{Z}-SFR$ plane (i.e. the ``fundamental
metallicity relation'', e.g. \citealt{Mannucci+10, Lara-Lopez+10}), we do not expect galaxies {\it within a given bin} in the $\rm M_*-SFR$ parameter space 
to show significant variations in metallicity. Therefore, the main effect of imposing SNR constraints on the [OIII] and [NII] emission lines 
is that our sample selection does not allow us to populate certain regions of the $\rm M_*-SFR$ parameter space, but we are
not introducing differential biases within the galaxy bins that are included in our sample.

A general caveat for this study is that 
at the median redshift of our sample,
$z_{\rm med}$ = 0.073, the 3 arcsec-diameter of the SDSS spectroscopic fibre samples only the galaxy emission
within $r\lesssim2.1$~kpc. 
Moreover, the spectroscopic fibre samples a wide range of different projected sizes within our galaxy sample. Aperture effects
are an obvious limitation of our study that will need to be taken into account in future works and that can be tackled properly
only by using spatially-resolved observations. In order to be able to study outflows in ``normal'' galaxies (i.e. in galaxies that
are neither extreme starbursts nor AGN hosts) we need to wait for the completion of ongoing large IFS surveys such as
the CALIFA, SAMI and MaNGA surveys, although these will still have a statistics that is orders of magnitudes smaller than that provided by the SDSS.
Another potential source of bias is given by galaxy inclination, which can have an important effect on the detectability of outflows (e.g. \citealt{Chen+10}).
However, in terms of outflow properties, we expect the stacking of thousands of spectra to average out the effects of inclination, and so we do not expect
our results to be greatly affected.

Our final selection consists of 157,639 galaxy spectra, whose distribution
in the ${\rm M_{*}-SFR}$ diagram
is shown in Fig.~\ref{fig:grid_sf}. 
The stellar mass of each galaxy was derived from a fit to the observed spectral energy 
distribution (SED), obtained with SDSS broad-band photometry \citep{Brinchmann+04, Salim+07}, 
which therefore takes into account the {\it total} emission from
the galaxy. The SFR estimate is instead based on the H$\alpha$ intrinsic luminosity following the procedure described by \cite{Brinchmann+04},
without applying any aperture correction to the SDSS fibre-limited spectrum: the SFRs employed in this work 
are therefore the ``fibre SFRs'' provided in the MPA-JHU catalogue. Such choice is justified by the fact that, 
in this study, we infer the properties of the gaseous and stellar kinematics by using exclusively
the information provided by the fibre-limited SDSS spectroscopy, hence we do not have information on the gas (and so on possible outflows)
on scales larger than those sampled by the fibre. Therefore, we believe that, given the obvious limitations of SDSS spectroscopy 
(which can be overcome only with IFS studies), it is more sensible to look for possible correlations between the properties
of ionised outflows (on the scales sampled by the SDSS fibre) and the fibre SFRs, rather than the aperture-corrected SFRs.
For the stellar mass instead we have chosen to use the ``total'' M$_*$ because galactic outflows are expected to feel the effect of the entire
galaxy gravitational potential well, which is linked to the {\it total} stellar mass.

\subsection{Stacking procedure}\label{sec:stacking}

\begin{figure}[tb]
\centering
    \includegraphics[clip=true,scale=.6,angle=270]{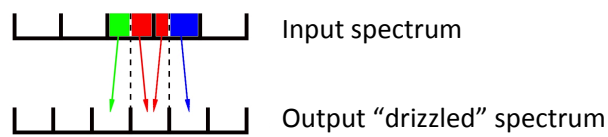}\\
     \caption{Sketch of the ``Drizzle'' method applied to the SDSS spectra.}
   \label{fig:drizzling}
\end{figure}

The ${\rm M_{*}-SFR}$ parameter space sampled by our SDSS galaxy selection 
is divided into a fine grid of 50 bins, as shown in Fig.~\ref{fig:grid_sf}, and the spectra within each
bin are stacked together to obtain very high signal-to-noise composite spectra.
Each bin is required to include a minimum of 50 sources, but most bins contain over 100 galaxies, and the central
ones up to $\sim20,000$. Throughout this paper, we will code each stack by its average log(SFR) and
log(M$_{*}$) (ID=log(SFR)\_log(M$_{*}$)).

We also show in Fig.~\ref{fig:grid_sf} the position of the main sequence of local star-forming
galaxies and its $\pm0.3$~dex scatter \citep{Salim+07, Peng+10}. 
Since we use fibre-limited SFRs, the reference MS for this study differs slightly from
the MS relationship derived by \cite{Peng+10}, who instead used aperture-corrected SFRs.
The main effect of using SFRs that are not corrected for the SDSS aperture is a downward shift
of the MS relation by $\rm\Delta log(SFR)=-0.4$~dex. This shift was estimated by selecting, in our sample,
galaxies with $0.02<z<0.085$ (to match the redshift range of \citealt{Peng+10}) and $\rm9.0<log(M_{*}[M_{\odot}])<11.0$
(i.e. the stellar mass range where the mean relationship between log(SFR) and log(M$_*$) is linear) and by fitting
the mean trend between log(SFR) and log(M$_*$) with a linear relationship, whose slope was constrained to be the
same as \cite{Peng+10}.
However, we stress that the aim of this work is not to provide
a ``general'' fit to the local MS relation of star-forming galaxies, and that 
the MS relationship shown in Fig.~\ref{fig:grid_sf}
should only be used as a reference for this galaxy sample.
We refer to the works by \cite{Peng+10,Schreiber+15} and \cite{Gavazzi+15}
for more detailed discussion about the MS of star forming galaxies.

Prior to stacking, the galaxy spectra in each bin are de-reddened and normalised to their
extinction-corrected H$\beta$ line flux, which is related to the star formation
rate in the galaxy. The reddening $E(B-V)$ is estimated from the observed
H$\alpha$-to-H$\beta$ flux ratio, by using the Galactic extinction curve
as described in \cite{CCM89} and by assuming $R_V=A_V/E(B-V)=3.1$.
The spectra are then shifted to the rest-frame by using the redshifts provided by the Princeton-1D spectroscopic 
pipeline\footnote{These are heliocentric redshifts 
derived from a combination of stellar absorption lines and nebular emission lines. For additional information we refer the reader to D. Schlegel's Princeton/MIT SDSS
Spectroscopy webpage (\texttt{http://spectro.princeton.edu}) and to \cite{Bolton+12}.}
and contextually re-binned
by using a technique based on ``Drizzle'', the method
for linear reconstruction of under-sampled images \citep{Fruchter+Hook02}.
The drizzling consists in mapping the original redshifted SDSS spectrum into
a rest-frame pixel grid (the output spectrum) by taking into account 
their different pixel sizes and relative shift, as sketched in Fig.~\ref{fig:drizzling}. 
In practice, the flux contained in the input pixel is
averaged into the output pixel with a weight that is proportional to the area of
overlap between the two pixels.
The resulting spectra have a wavelength separation of 0.8 \AA{} pixel$^{-1}$, i.e.
slightly finer than the original SDSS spectra.
The rest-frame galaxy spectra in each bin are finally averaged together to produce a
single composite spectrum (the final stack), which is then normalised to its mean.

\subsection{Stellar subtraction}\label{sec:stellar_sub}

\begin{figure*}[tbp]
\centering
    \includegraphics[width=0.49\columnwidth,angle=90]{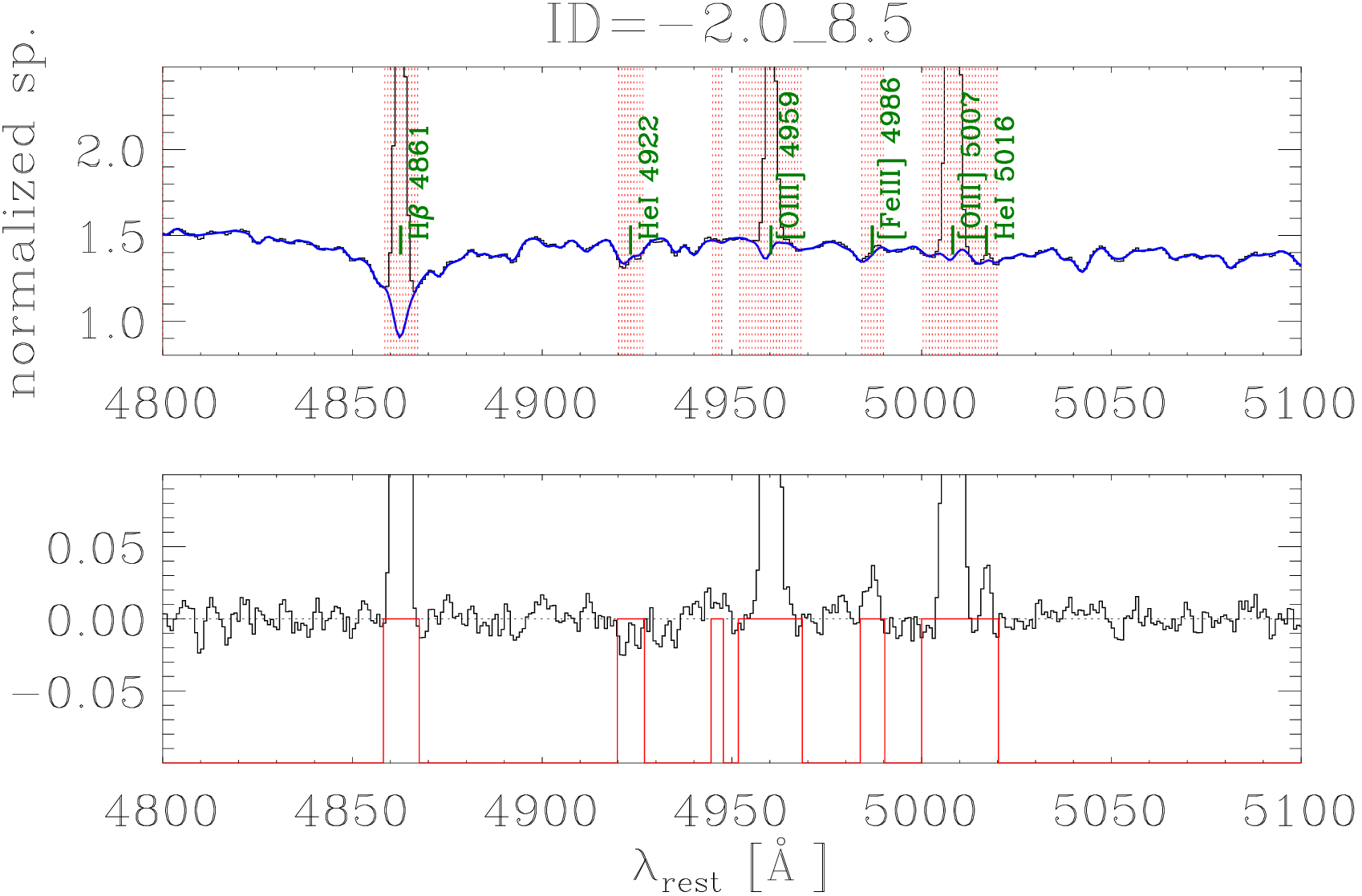}\quad	
    \includegraphics[width=0.49\columnwidth,angle=90]{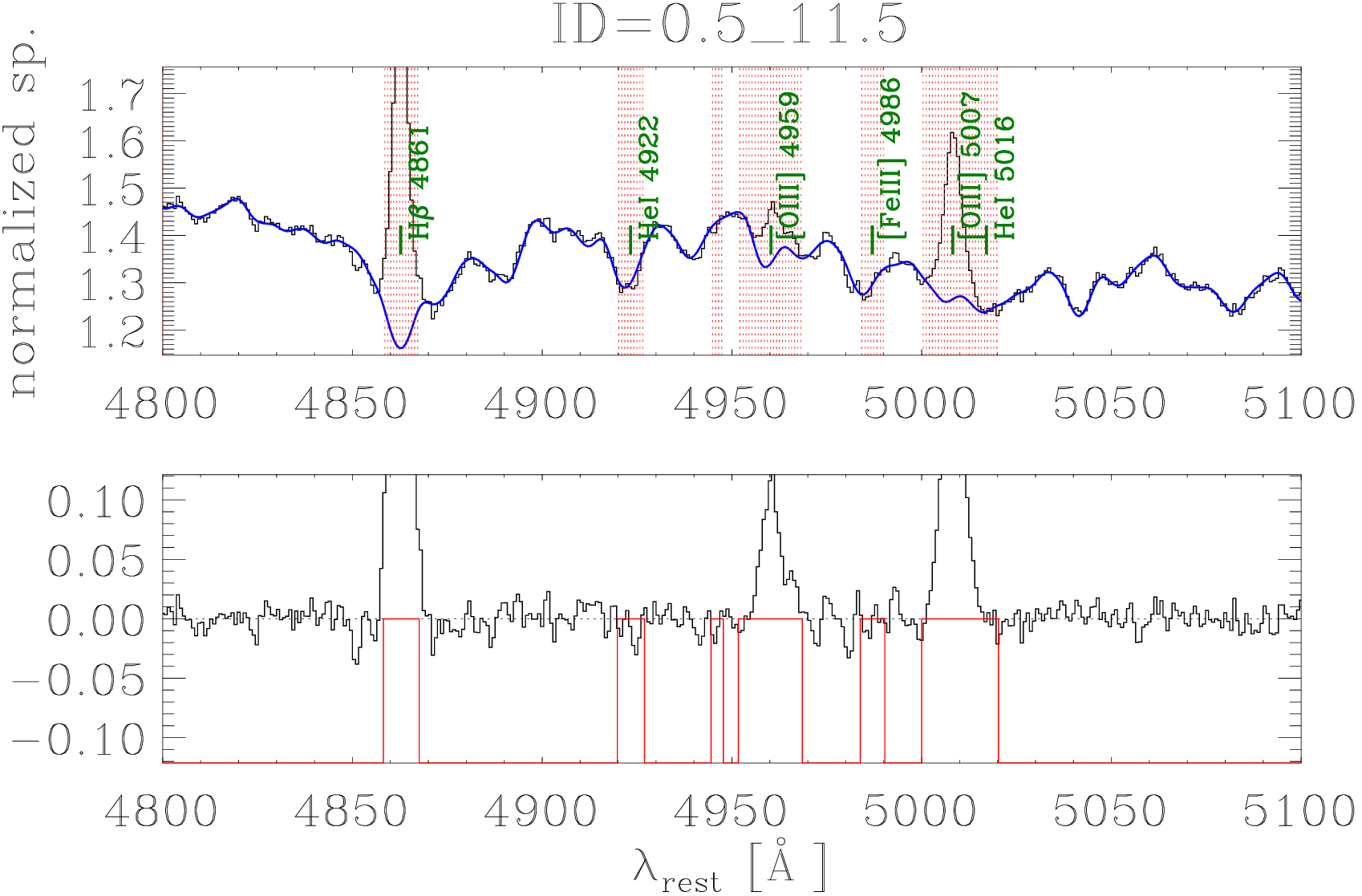}\\
    \vspace{0.8cm} 
    \includegraphics[width=0.49\columnwidth,angle=90]{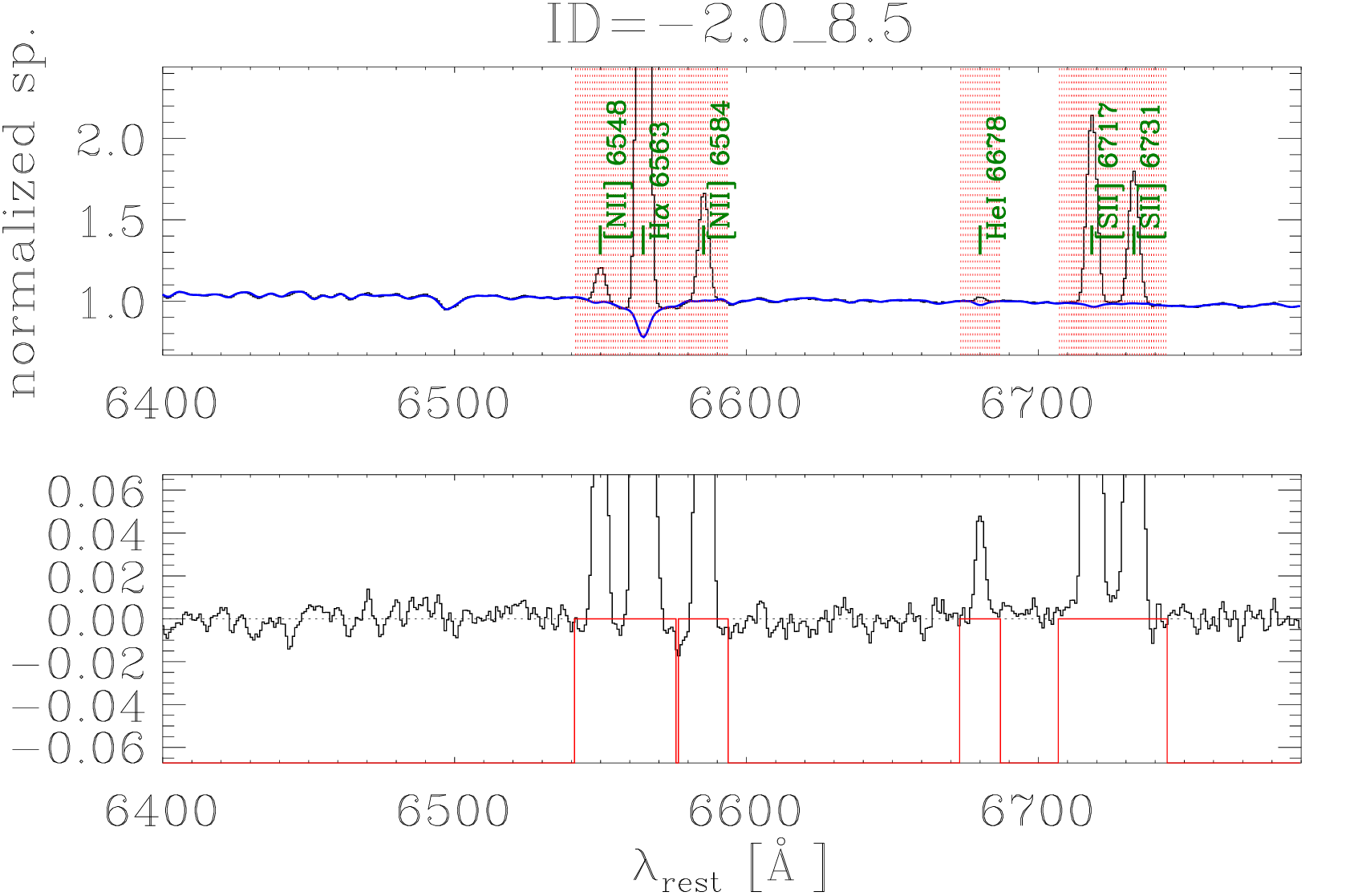}\quad
    \includegraphics[width=0.49\columnwidth,angle=90]{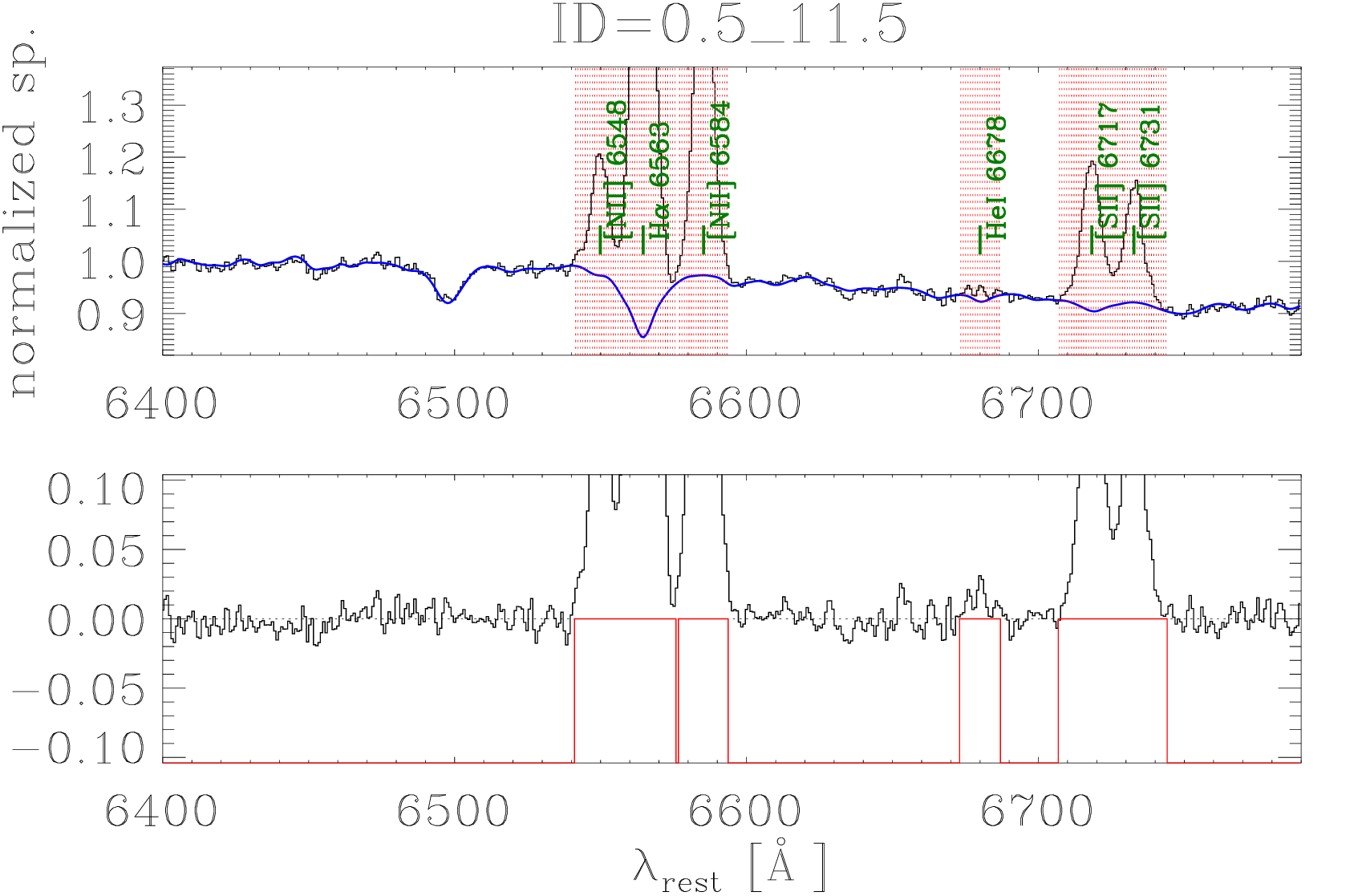}\\
  \vspace{0.2cm} 
     \caption{Stellar continuum fit and subtraction over two wavelength ranges including the 
	[OIII]$\lambda5007$ emission line ({\it top panels}) and the H$\alpha$+[NII] emission lines ({\it bottom panels}), performed
	for two galaxy stacks: ID=-2.0\_8.5 ({\it left}) and ID=0.5\_11.5 ({\it right}). 
     In each plot, the upper panel shows the stacked spectrum (black) and
     the best fit to the stellar continuum (blue), while the lower panel shows the residual
     spectrum. The red shaded regions mark the
     spectral windows containing nebular features and therefore masked from the stellar continuum fit.}
   \label{fig:consub}
\end{figure*}

The subtraction of the stellar continuum emission from the stacked spectra
is of crucial importance for this study, which is aimed at
detecting faint wings of nebular emission lines. For the purpose of stellar subtraction, we
select as stellar templates a large set of 
synthetic spectra of single stellar populations (SSPs) computed with 
the P\'EGASE-HR v3.0 code\footnote{\texttt{http://www2.iap.fr/pegase/pegasehr/}} \citep{LeBorgne+04}, 
with both low metallicities
($\mathcal{Z}=0.004$) and solar metallicities ($\mathcal{Z}=0.02$). The P\'EGASE-HR spectral templates cover
the wavelength range $\lambda\in[4000, 6800]$~\AA{}, with a spectral resolving power of
$R=10,000$ at $\lambda= 5500$~\AA{}, and are computed for a wide range of ages
(from 10 Myr to 2 Gyr), by assuming a Salpeter initial mass function (IMF, \citealt{Salpeter55}).

We fit the P\'EGASE-HR templates to the nebular line-free regions of the composite galaxy spectra by using
the IDL implementation of the penalised pixel-fitting (pPXF) method developed by \cite{Cappellari+Emsellem04}.
The stellar continuum fitting and subtraction is performed in two wavelength
ranges, i.e. $\lambda\in[4800, 5100]$~\AA{} and 
 $\lambda\in[6200, 6790]$~\AA{}, selected to include the [OIII]$\lambda$5007
 and the H$\alpha$ +[NII]$\lambda\lambda$6548,6583 nebular emission lines,
 which are the targets of this study. 
The nebular emission lines and ISM absorption features are masked from the 
stellar fit to prevent it from being affected by non-stellar features.
When the fit is provided with a large enough number of stellar features in the non-masked portion
of the spectrum (which is always true in the spectral region we are interested in),
this procedure can accurately determine the stellar contamination under the masked 
nebular lines.
Figure~\ref{fig:consub} shows the stellar continuum fit and subtraction
performed for two different stacks. We note that the pPXF procedure, 
 together with the high resolution P\'EGASE-HR templates, produces an excellent fit
to the stellar continuum underlying the nebular features in the galaxy stacks.

\subsection{SDSS spectral instrumental profile}\label{sec:instr_prof}

\begin{figure}[tbp]
\centering
    \includegraphics[clip=true,trim=1.2cm 2cm 0cm 3cm,angle=90,width=\columnwidth]{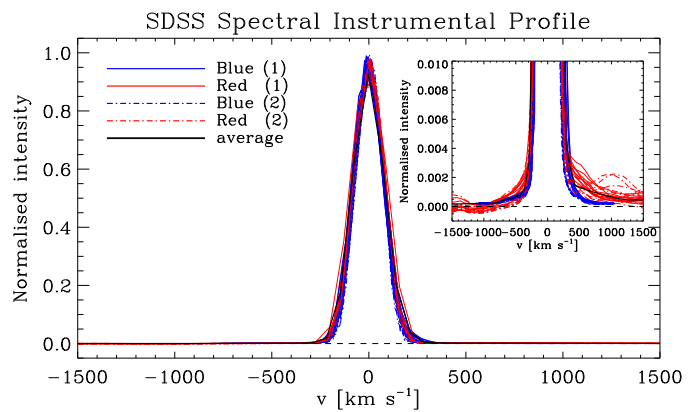}\\
     \caption{Instrumental profile of the SDSS spectrographs, estimated 
     from bright, non-blended emission lines in the arc lamp images, selected within the wavelength range of interest for this study
     (i.e. HgI $\lambda$5460.753~\AA{} for the blue cameras and NeI $\lambda$6598.953~\AA{} for the red cameras). 
     For each camera (blue/red) of the two spectrographs (1/2), we overplotted
     ten representative profiles from as many different frames (corresponding to different plates) obtained by averaging the arc lamp spectra
     over all 320 fibres. The black solid line corresponds to the mean profile.     
     The inset highlights the presence of low-level instrumental wings (corresponding to $\lesssim3/1000$ of the peak), 
     which are more prominent for the red cameras.
     }
   \label{fig:res}
\end{figure}

Our investigation of the kinematics of the ionised gas in local star forming galaxies, as traced by the [OIII]$\lambda$5007
and H$\alpha$+[NII]$\lambda\lambda$6548,6583 nebular emission lines, is aimed at finding broad wings
of the LoSVD of the ionised gas that may be signature of outflows. 
This is done by comparing, in each galaxy stack, the LoSVD of the gas with the LoSVD of the stars.
As it will be explained in $\S$~\ref{sec:losvd_stars}, the LoSVD of the stars is derived by fitting the stellar
continuum around $\lambda=5500$~\AA{}, i.e. in a spectral region where the stellar fit
produces the best results thanks to the large number of stellar absorption lines and the scarcity of nebular (ISM) features.
However, in order to extract the real kinematics of stars and gas from the observed line profiles (either from
the nebular emission lines, in the case of the gas, or from the stellar features seen in absorption, in the case of the stars),
the LoSVDs of gas and stars must be deconvolved from the instrumental spectral profile. 
Therefore, it is of primary importance for our study to determine the instrumental profile of the SDSS spectrographs.

The presence of scattered light in the spectrometer may affect the spectral 
instrumental profile, resulting in extended wings. For the purpose of our investigation, we use
the arc lamp calibration spectra to retrieve and 
analyse the spectral instrumental profile of 
the original SDSS spectrographs (adopted for SDSS-II). 
The average SDSS instrumental profile profile, shown in Fig.~\ref{fig:res}, is impressively clean from wings down to
$\lesssim3/1000$ of the peak, where some low-level instrumental wings start to appear. This effect is however negligible for our study, since it clearly cannot
account for the much stronger broad (and asymmetric) line wings that we detect in our galaxy stacks, as it will be shown in $\S$~\ref{sec:results}.

\cite{Smee+13} analyse, in their Figure 35, the spectral resolving power ($R(\lambda) = \lambda/(2.35*\sigma_{\lambda})$) of 
the original SDSS spectrographs as a function of wavelength. The resolving power of SDSS is quite difficult to constrain 
between $\lambda=5000$~\AA{} and $\lambda=6000$~\AA{}, because in this spectral region $R$ does not depend monotonically on wavelength,
due to the transition from the blue to the red camera of the spectrographs (and some overlap between the two). In this 
study we adopt $R=1850$ (corresponding to ${\rm \sigma_v = 69~km~s^{-1}}$), which
is the average spectral resolving power inferred from the arc lamp images (Figure~\ref{fig:res}) and 
is on the lower side of (but still well within) the nominal range quoted in the literature for the SDSS spectrographs
 (i.e. $R\in[1800, 2000]$).

\subsection{The LoSVD of the stars}\label{sec:losvd_stars}

\begin{figure*}[tbp]
	\centering
    \includegraphics[width=1.4\columnwidth]{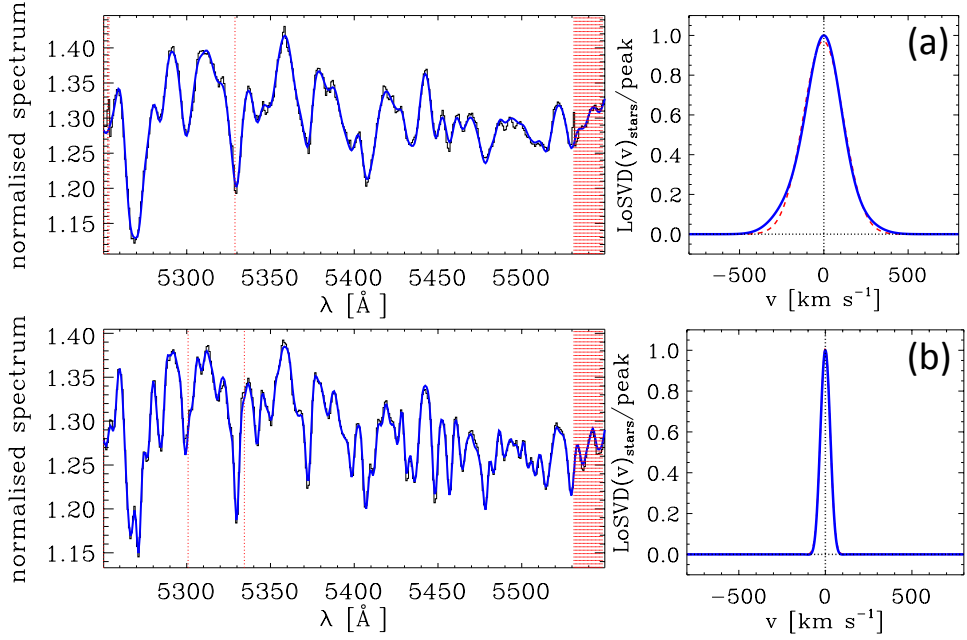}\\
   \caption{Stellar continuum fit over the range 
   		${\rm \lambda \in [5250, 5550]~\AA{}}$ ({\it left}) and resulting stellar LoSVD 
   		({\it right}) derived for two galaxy stacks, i.e. $\rm ID=-0.5\_11.0$ ({\it panel a}) 
		and $\rm ID=-1.5\_9.5$ ({\it panel b}). 
		In the left panels, the stacked spectrum is plotted in black, and the best fit to 
		the stellar continuum is in blue. 
		The red dashed regions mark the spectral windows masked from the stellar
		continuum fit because of the presence of nebular features.
		In the right panels, the 
		red dashed line indicates the LoSVD profile obtained
		by using, in the pPXF procedure, a simple Gaussian function to model the stellar features,
		while the blue solid line shows the stellar LoSVD profile obtained by allowing the use of Hermite polynomials
		to account for possible asymmetries of 
		the stellar absorption features \citep{Cappellari+Emsellem04}.
		The stellar LoSVD obtained for the stack $\rm ID=-0.5\_11.0$ shows a clear blue asymmetry.
		The fit is excellent (${\chi_{\rm red}^2}$ = 0.8) for both stacks.}
   \label{fig:appendixB}
\end{figure*}

The pPXF method developed by \cite{Cappellari+Emsellem04} allows us not only to
accurately fit the stellar continuum and subtract it from
the stacked spectra of galaxies, but also to extract the stellar kinematics.
The pPXF procedure adopts a Gauss-Hermite parametrisation 
for the stellar line-of-sight velocity distributions, which accounts for
possible asymmetries in the stellar absorption line profiles.

To infer the stellar LoSVDs, we first convolve the stellar templates (i.e. the P\'EGASE-HR synthetic spectra)
with the SDSS instrumental spectral profile obtained in $\S$~\ref{sec:instr_prof}. When performing
the stellar continuum fit by using these stellar templates (i.e. the ones convolved with the instrumental profile), the pPXF
procedure yields the ``real'' stellar LoSVD, i.e. the stellar LoSVD deconvolved from the instrumental profile.
As already mentioned in $\S$~\ref{sec:instr_prof}, in order to extract the stellar kinematics, it is preferable
to perform the continuum fitting around $\lambda\sim5500$~\AA{}, because this spectral region is abundant in stellar absorption
features and relatively free of nebular emission lines. For this reason, to infer the stellar LoSVDs, we fit the stellar continuum
in wavelength ranges selected between $\lambda=5100$~\AA{} and $\lambda=5850$~\AA{} or 
between $\lambda=6050$~\AA{} and $\lambda=6250$~\AA{}. The range $\lambda\in[5850, 6050]$~\AA{} is avoided because
of the presence of the NaID $\lambda\lambda$5890,5896~\AA{} resonance absorption doublet, which is contaminated by the galaxy ISM.

An example of continuum fitting aimed at extracting the stellar kinematics is reported in Fig.~\ref{fig:appendixB} for two different galaxy stacks,
i.e. $\rm ID=-0.5\_11.0$ and $\rm ID=-1.5\_9.5$. We note that the $\rm ID=-0.5\_11.0$ stack 
shows a significant asymmetry in its stellar LoSVD. The presence of
asymmetries in the stellar LoSVDs has been evidenced also in 
other stacks and will be discussed 
in $\S$~\ref{sec:asym}.

\subsection{The LoSVD of the ionised gas}\label{sec:losvd_fit}

\begin{figure*}[tbp]
	\centering
    \includegraphics[clip=true,trim=1.6cm 2.cm 0.2cm 3cm,angle=90,width=\columnwidth]{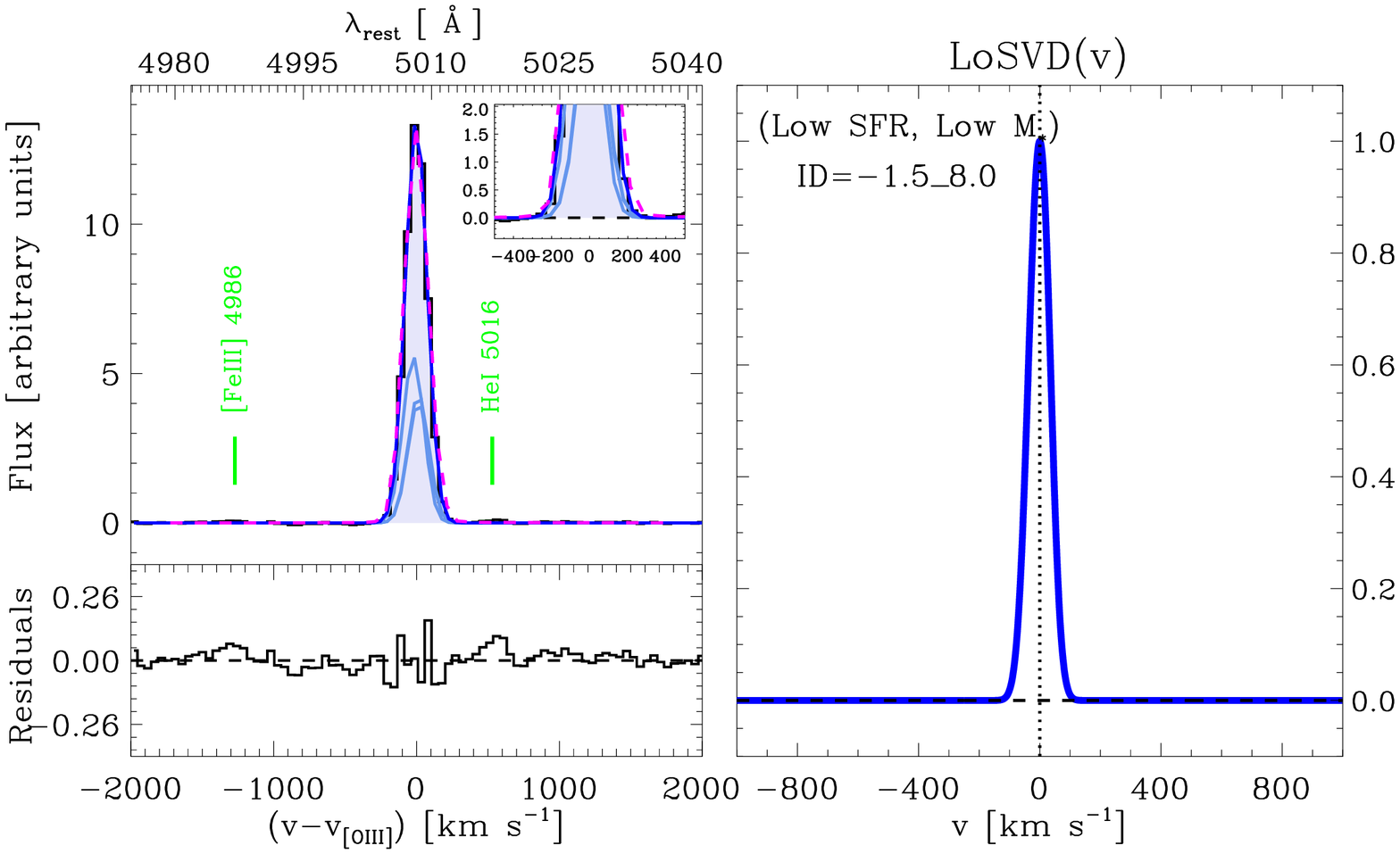}\quad
    \includegraphics[clip=true,trim=1.6cm 2.cm 0.2cm 3cm,angle=90,width=\columnwidth]{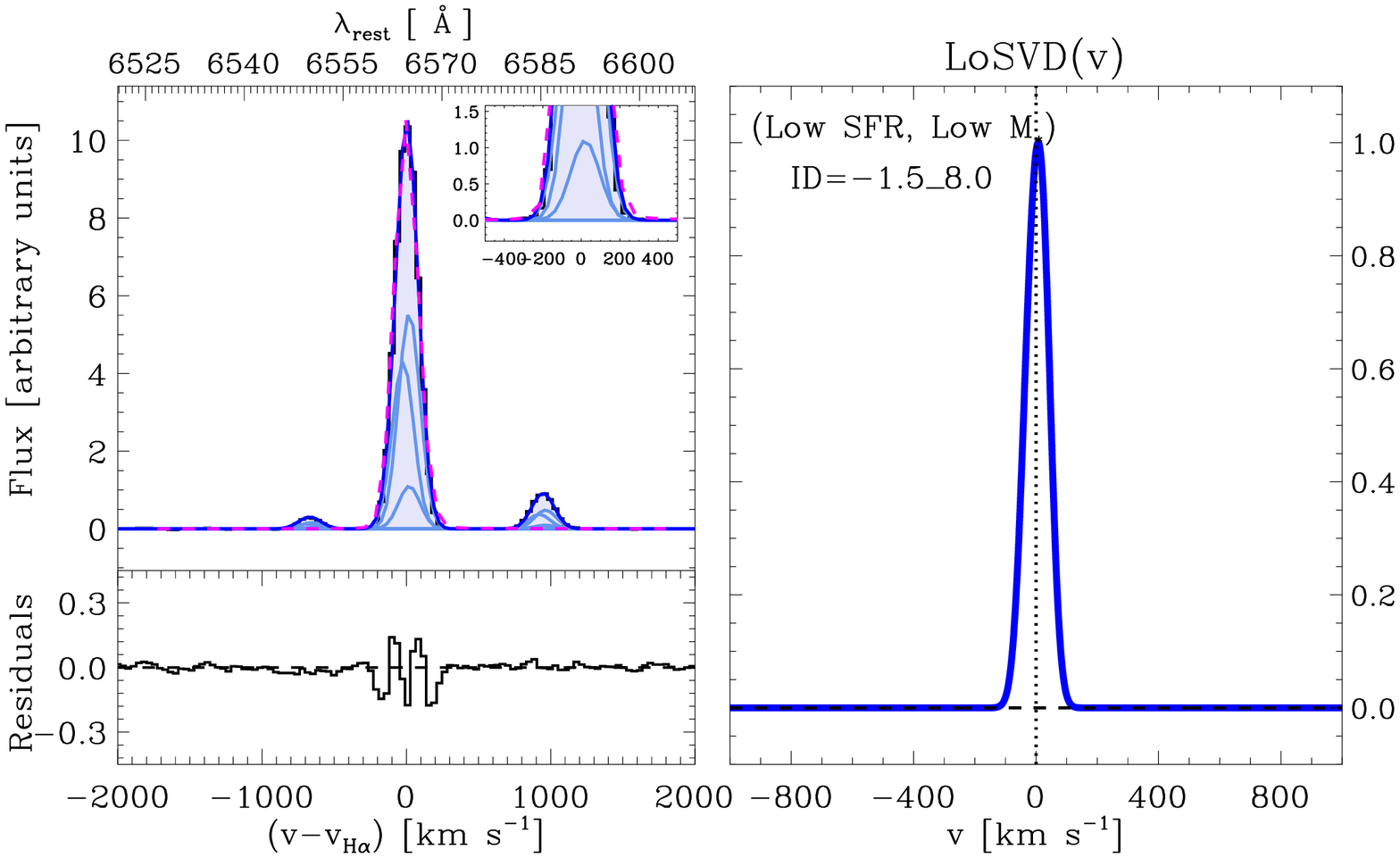}\\
    \includegraphics[clip=true,trim=1.6cm 2.cm 0.2cm 3cm,angle=90,width=\columnwidth]{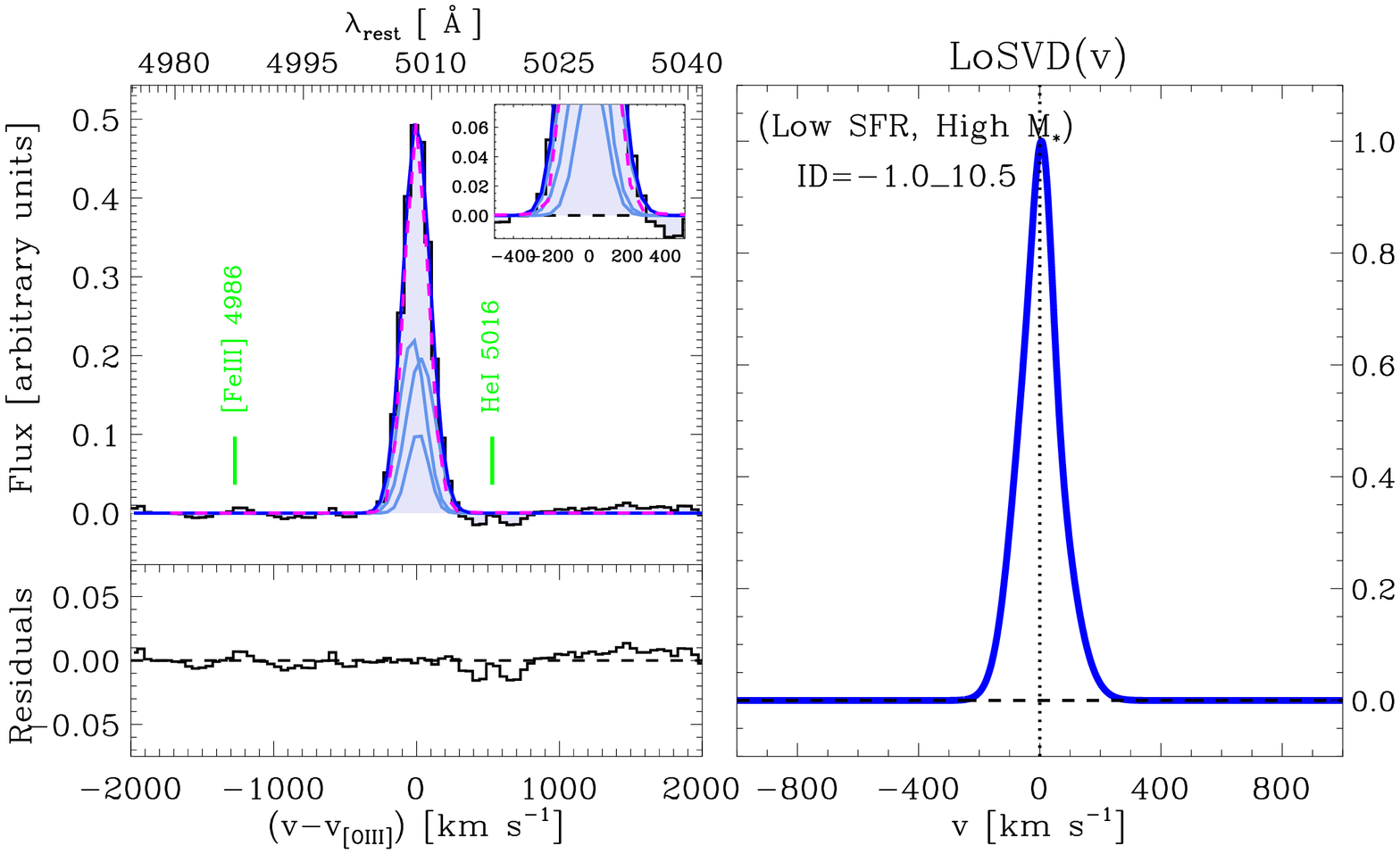}\quad
    \includegraphics[clip=true,trim=1.6cm 2.cm 0.2cm 3cm,angle=90,width=\columnwidth]{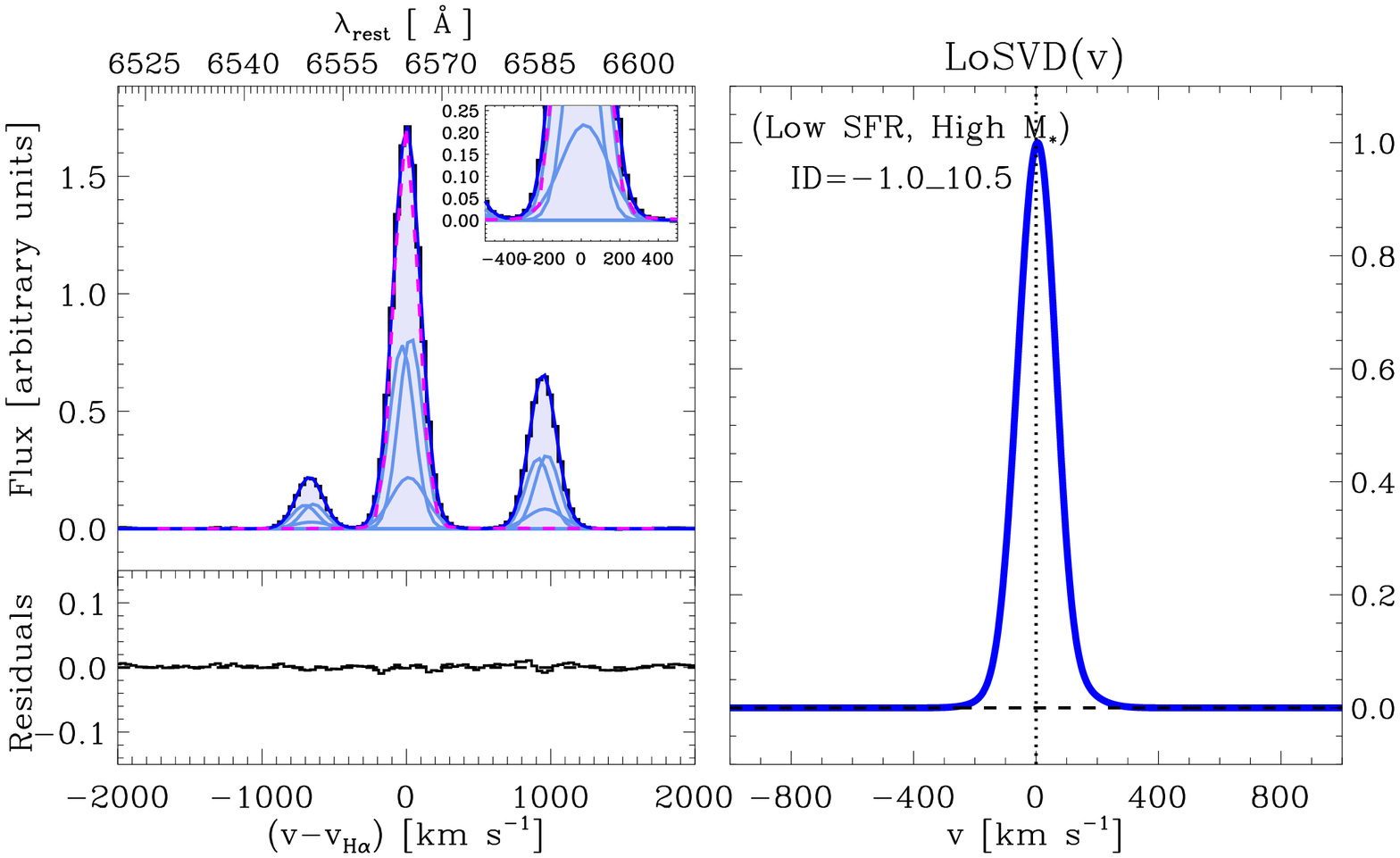}\\
    \includegraphics[clip=true,trim=1.6cm 2.cm 0.2cm 3cm,angle=90,width=\columnwidth]{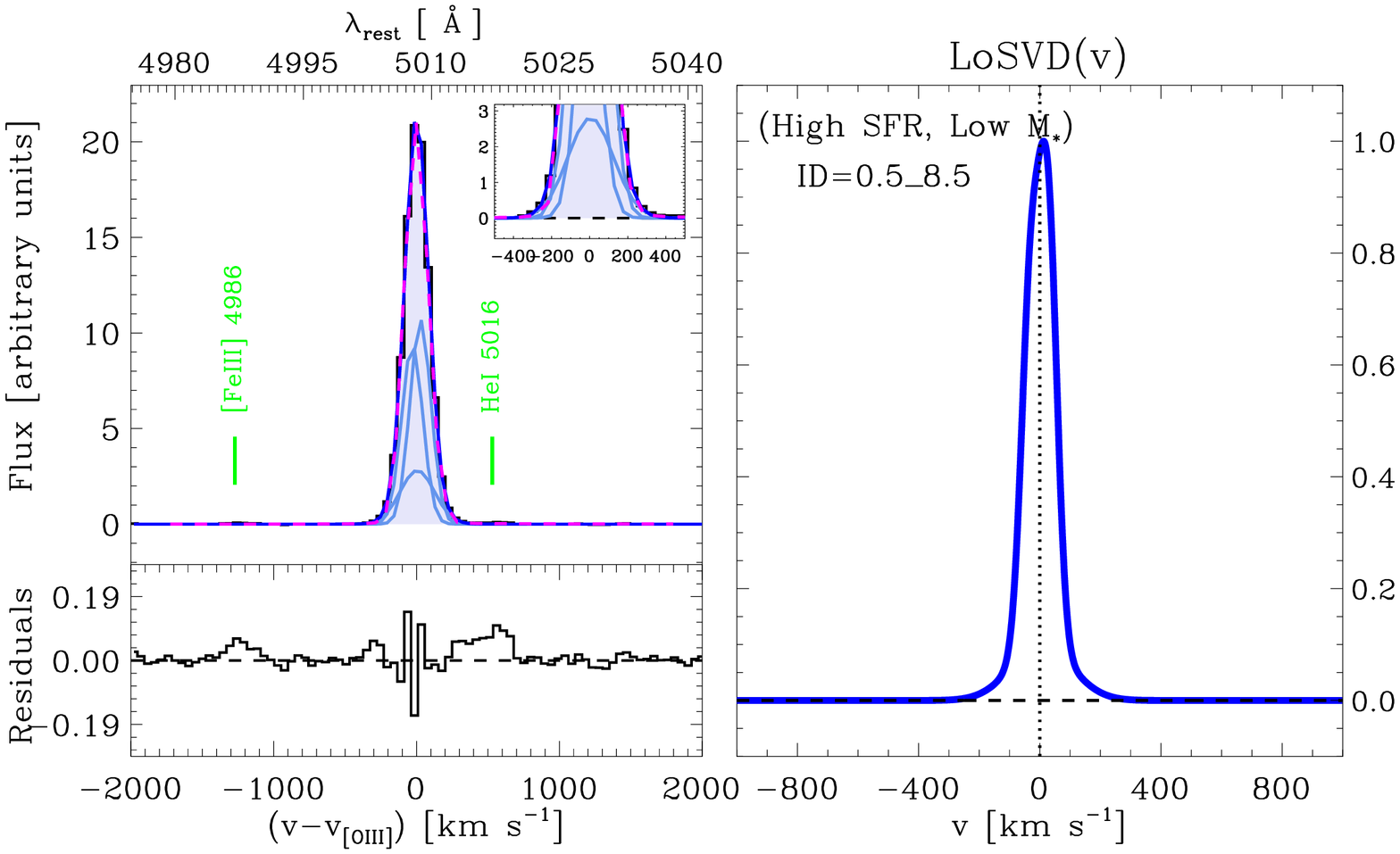}\quad
    \includegraphics[clip=true,trim=1.6cm 2.cm 0.2cm 3cm,angle=90,width=\columnwidth]{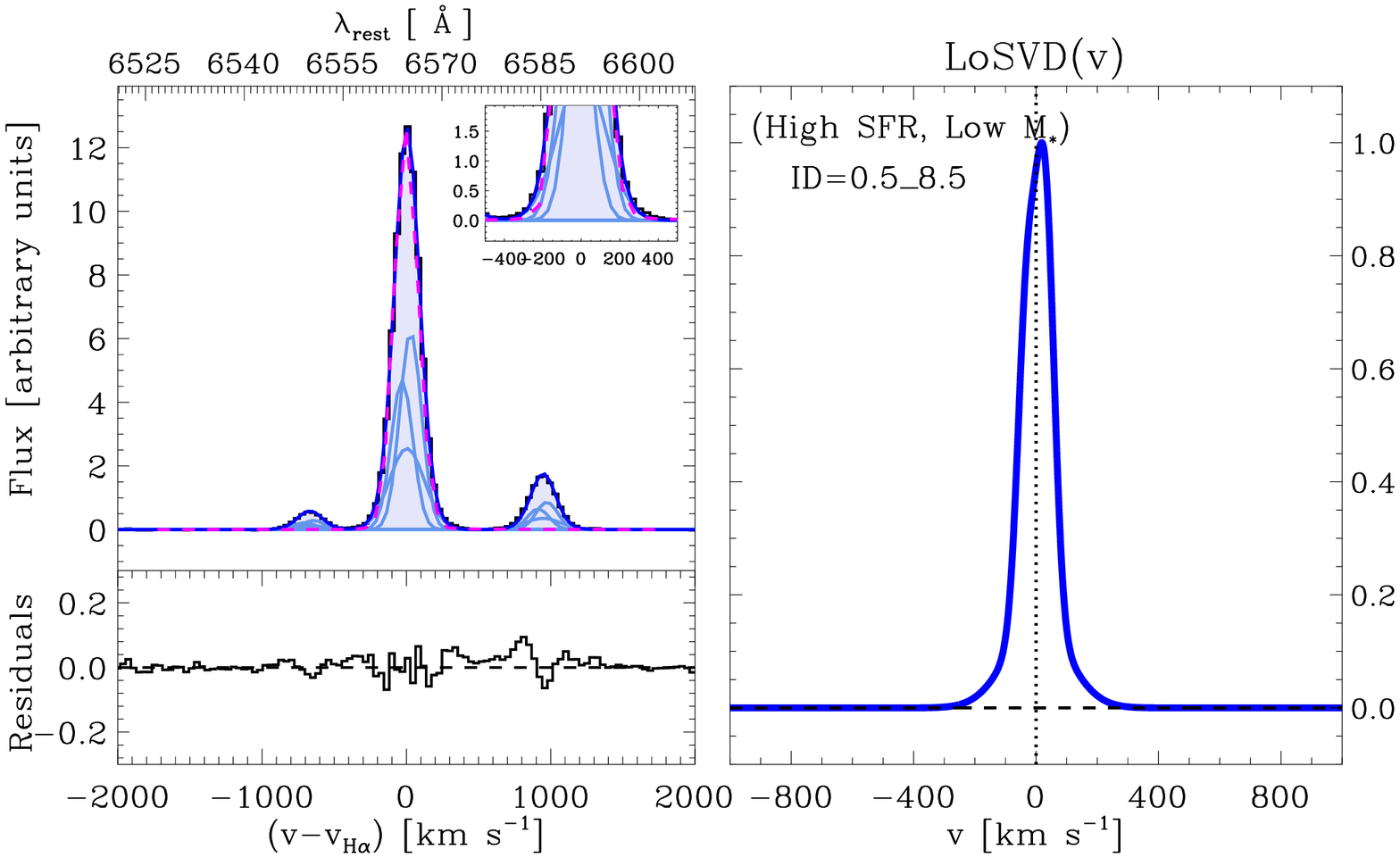}\\
    \includegraphics[clip=true,trim=1.6cm 2.cm 0.2cm 3cm,angle=90,width=\columnwidth]{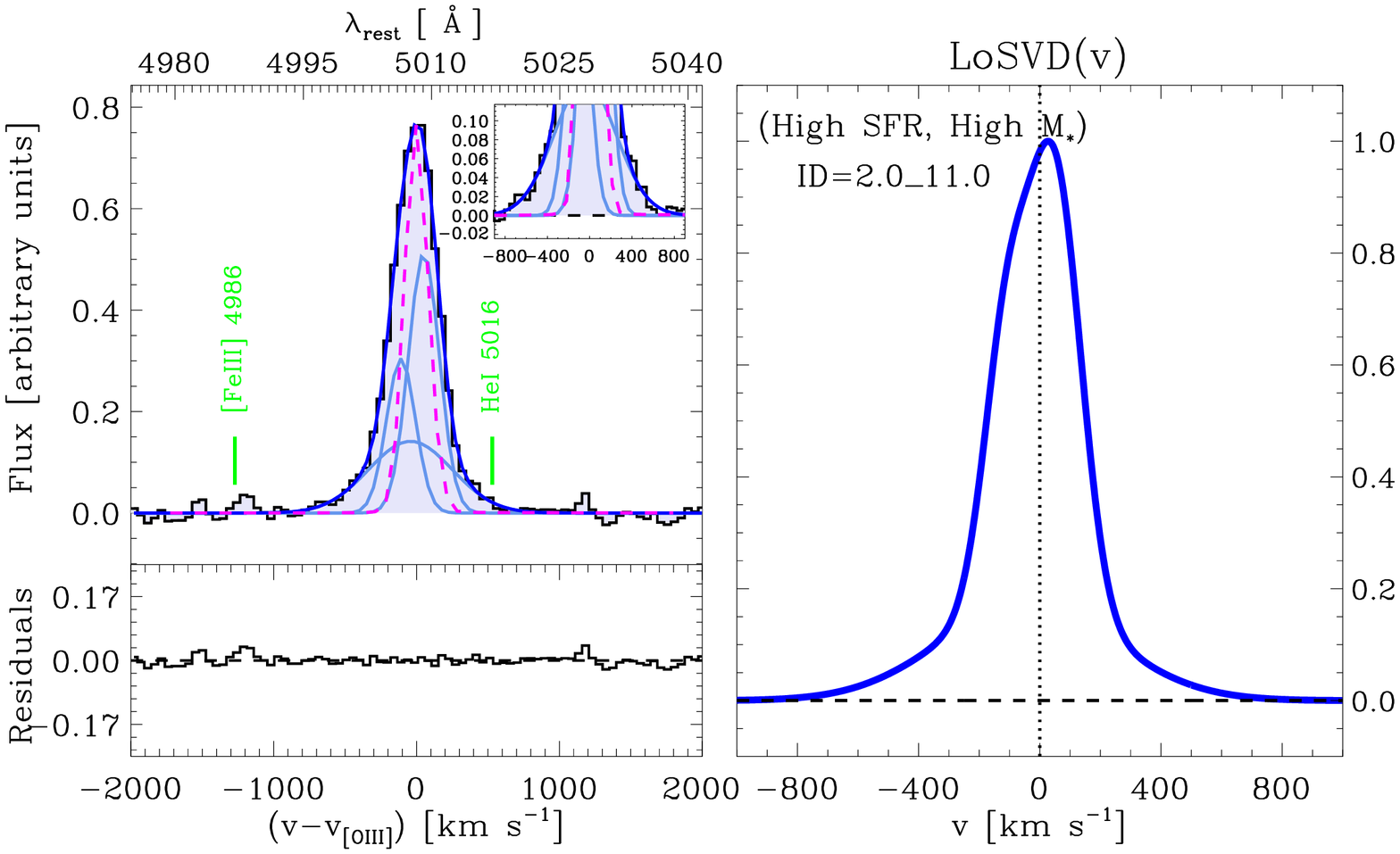}\quad
    \includegraphics[clip=true,trim=1.6cm 2.cm 0.2cm 3cm,angle=90,width=\columnwidth]{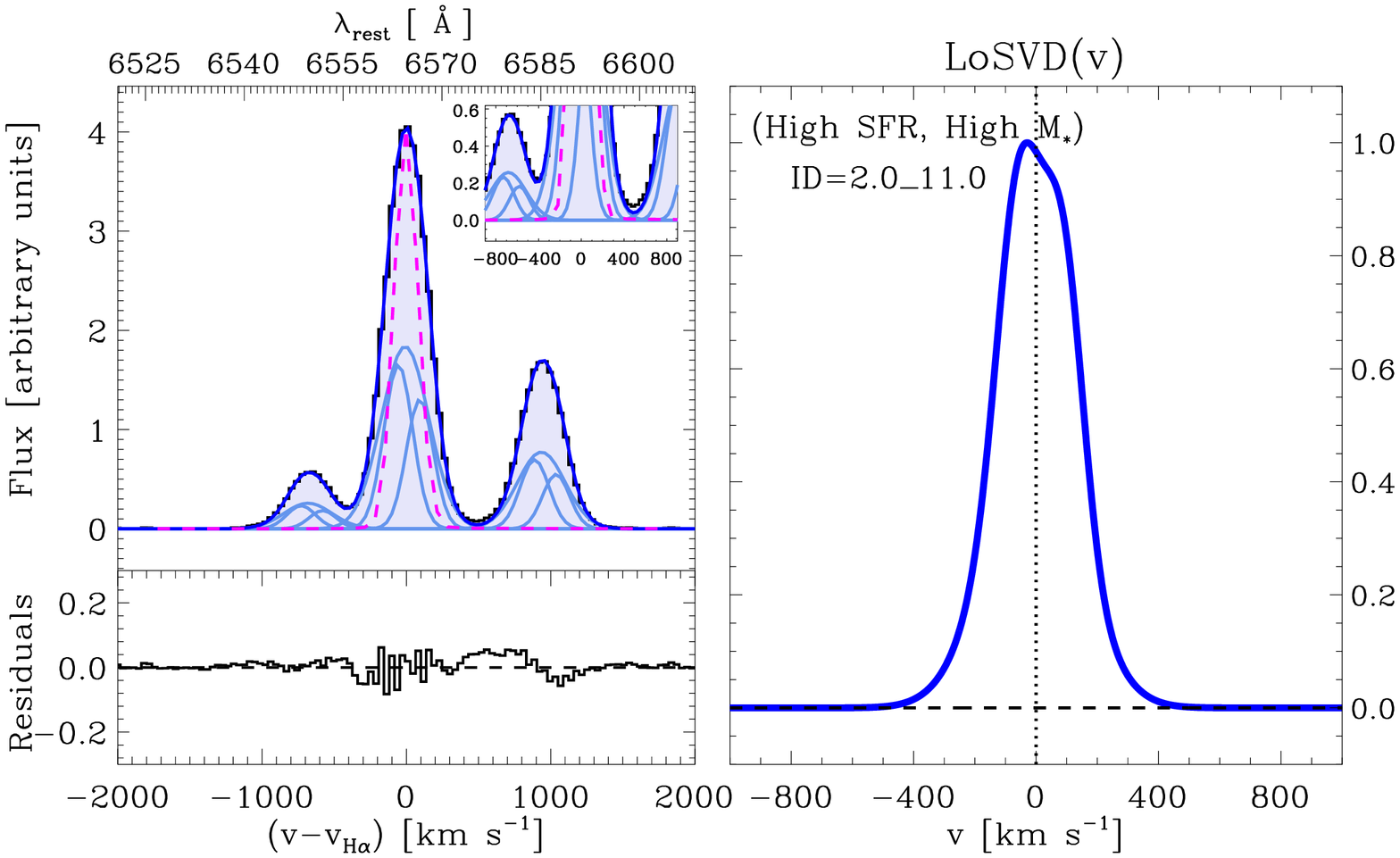}\\
    \caption{Examples of spectral fits to the [OIII]$\lambda$5007 ({\it left column}) and H$\alpha$+[NII] ({\it right column}) emission line
     		profiles for four representative galaxy bins with different SFR and M$_*$. From top to bottom: ID=-1.5\,8.0 (low SFR and low M$_*$),
		ID=-1.0\,10.5 (low SFR and high M$_*$), ID=0.5\,8.5 (high SFR and low M$_*$), and ID=2.0\,11.0 (high SFR and high M$_*$).
		In each plot, the {\it left} panel shows the fit to the data (solid curves, the total fit is in blue, the multiple Gaussian components in a 
		lighter shade of blue). For comparison, we have overplotted to the data the average SDSS instrumental resolution profile
		(magenta dashed curve, from Fig.~\ref{fig:res}). In the {\it right} panel of each plot we show the final LoSVD of 
		the ionised gas resulting from the fit, hence deconvolved from the SDSS instrumental profile (details in $\S$~\ref{sec:losvd_fit}).
		}
   \label{fig:fit_1}
\end{figure*}

\begin{figure*}[tbp]
	\centering
    	\includegraphics[clip=true,trim=1.6cm 2.cm 0.2cm 3cm,angle=90,width=\columnwidth]{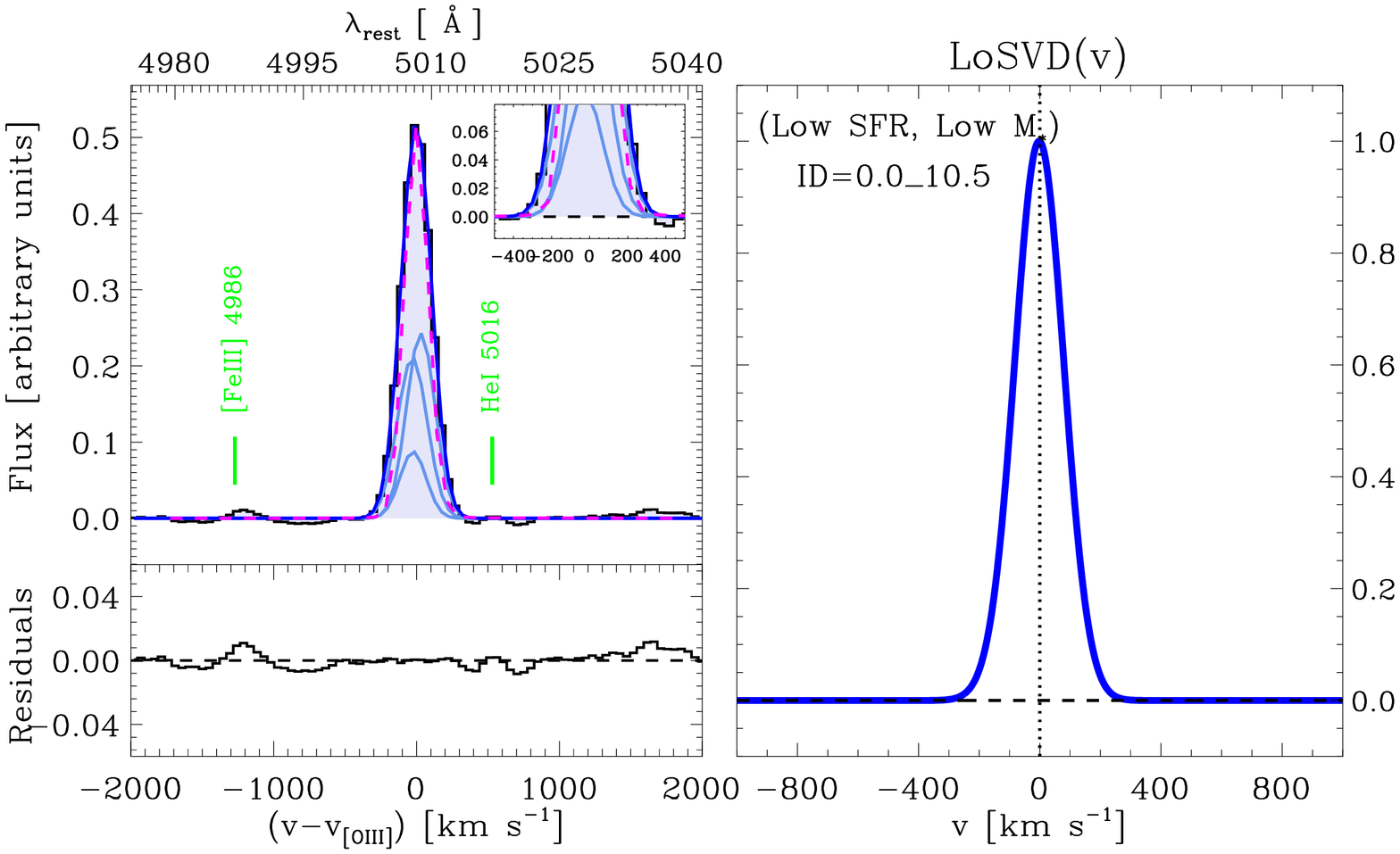}\quad
    	\includegraphics[clip=true,trim=1.6cm 2.cm 0.2cm 3cm,angle=90,width=\columnwidth]{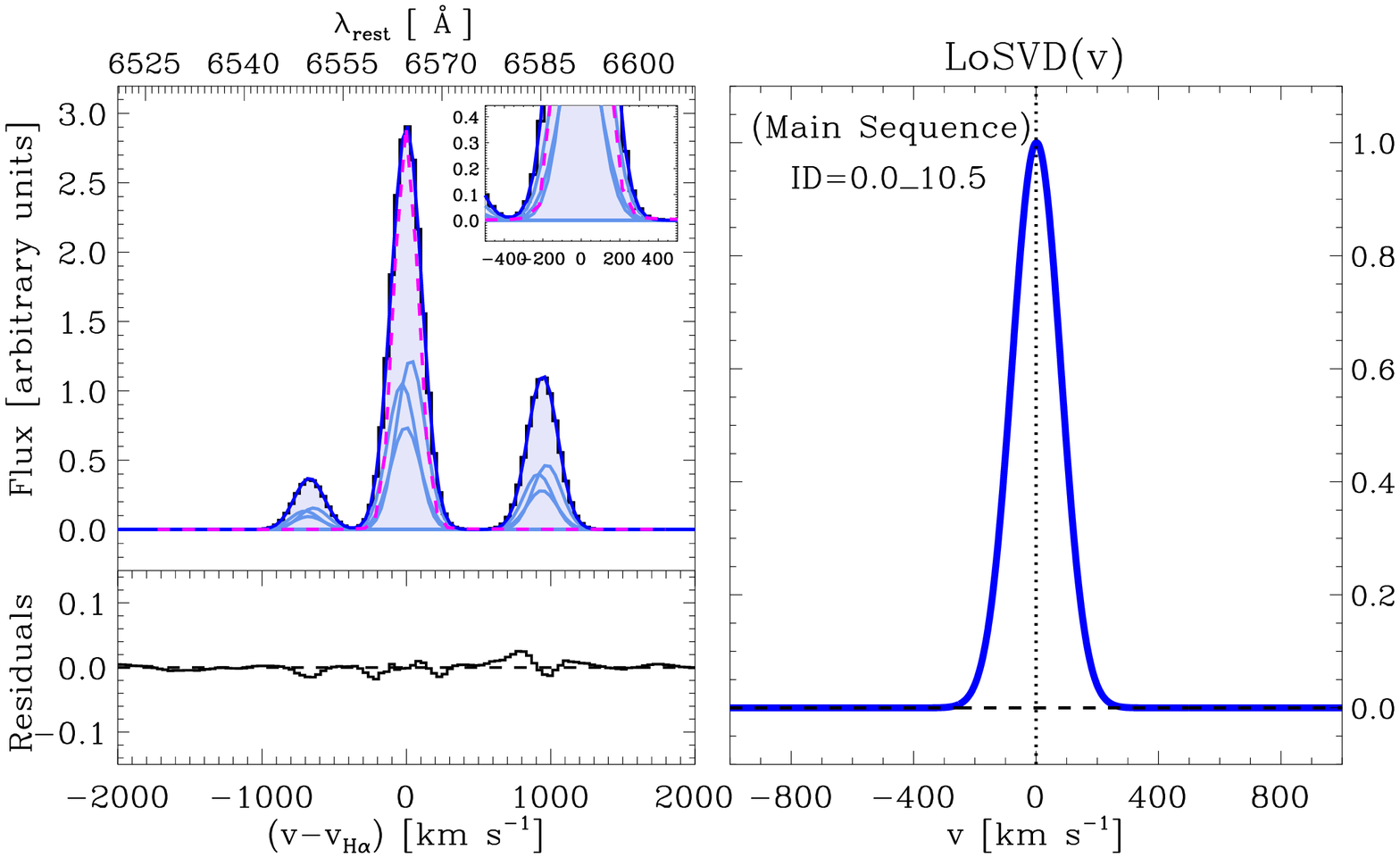}\\
     \caption{Same as Fig.~\ref{fig:fit_1} but for a typical bin on the MS, with intermediate SFR and M$_*$  (ID=0.0\,10.5).}
   \label{fig:fit_2}
\end{figure*}

To infer the LoSVD of the ionised gas we fit, for each stack, the 
[OIII]$\lambda$5007, H$\alpha$ and [NII]$\lambda\lambda$6548,6583 emission lines
with a function that results from the convolution
of the SDSS instrumental profile, $\mathcal{R}(v)$, with another function $\mathcal{L}(v)$, which by
definition corresponds to the ``real'' LoSVD of ionised gas. 
To account for potential broad wings of the LoSVD, possibly tracing
outflows of ionised gas, $\mathcal{L}(v)$ is parametrised by the sum of
three Gaussians, $l_k(v)$ ($k=$1,2,3), each centred at a velocity $v_{0,k}$, with 
area $a_k$ and standard deviation $\sigma_k$, i.e. 
\begin{equation}\label{eq:losvd_formula}
\mathcal{L}(v) = \sum_{k=1}^{3}  l_{k} (v) = \sum_{k=1}^{3} a_k\,n_{k}\,e^{-(v-v_{0,k})^2/2\sigma_k^2},
\end{equation}
where $n_{k}= 1/\sqrt{2\pi \sigma_{k}^2}$ is the normalisation factor.
Therefore, the ``observed'' LoSVD, $F(v)$, resulting from the convolution of
the ``real'' LoSVD ($\mathcal{L}(v)$) and the SDSS spectral resolution profile 
($\mathcal{R}(v)$), is given by:
\begin{equation}
{F(v) = \mathcal{L}(v) \ast \mathcal{R}(v) =  \sum_{k=1}^{3} l_{k} (v) \ast \mathcal{R}(v) =  \sum_{k=1}^{3} f_{k} (v)},
\end{equation}
where we have used the distributive property of the convolution (indicated by the 
symbol $\ast$) and we have defined $f_{k} (v) \equiv  l_{k} (v) \ast \mathcal{R}(v)$.
As a result, since both $l_{k}(v)$ and $\mathcal{R}(v)$ are Gaussians, $f_{k}(v)$ has the form:
\begin{equation}
f_{k} (v) = a_k\,N_{k} \,e^{-(v-v_{0,k})^2/2(\sigma_k^2+\sigma_R^2)},
\end{equation}
with ${N_{k}}= 1/\sqrt{2\pi (\sigma_{k}^2+\sigma_R^2)}$.

 An emission line centred at a velocity ${v_i}$, 
corresponding to a central wavelength $\lambda_i$, 
 will be fitted with a final function of the form:
\begin{equation}
{\mathcal{F}_i(\lambda) = F(v) \ast [\varphi_i~\delta(v-v_i)] = F(v)  \ast \Biggl[\varphi_i~\delta \Bigl( c\frac{\lambda-\lambda_i}{\lambda_i} \Bigr) \Biggl] },
\label{eq:function_fit}\end{equation}
that is, the observed LoSVD, $F(v)$, convolved with a Dirac delta function centred at the vacuum wavelength
of the line ${\lambda_i}$
and whose integral ${\varphi_i}$ is proportional to the line flux.
Following Eq.~\ref{eq:function_fit}, the final function ${\mathcal{F}_{[OIII]}(\lambda)}$ that we
employ to fit
the [OIII]$\lambda$5007 emission line profile in the composite spectra is:
\begin{multline}
{\mathcal{F}_{[OIII]}(\lambda)} = \\ {\varphi_{[OIII]} \sum_{k=1}^{3} a_k N_k\,e^{-\bigl[\frac{c(\lambda-\lambda_{[OIII]})}{\lambda_{[OIII]}}-v_{0,k}\bigr]^2\bigl/2\bigl(\sigma_k^2+\sigma_R^2\bigr)}}.
\end{multline}
Similarly, we fit the H$\alpha$+[NII] emission lines with a function of the form:
\begin{multline*}
{\mathcal{F}_{tot}(\lambda) = \mathcal{F}_{H\alpha}(\lambda)+\mathcal{F}_{[NII]6548}(\lambda)+\mathcal{F}_{[NII]6583}(\lambda)} = 
 \\  {\varphi_{H\alpha}\sum_{k=1}^{3} a_k N_k\,e^{-\bigl[\frac{c(\lambda-\lambda_{H\alpha})}{\lambda_{H\alpha}}-v_{0,k}\bigr]^2\bigl/2\bigl(\sigma_k^2+\sigma_R^2\bigr)}} + \\
  {\varphi_{[NII]}\sum_{k=1}^{3} a_k N_k\,e^{-\bigl[\frac{c(\lambda-\lambda_{H\alpha})}{\lambda_{H\alpha}} -  \frac{c(\lambda_{[NII]}-\lambda_{H\alpha})}{\lambda_{H\alpha}}- v_{0,k}\bigr]^2\bigl/2\bigl(\sigma_k^2+\sigma_R^2\bigr)}} + \\
   {3\varphi_{[NII]}\sum_{k=1}^{3} a_k N_k\,e^{-\bigl[\frac{c(\lambda-\lambda_{H\alpha})}{\lambda_{H\alpha}} -  \frac{c(\varrho\lambda_{[NII]}-\lambda_{H\alpha})}{\lambda_{H\alpha}}- v_{0,k}\bigr]^2\bigl/2\bigl(\sigma_k^2+\sigma_R^2\bigr)}}.
\end{multline*}
In this equation, [NII] indicates the bluest line of the [NII] doublet, i.e. the [NII]$\lambda$6548 line,
and $\varrho$ denotes the theoretical ratio between the rest-frame vacuum wavelengths of the [NII]$\lambda$6548 and
[NII]$\lambda$6583 lines as reported in the NIST database.

The kinematics (velocity and widths) of all components fitting the H$\alpha$+[NII]
complex is set to be equal for the three transitions. Such a constrained approach is necessary 
to separate the partially blended H$\alpha$ and
[NII] emission lines and it improves the quality and reliability of the fit, as pointed out by
\cite{Westmoquette+12}. Moreover this method is justified by the fact that in star forming galaxies
and in absence of strong shocks, the H$\alpha$ and [NII] emission lines arise from adjacent regions.

The fit, whose input parameters are only the instrumental resolution (${\sigma_R}$)
and the (vacuum) central wavelengths of the different transitions (${\lambda_i}$), provides
us with the relative line fluxes ${\varphi_i}$ and with other nine parameters, i.e.
$v_{0,k}$, $\sigma_k$ and
$a_k$ (with the index ${k = 1,2,3}$ denoting the three Gaussian components 
used to model the observed LoSVD) from which, using Eq.~\ref{eq:losvd_formula}, we retrieve
${\mathcal{L}(v)}$, i.e. the real LoSVD of the ionised gas.
Figures \ref{fig:fit_1} and \ref{fig:fit_2} show example fits to the [OIII]$\lambda$5007 and 
H$\alpha$+[NII] emission lines as well as the resulting LoSVDs, for five representative galaxy stacks,
having different SFR and M$_*$ properties (in particular, low and high SFR, low and high M$_*$, and
intermediate SFR and M$_*$). In Figs.~\ref{fig:fit_1} and \ref{fig:fit_2} 
we also show for comparison, over-plotted to the observed spectral emission line profiles, the 
average SDSS instrumental profile (from Fig.~\ref{fig:res}), to show that the
broad asymmetric wings we detected in a few of the stacks (e.g. ID=2.0\,11.0) are not instrumental.


\section{Results}\label{sec:results}

\begin{figure*}[p]
    \includegraphics[angle=90,height=0.40\textheight]{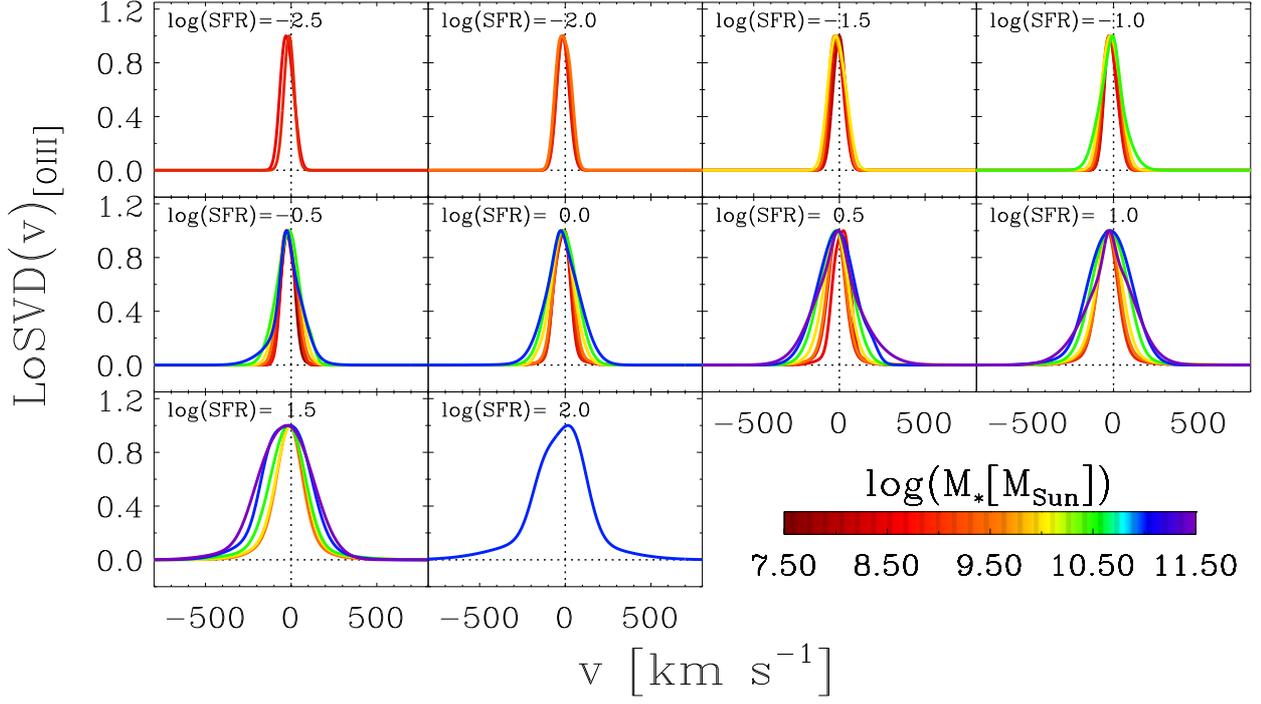}\\
     \caption{Line-of-sight velocity distribution of the ionised gas as traced by the
     		[OIII]$\lambda$5007 emission line. In each panel we have over-plotted the LoSVDs of galaxy bins identified
		by the same (average) SFR and by different (average) stellar masses, 
		colour-coded by their M$_{*}$.}
   \label{fig:losvd_oiii_2}
\end{figure*}

\begin{figure*}[p]
    \includegraphics[angle=90,height=0.40\textheight]{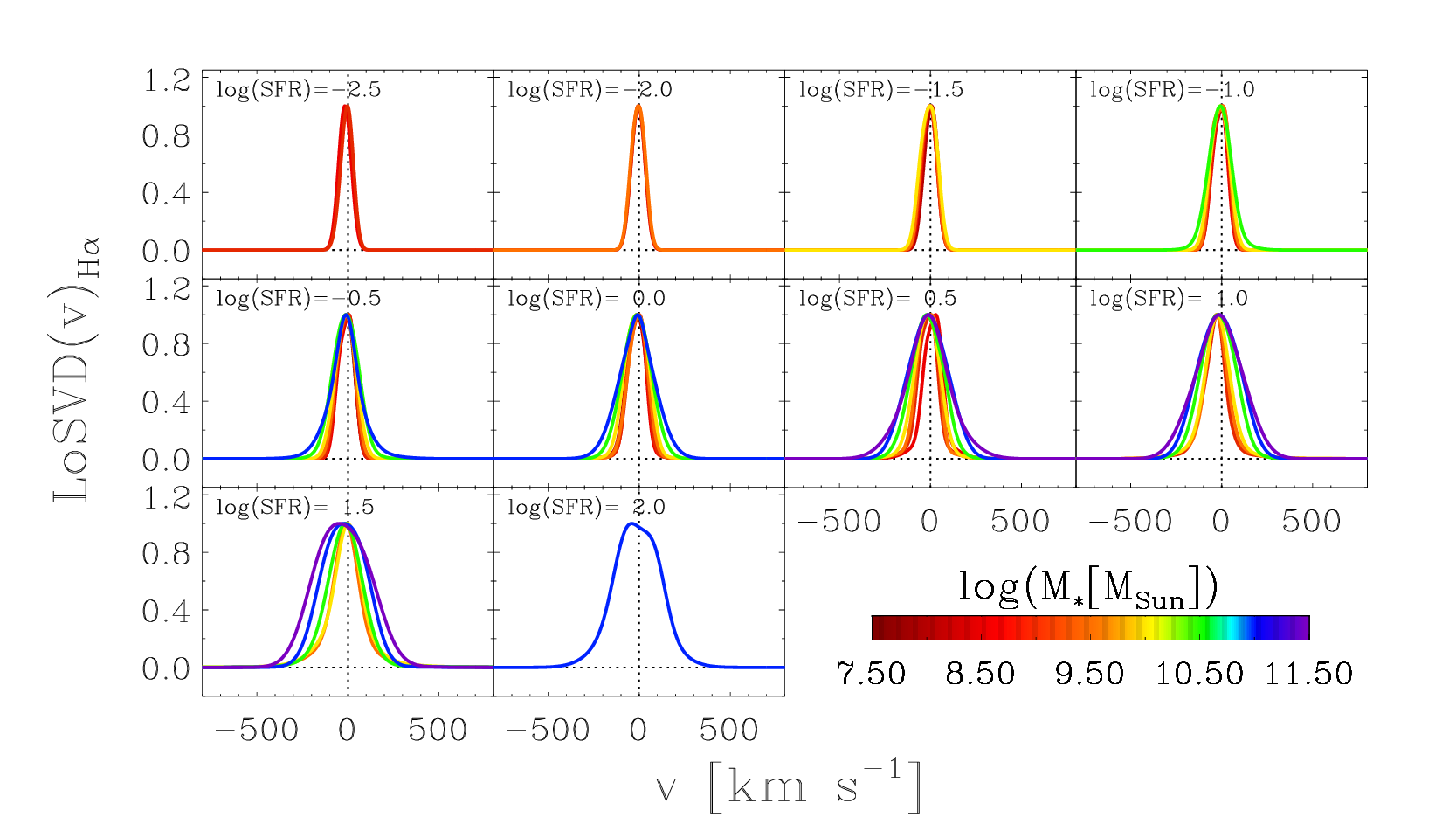}\\
     \caption{Line-of-sight velocity distribution of the ionised gas as traced by the
     		H$\alpha$+[NII] emission lines. In each panel we have over-plotted the LoSVDs of galaxy bins identified
		by the same (average) SFR and by different (average) stellar masses, 
		colour-coded by their M$_{*}$.}
   \label{fig:losvd_ha_2}
\end{figure*}

\begin{figure*}[p]
    \includegraphics[angle=90,height=0.40\textheight]{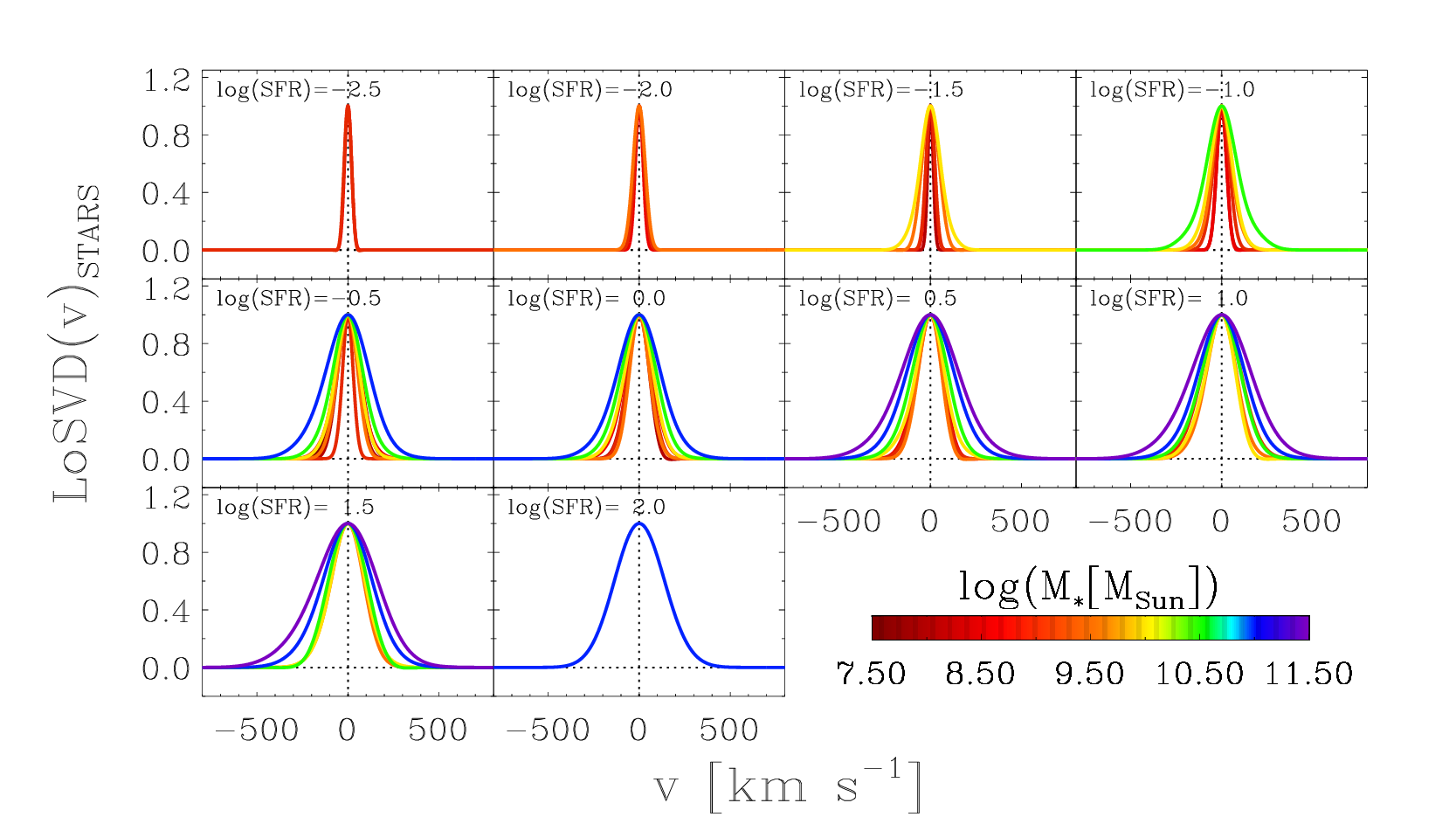}\\
     \caption{Line-of-sight velocity distribution of the stars. In each panel we have over-plotted the LoSVDs of galaxy bins identified
		by the same (average) SFR and by different (average) stellar masses, 
		colour-coded by their M$_{*}$.}
   \label{fig:losvd_stars_2}
\end{figure*}

\begin{figure*}[p]
    \includegraphics[angle=90,height=0.40\textheight]{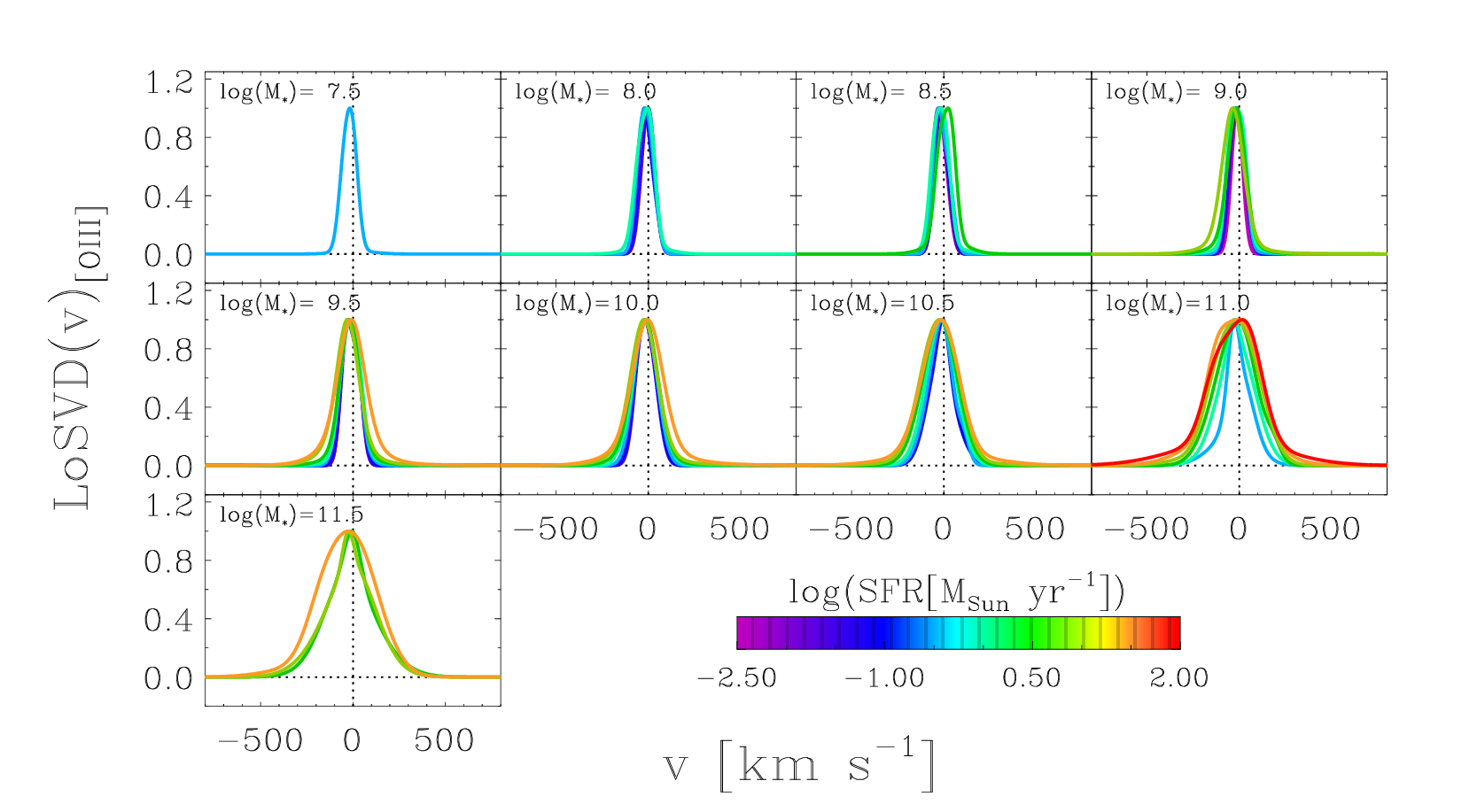}\\
     \caption{Line-of-sight velocity distribution of the ionised gas as traced by the
     		[OIII]$\lambda$5007 emission line. In each panel we have 
		over-plotted the LoSVDs of galaxy bins identified by the same (average) 
		M$_{*}$ and by different (average) SFRs, colour-coded by their SFR.}
   \label{fig:losvd_oiii_1}
\end{figure*}

\begin{figure*}[p]
    \includegraphics[angle=90,height=0.40\textheight]{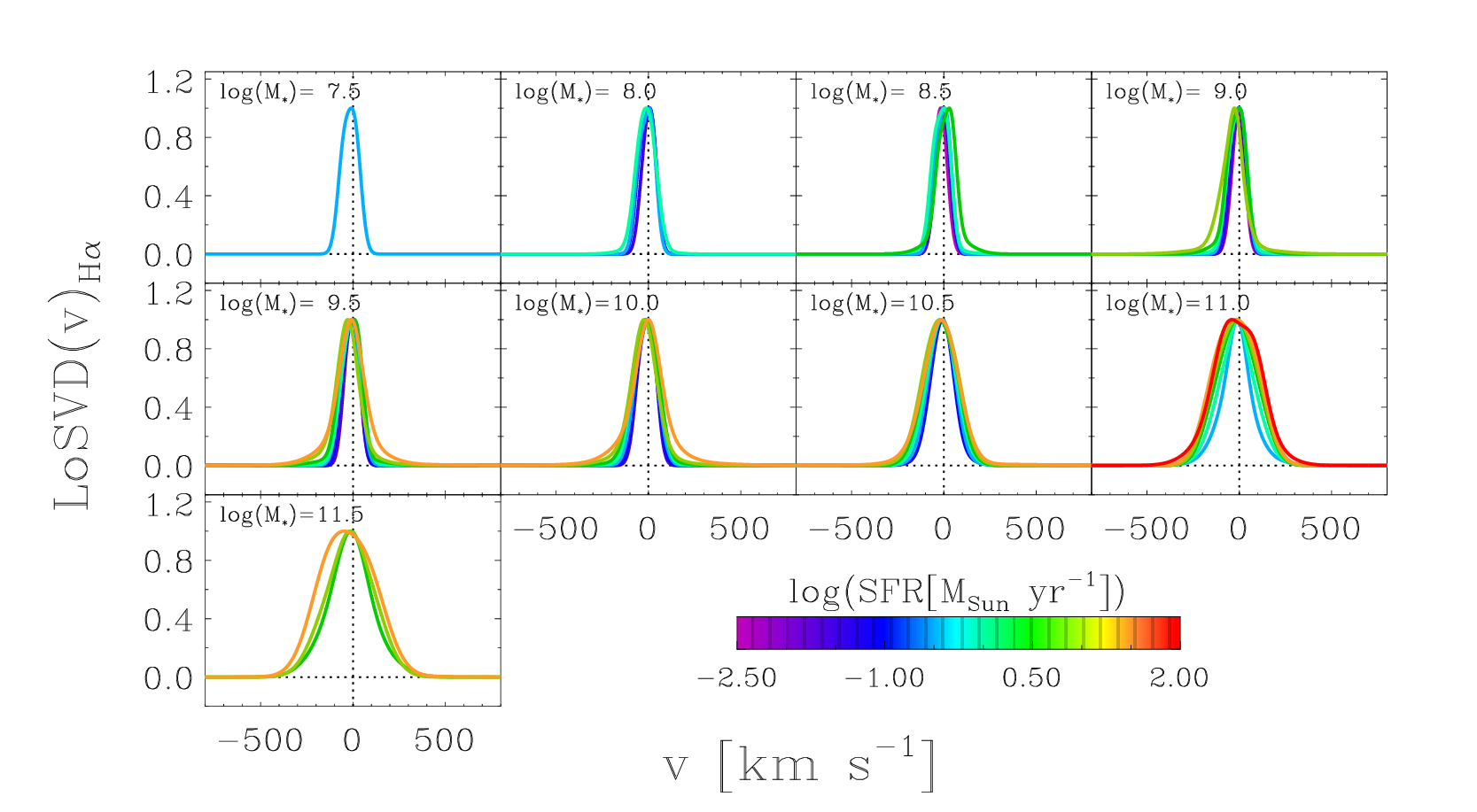}\\
     \caption{Line-of-sight velocity distribution of the ionised gas as traced by the
     		H$\alpha$+[NII] emission lines. In each panel we have 
		over-plotted the LoSVDs of galaxy bins identified by the same (average) 
		M$_{*}$ and by different (average) SFRs, colour-coded by their SFR.}
   \label{fig:losvd_ha_1}
\end{figure*}

\begin{figure*}[p]
   \includegraphics[angle=90,height=0.40\textheight]{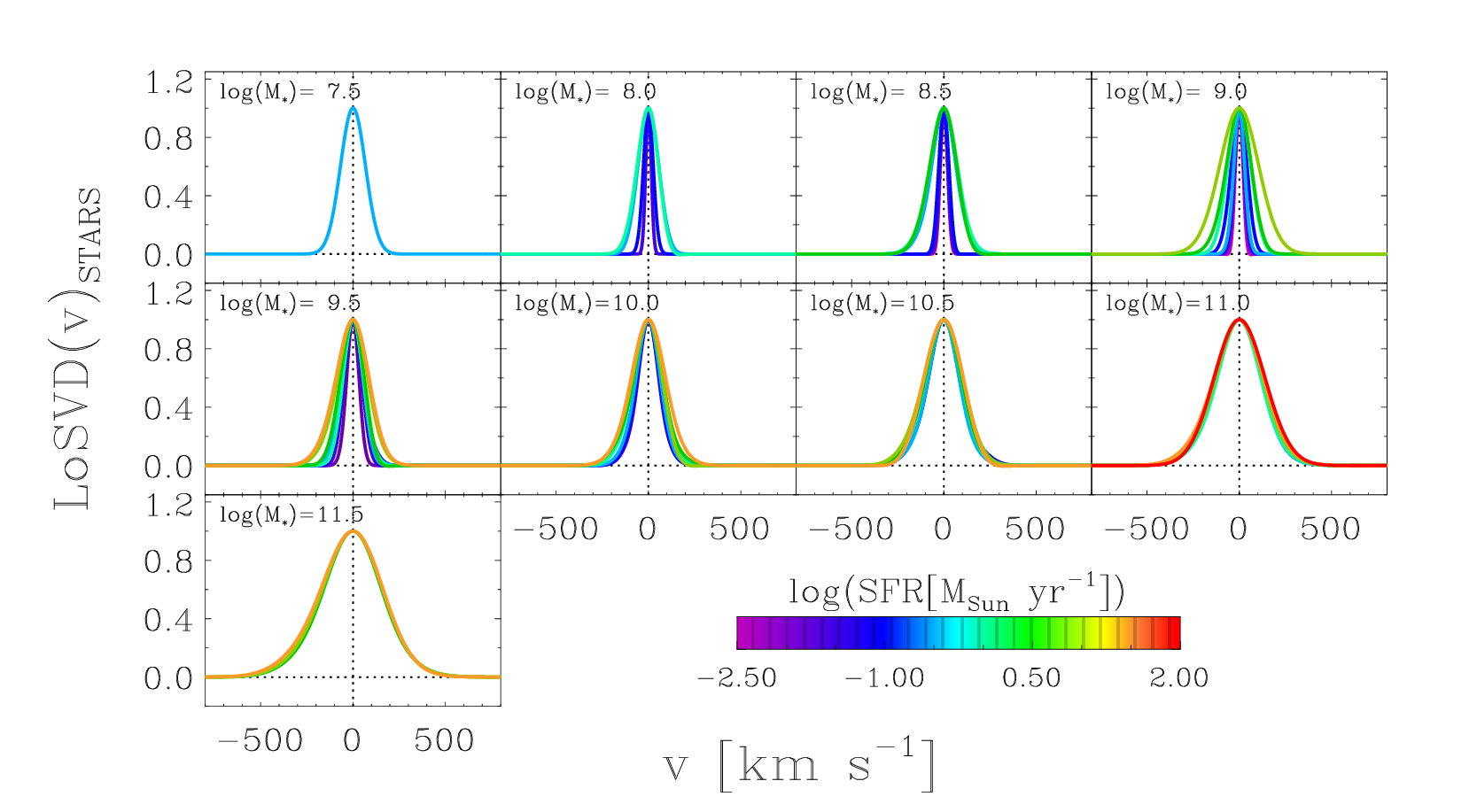}\\
     \caption{Line-of-sight velocity distribution of the stars. In each panel we have 
		over-plotted the LoSVDs of galaxy bins identified by the same (average) 
		M$_{*}$ and by different (average) SFRs, colour-coded by their SFR.}
   \label{fig:losvd_stars_1}
\end{figure*}

In this section we present the results of our analysis of the dynamical properties of ionised gas and stars
based on stacked SDSS spectra of local star-forming galaxies, aimed at identifying the presence of galactic outflows.
Throughout the remainder of the paper and in the figures, for the sake of simplicity, we will replace ``[OIII]$\lambda$5007'' and 
``H$\alpha$+[NII]$\lambda\lambda$6548,6583'' with ``[OIII]'' and ``H$\alpha$'', respectively.
The reader should however keep in mind that ``H$\alpha$'' actually refers to the H$\alpha$ and [NII] line complex {\it as a whole}, 
since the LoSVDs were derived by fitting the H$\alpha$ and [NII] lines simultaneously, as explained in $\S$~\ref{sec:losvd_fit}.
The adopted strategy is briefly summarised in $\S$~\ref{sec:strategy}. In $\S$~\ref{sec:kinematics} we analyse the general properties
of the LoSVDs of gas and stars; in $\S$~\ref{sec:asym} we study the presence of asymmetric wings in the LoSVD profiles and
finally in $\S$~\ref{sec:outflows} we discuss the outflow properties.

\subsection{The novelty of our approach}\label{sec:strategy}

Before presenting our results,
we first shortly summarise what is new in our approach to the study of galactic outflows, 
by emphasizing the advantages and strengths of this method.
For the first time, we study galactic outflows by directly comparing the 
LoSVD of the ionised gas (as traced by the [OIII], H$\alpha$ and [NII] nebular emission lines) with the
LoSVD of the stars. This has never been attempted
before, primarily because previous studies could not reach the signal-to-noise needed
to extract the stellar LoSVDs from the data, and (if any) they just relied on the comparison between
the velocity dispersion of gas and stars. 

Furthermore, this technique holds the potential of detecting the signature
of outflows in sources spanning a wide dynamic range of physical parameters as shown in Fig.~\ref{fig:grid_sf}, and
in particular in faint, low mass (i.e. ${\rm M_{*}}$ as low as ${\rm \sim 2\times 10^{7}~M_{\odot}}$) 
and moderately star-forming (i.e. SFR as low as ${\rm \sim 2\times 10^{-3}~M_{\odot}~yr^{-1}}$) galaxies. 
The main limitations when probing outflows in galaxies having such low stellar masses and low star formation rates
with SDSS spectroscopy
are both the low signal-to-noise and low spectral resolution of the associated spectra. 
In this study, however, we overcome both these limitations. Firstly, we use the
stacking technique to significantly improve the signal-to-noise in the
spectra ($\S$~\ref{sec:stacking}). Secondly, we minimise problems associated with the limited spectral resolution 
by deconvolving the ``real'' line-of-sight velocity distribution of gas and stars from the instrumental spectral profile
in the observed resolution-limited composite spectra (as described in
$\S$~\ref{sec:instr_prof}-$\S$~\ref{sec:losvd_fit}). This is 
made possible thanks to the high signal-to-noise of the stacked spectra,
especially for those galaxy bins in which the observed line profiles are narrow (i.e. line widths are close to the 
instrumental resolution).
In particular, since, in emission lines, 
outflows are traced by a broad 
component superimposed on a narrow component, 
our deconvolution technique in principle allows us
to detect such a broad component, provided it is broad enough, even if the bulk 
of the (narrow) emission line is not resolved. 

\subsection{Properties of the kinematics of gas and stars}\label{sec:kinematics}
\subsubsection*{LoSVD profiles}

In the following we analyse variations in the LoSVD profiles of (ionised) gas and stars
as a function of stellar mass and star formation rate.
We stress that, since the galaxy spectra have been stacked in bins of ${\rm M_{*}}$ and
SFR, the trends with ${\rm M_{*}}$ and SFR can be investigated {\it independently},
thus breaking the degeneracy between SFR and M$_*$ that holds for star forming galaxies on the MS.
For each stack, we adopt the velocity corresponding to the peak of the {\it stellar} LoSVD 
as a reference zero velocity for the LoSVD of both the stars
and the gas.

Figures~\ref{fig:losvd_oiii_2},  \ref{fig:losvd_ha_2} and \ref{fig:losvd_stars_2} show
the variations of the LoSVD profiles of gas and stars as a function of ${\rm M_{*}}$ 
at a fixed SFR. The most obvious trend that emerges from these figures is 
the increasing width of the LoSVDs of both gas and stars with stellar mass at a given
SFR. 
This is simply a consequence of the fact that, {\it to first order},
the velocity dispersion of both gas and stars primarily traces 
virial motions that are related to the dynamical mass of the galaxy. 
The dynamical mass of a galaxy is, to first order, proportional
to its stellar mass, although also gas and dust contribute to it. 
Therefore, broader LoSVDs are expected at higher stellar masses,
simply because higher stellar masses are generally associated with higher dynamical masses.
A comparison between Fig.~\ref{fig:losvd_oiii_2} and \ref{fig:losvd_ha_2} shows that
the [OIII] and the H$\alpha$ LoSVD profiles are overall consistent. 

The stellar LoSVDs in Fig.~\ref{fig:losvd_stars_2} show some interesting properties:
they are overall quite broad, and have wings that extend up to several hundreds of
km~s$^{-1}$ in velocity, similar to the ionised gas traced by [OIII] and H$\alpha$, in particular
for bins with ${\rm -0.5\leq log(SFR [M_{\odot}~yr^{-1}])\leq 1.5}$. This is in
part a consequence of the velocity tail of the stellar distribution, but additional effects may be present,
and these will be discussed in detail later in this section.

Figures~\ref{fig:losvd_oiii_1}, \ref{fig:losvd_ha_1} and \ref{fig:losvd_stars_1} are complementary to
Figs.~\ref{fig:losvd_oiii_2},  \ref{fig:losvd_ha_2} and \ref{fig:losvd_stars_2}: they show the
trends of LoSVD profiles of gas and stars as a function of SFR for fixed M$_*$.
We first focus on Figs.~\ref{fig:losvd_oiii_1} and  \ref{fig:losvd_ha_1}: at a given stellar mass,
the LoSVDs of [OIII]- and H$\alpha$-emitting gas broaden at higher SFRs.
Such broadening is particularly evident for the ``wings'' of the line-of-sight velocity distributions, especially
at stellar masses between ${\rm 9.0\leq log(M_{*} [M_{\odot}] )\leq 11.0}$ for the [OIII] LoSVDs, and between
${\rm 9.0 \leq log(M_{*} [M_{\odot}])\leq 10.0}$ for the H$\alpha$ LoSVDs. As we will discuss later in
this section, the overall broadening of the LoSVDs of ionised gas observed with increasing SFR at fixed M$_*$
could trace various stellar feedback-related phenomena, such as turbulence in the disk gas and
outflows.

A dependency of the width of the LoSVD on the SFR is also seen in the stars (Figure~\ref{fig:losvd_stars_1}),
although in a smaller measure and especially at lower stellar masses (i.e. ${\rm log(M_{*} [M_{\odot}])\leq10.5}$) than, for example, in 
the [OIII] LoSVDs (Figure~\ref{fig:losvd_oiii_1}).  
In the case of the stars, the LoSVD broadening 
with SFR (for constant ${\rm M_{*}}$) may be
associated with an increase of the total dynamical mass of the systems. In particular, 
the increase in width of the stellar LoSVDs is possibly tracing an increase of the total
gas content in galaxies with higher SFRs, which is in turn a consequence of
the Schmidt--Kennicutt relation between SFR and total gas mass surface density
\citep{Schmidt1959, Kennicutt98}. However, this trend might be affected by a
potential observational bias: the average projected size of galaxies decreases
for higher SFRs (higher SFRs correspond, on average, to more distant sources because
of selection effects), and this may have an effect on the stellar velocity dispersion measured 
within the SDSS fibre.

The line-of-sight velocity distributions of [OIII] and H$\alpha$ (Figures~\ref{fig:losvd_oiii_2},  
\ref{fig:losvd_ha_2}, \ref{fig:losvd_oiii_1}, and \ref{fig:losvd_ha_1})
exhibit asymmetries between the blue and the red side of the profiles. By eye inspection, such asymmetries
appear to depend on the SFR, but a more quantitative analysis will be presented
 in $\S$~\ref{sec:asym}. We note that,
quite surprisingly, also the stellar LoSVDs in Fig.~\ref{fig:losvd_stars_2} and
\ref{fig:losvd_stars_1} appear, although to a smaller extent, asymmetrical; such an effect can be
best appreciated in bins with ${\rm log(SFR~[M_{\odot} yr^{-1}]) =0.5}$ and
${\rm log(SFR~[M_{\odot} yr^{-1}]) =1.5}$ (Figure~\ref{fig:losvd_stars_2}). 
Asymmetries of the LoSVDs of gas and stars will be investigated in $\S$~\ref{sec:asym}.

\subsubsection*{LoSVD parameters: $\sigma$ and percentile velocities}

The next step is to extrapolate physically meaningful quantities from the
LoSVDs of stars and gas, and investigate the relationships with galaxy properties.
Errors on the LoSVD parameters are obtained using the bootstrap technique \citep{Efron79}.
The bootstrap method consists in randomly resampling (with replacements) our
galaxy stacks, by generating, for each bin, resampled stacks with sample size (here indicated with $S$) equal to the original stack.
More specifically, for each bin in the $\rm M_{*}-SFR$ parameter space (Figure~\ref{fig:grid_sf}),
we produce $N\gg1$ resampled stacks, each of which is obtained by randomly 
selecting $S$ galaxy spectra from the original sample of $S$ spectra included in that bin 
and by co-adding them to produce a new ``resampled'' stack. 
The difference from the original stack is that, in the resampled stacks, the $S$ spectra to be co-added together are
extracted randomly from the parent sample of $S$ galaxy spectra by allowing repetitions, hence the spectrum of a given
galaxy can be selected either once, or more than once, or never. 
After producing $N$ resampled stacks for each bin, we perform for each of them 
the stellar continuum subtraction and the [OIII] and H$\alpha$+[NII] line
fitting, exactly in the same way as for the original stacks (see $\S$~\ref{sec:stellar_sub}, $\S$~\ref{sec:losvd_stars} and $\S$~\ref{sec:losvd_fit}),
in order to retrieve the LoSVDs of stars and gas.
Therefore, the bootstrap provides, for each bin and for a given LoSVD
parameter of interest ($p_i$), a distribution of $N$ values. 
The variance of this distribution of $N$ parameters
(following on from the variance of the galaxy samples) is used to estimate the error on the measure of $p_i$.
The choice of the number $N$ of resampled stacks for each bin is a compromise
between statistics and computational effort. We set $N=10$ for the most crowded
bins, i.e. those including $S\geq10,000$ galaxy spectra, $N = 50$ for bins with
$5,000 \leq S < 10,000$, and $N = 100$ for the remaining ones, i.e. those
with less than 5,000 objects.

\begin{figure}[tbp]
   \includegraphics[clip=true, trim=5.75cm 1.cm 2.2cm 1.8cm,angle=90,width=1.\columnwidth]{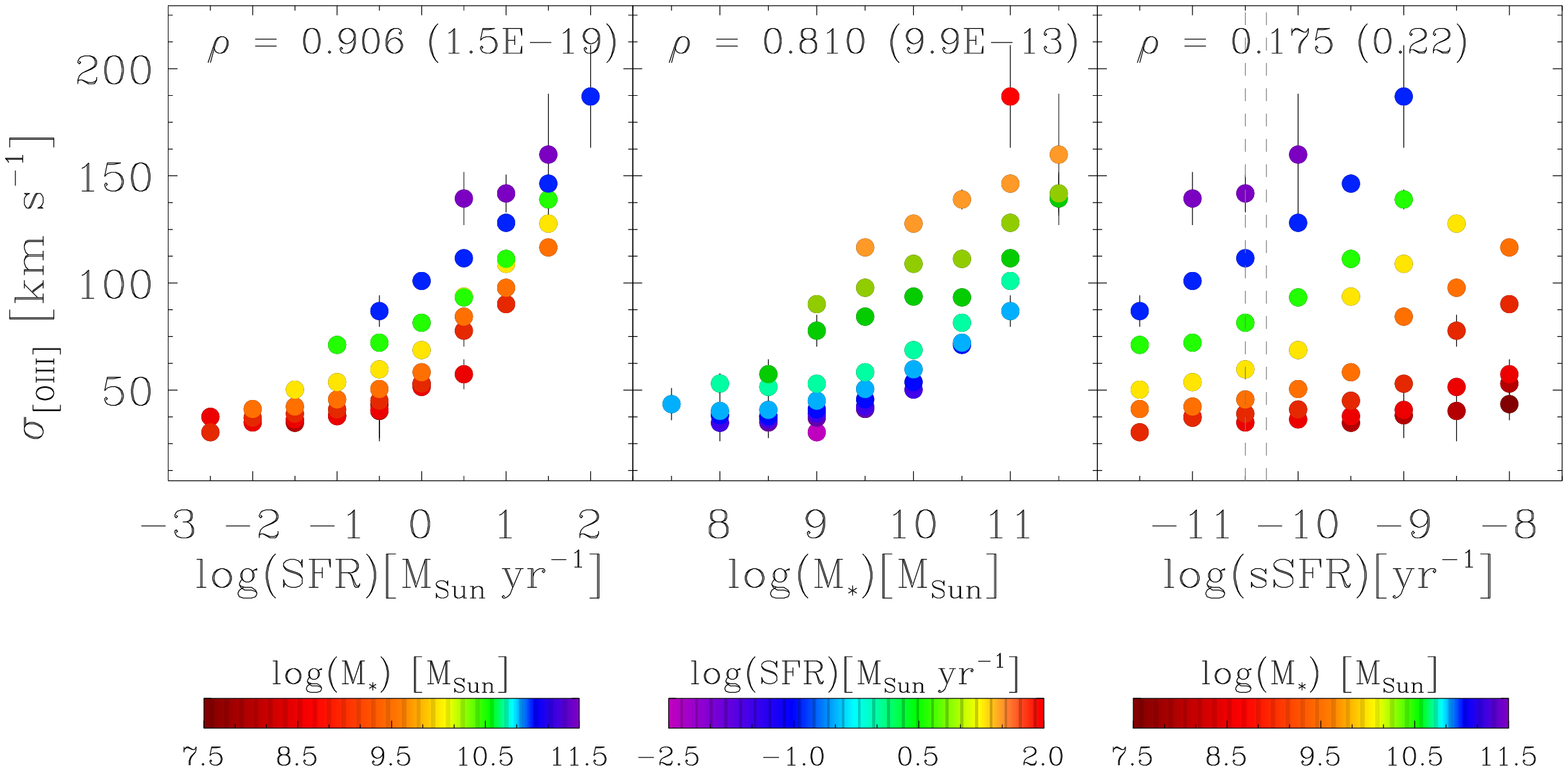}\\
   \includegraphics[clip=true, trim=5.75cm 1.cm 2.2cm 1.8cm,angle=90,width=1.\columnwidth]{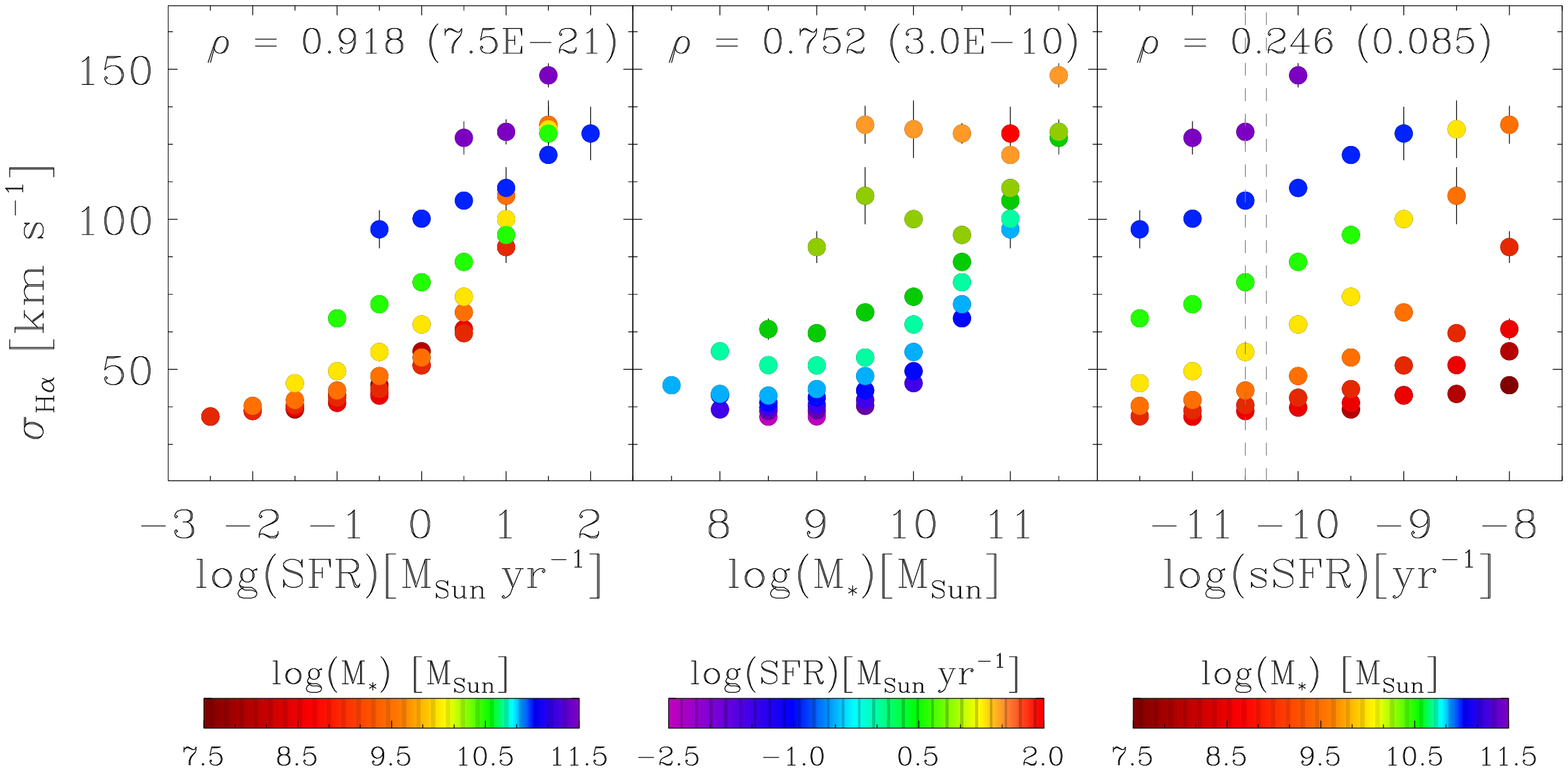}\\
   \includegraphics[clip=true, trim=1.8cm 1.cm 2.2cm 1.8cm,angle=90,width=1.\columnwidth]{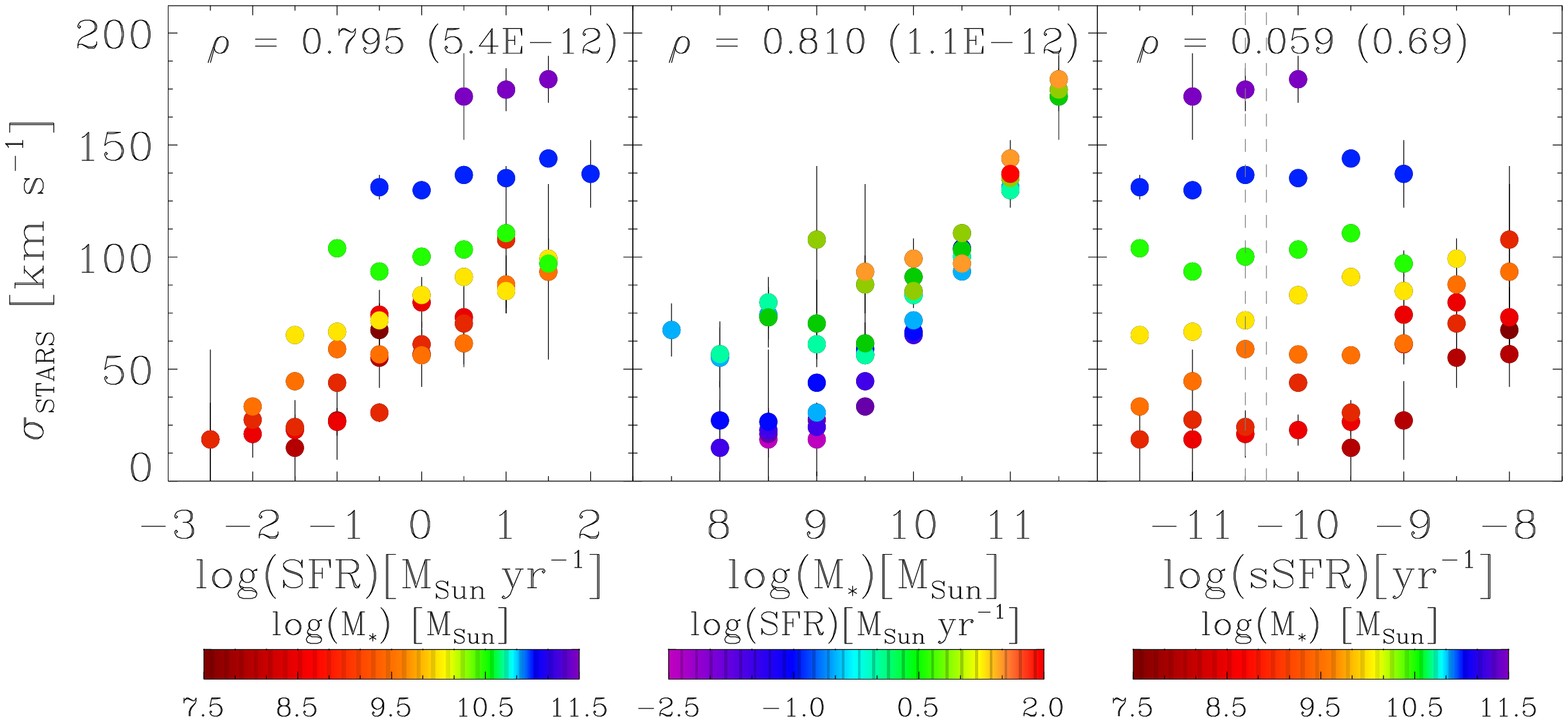}
   \caption{Line-of-sight velocity dispersion of the ionised gas as traced by the 
   		[OIII] (\emph{top panel}) and H$\alpha$ (\emph{middle panel}) emission lines, and of the
		stars (\emph{bottom panel}). The errors were obtained by applying the bootstrap method, as explained
		in the text. The grey dashed lines in the third column mark the sSFR range corresponding to the MS
		of local star-forming galaxies between ${\rm 9.0 \leq log(M_{*} [M_{\odot}]) \leq 11.0}$.
		The Spearman rank correlation coefficient ($0<\rho<1$, higher values of $\rho$ indicate
		stronger correlation) is reported in each plot, along with the corresponding two-sided p-value as 
		given in parenthesis (if the p-value is $\leq\alpha$,
		where $\alpha=0.05$ is the level of significance, the observed correlation is statistically significant).
		}
   \label{fig:sigma}
\end{figure}

The simplest parameter to study is the line-of-sight velocity dispersion ($\sigma$), given by:
\begin{equation}
\sigma = \biggl(\int(v - v_c)^2  \mathcal{L}(v) dv\biggr)^{1/2},
\end{equation}
where the line-of-sight velocity distribution described by ${\mathcal{L}(v)}$ is normalised
such as ${\int\mathcal{L}(v) dv = 1}$, and $v_c$ is
the centroid of the distribution, i.e. ${v_c = \int v~\mathcal{L}(v) \textit{d} v}$.
The relationships between $\sigma$ (of both the ionised gas, as traced by [OIII] and H$\alpha$ and the stars), and galaxy
properties such as SFR, M$_*$ and sSFR
(sSFR = SFR/${\rm M_{*}}$, i.e. star formation rate per stellar mass unit)
are shown in Fig.~\ref{fig:sigma}. 

For each plot in Fig.~\ref{fig:sigma}, 
we calculated the non-parametric Spearman
rank correlation parameter ($\rho\in(0,1)$\footnote{Calculated by using the IDL function \texttt{R\_CORRELATE}.}) to test the hypothesis of correlation between
the line-of-sight velocity dispersion of gas and stars and the galaxy properties (SFR, M$_*$, sSFR). 
In our case, non-parametric tests such as the Spearman rank test are more appropriate than the standard
Pearson correlation coefficient ($r$), since we are dealing
with binned data and the relationship between $\sigma$ and galaxy properties (if any) could be non-linear. 
A high value of $\rho$ (e.g., $\rho\gtrsim0.5$) combined with a low p-value ($\leq\alpha$, where $\alpha$ is the level of significance, which we set
$\alpha=0.05$) indicates the presence of 
a statistically significant correlation. In the calculation of the p-value, 
we allow for both positive and negative correlations (``two-sided''). 
The results of the Spearman rank test are reported in Fig.~\ref{fig:sigma}.

As illustrated by the clean relationships in Fig.~\ref{fig:sigma}, our method is particularly
effective at uncovering general trends between the dynamical properties of gas and stars and the galaxy properties.
Moreover, Fig.~\ref{fig:sigma} allows us to appreciate some differences between the dynamical behaviour of ionised 
gas and stars as a function of SFR and M$_*$.
In absolute values, the velocity dispersions of gas and 
stars are roughly consistent, with
the bulk of the galaxy bins showing values in the range ${\rm 40 \lesssim \sigma~[ km~s^{-1} ] \lesssim 150}$.
However, for the highest stellar mass bins, i.e. ${\rm log(M_{*} [M_{\odot}]) \gtrsim 10.0}$, the velocity dispersions measured
for the stars are slightly larger than the ones measured for the gas, especially when using H$\alpha$ as a gas tracer.
This small effect can be explained by the contribution of stellar bulges in more massive galaxies.
\cite{Gavazzi+15} have recently shown that $>40$\% of isolated local star-forming galaxies with 
M$_*>10^{9.5}$~M$_{\odot}$ host stellar bulges (without distinguishing between classical bulges and pseudo-bulges), whereas
the fraction of stellar bulges at  M$_*<10^{9.5}$~M$_{\odot}$ is only $\sim5$\%.
When a galaxy has a prominent stellar bulge, its stellar LoSVD probes both the galactic disk and the bulge, whereas the
LoSVD of the ionised gas only traces the dynamics in the disk (and, possibly, a ionised outflow), 
which is more subject to projection effects than the bulge. In particular, disks viewed face-on contribute 
much less to the broadening of the LoSVD than edge-on disks. Therefore, it is possible that, by averaging 
together many spectra of massive galaxies with stellar
bulges, observed from different viewing angles, the resulting line-of-sight velocity dispersions are slightly higher for the
stars than for the gas, because of the additional contribution from the stellar bulge.

By inspecting the plots in the first column of Fig.~\ref{fig:sigma},
we note that $\sigma_{\rm [OIII]}$ and $\sigma_{\rm H\alpha}$ are tightly correlated ($\rho > 0.9$) with the SFR, and the correlation
holds both at low and high stellar masses. The line-of-sight velocity dispersion of the stars is 
also correlated with the SFR, but more weakly than the gas ($\rho=0.795$), and the $\rm \sigma_{stars}-SFR$ relationship 
flattens at ${\rm log(M_{*} [M_{\odot}])\sim 10.5}$ ($\rho=0.029$, p-value=0.957).

The steady increase of gas velocity dispersion with SFR observed in Fig.~\ref{fig:sigma}
may trace a combination of various physical processes.
Since feedback-related mechanisms should affect only the gas kinematics, leaving the motions of (evolved) 
stellar populations unperturbed, we can use 
the stellar velocity dispersion as a reference to identify trends that are not related to feedback.
The positive correlation between $\sigma_{\rm stars}$ and SFR observed at low stellar masses
may be due to the presence of gas contributing to the dynamical mass, the gas being traced 
by the SFR via the S--K relation, and therefore it may be related 
to an increase of gas fraction (at fixed stellar mass). 
However other mechanisms may
produce a similar effect: in the first place, as already mentioned, 
a possible anti-correlation
between the projected size and the SFR in galaxies, in 
conjunction with the limited SDSS fibre aperture, may
affect the observed stellar LoSVD.
Moreover, mergers and galaxy interactions, whose fraction is expected
to increase at higher SFRs, can also have an impact on the observed 
LoSVDs, due to the combined motions of overlapping disks
or tidal motions of collision, as pointed out by \cite{Rupke+Veilleux13} in their study of 
ULIRGs. However, mergers 
are relatively unimportant in the local Universe: the local galaxy merger rate
estimated using SDSS data is about 0.01 Gyr$^{-1}$ \citep{Patton+Atfield08}. 
Assessing the relative importance of these different factors in
producing the observed trend of increasing $\sigma_{\rm stars}$
with SFR (at a given ${\rm M_{*}}$) goes beyond the scope of this paper and requires
spatially resolved observations. 
For what the present study is concerned, these results suggest caution in interpreting
the observed broadening of the ionised gas LoSVDs with SFR as {\it entirely} due to
feedback-related processes, since the effects described above (and affecting both stars and gas)
may be at work.

Moreover, as mentioned above, outflows of ionised gas are not the only feedback-related mechanisms
that can affect the observed gas velocity dispersions. 
Higher $\sigma_{\rm gas}$ may also 
trace star formation-induced turbulence in galactic disks \citep{Faucher-Giguere+13},
such as turbulence injected in the gas by SNe explosions, which would naturally 
depend on the SFR (e.g. \citealt{Shetty+Ostriker12,Hopkins+12}).
Indeed, observationally, vertical disk velocity dispersions in the gas (increased by turbulence) 
would be convolved with circular velocities due to
inclination effects, thereby affecting the sightline-averaged gas velocity dispersions. 
\cite{Faucher-Giguere+13} suggested that turbulence resulting from stellar feedback regulates the 
the rate of formation of giant molecular clouds (GMCs),
where most of star formation takes place, thereby having an important role in regulating the global
disk-averaged star formation efficiency in galaxies \citep{Faucher-Giguere+13}. 
The tendency of galaxies with higher SFRs (as traced by their IR luminosity) to show 
larger optical emission line widths was already evidenced by earlier observational studies \citep{Veilleux+95, Lehnert+Heckman96}.
However, more recent results by \cite{Arribas+14} indicate that star formation-driven turbulence
may have only a marginal role in increasing the velocity dispersion of ionised gas in galaxy disks. These authors suggested
that, at least in (U)LIRGs, galaxy interactions and AGNs may instead be 
the main responsible for the presence of dynamically hot gas disks.

Fig.~\ref{fig:sigma} shows also the line-of-sight velocity dispersion
of gas and stars as a function of M$_*$ (second column), with the bins colour-coded by their SFR, i.e. the 
diagram orthogonal to the one in the first column. As already pointed out,
a positive correlation is expected since the stellar mass is a proxy for the total
dynamical mass of a system, and higher dynamical masses result in
higher velocity dispersions for gas and stars. The scatter is mostly ascribable to
differences in SFR. For M$_{*}>10^{10}$~M$_{\odot}$, where the relationship between $\sigma_{\rm stars}$ and
SFR flattens (as shown in the first column plot), the correlation between $\sigma_{\rm stars}$ and ${\rm M_{*}}$ tightens.

For fixed stellar mass (indicated by the colour coding), the trends of $\sigma$
vs sSFR reported in the third column of Fig.~\ref{fig:sigma} are identical to the trends vs SFR, only shifted
by ${\rm -log(M_{*})}$. This is a natural consequence of binning galaxies in the
${\rm M_{*}-SFR}$ parameter space. We show in these plots the range of sSFR corresponding to the
MS of local star-forming galaxies (Figure~\ref{fig:grid_sf}), calculated between 
${\rm 9.0 \leq log(M_{*}) \leq 11.0}$\footnote{Without including the MS scatter of $\pm0.3$ dex in SFR.} following \cite{Peng+10} as
detailed in $\S$~\ref{sec:stacking}.
Such sSFR range is very narrow, and this constitutes a fundamental property of the galaxy population:
the sSFR, that is the ratio between ongoing star formation (i.e. SFR) and past-integrated star formation
(i.e. ${\rm M_{*}}$) is roughly constant in star-forming galaxies at a given epoch \citep{Peng+10} (however,
see also discussion in \cite{Gavazzi+15} and references therein according to which the MS relation changes 
slope above a turnover stellar mass, resulting in a decrease of sSFR with M$_*$ in normal star forming galaxies).

In conclusion, the widths of the [OIII] and H$\alpha$ LoSVDs are not the
most appropriate tools to investigate galactic outflows: although the correlations observed 
in Fig.~\ref{fig:sigma} between $\sigma$ and
SFR (or sSFR) are likely probing also galactic outflows, there are other effects that may affect these trends,
such as the previously discussed possible projection effects, star formation-induced turbulence in the disk, and the effects of variations in the total 
dynamical mass 
on the virial motions of stars and gas. In order to reduce the influence of virial motions and to shed light
on the effects of galactic outflows on the ionised gas dynamics, we need to focus our investigation on the {\it high
velocity tail} of the LoSVDs.

\begin{figure}[tb]
   \includegraphics[clip=true, trim=5.75cm 1.cm 2.2cm 1.8cm,angle=90,width=1.\columnwidth]{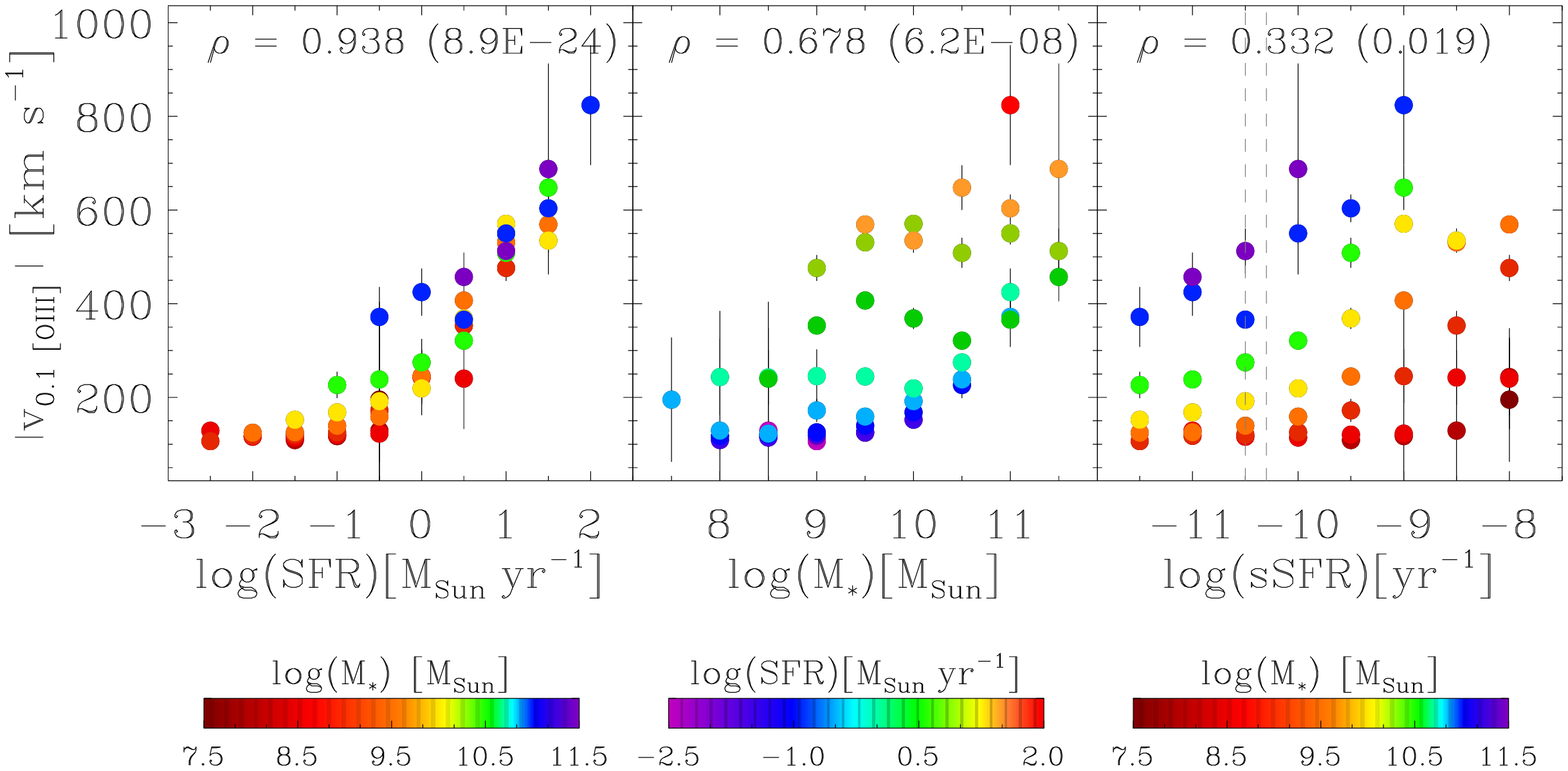}\\
   \includegraphics[clip=true, trim=5.75cm 1.cm 2.2cm 1.8cm,angle=90,width=1.\columnwidth]{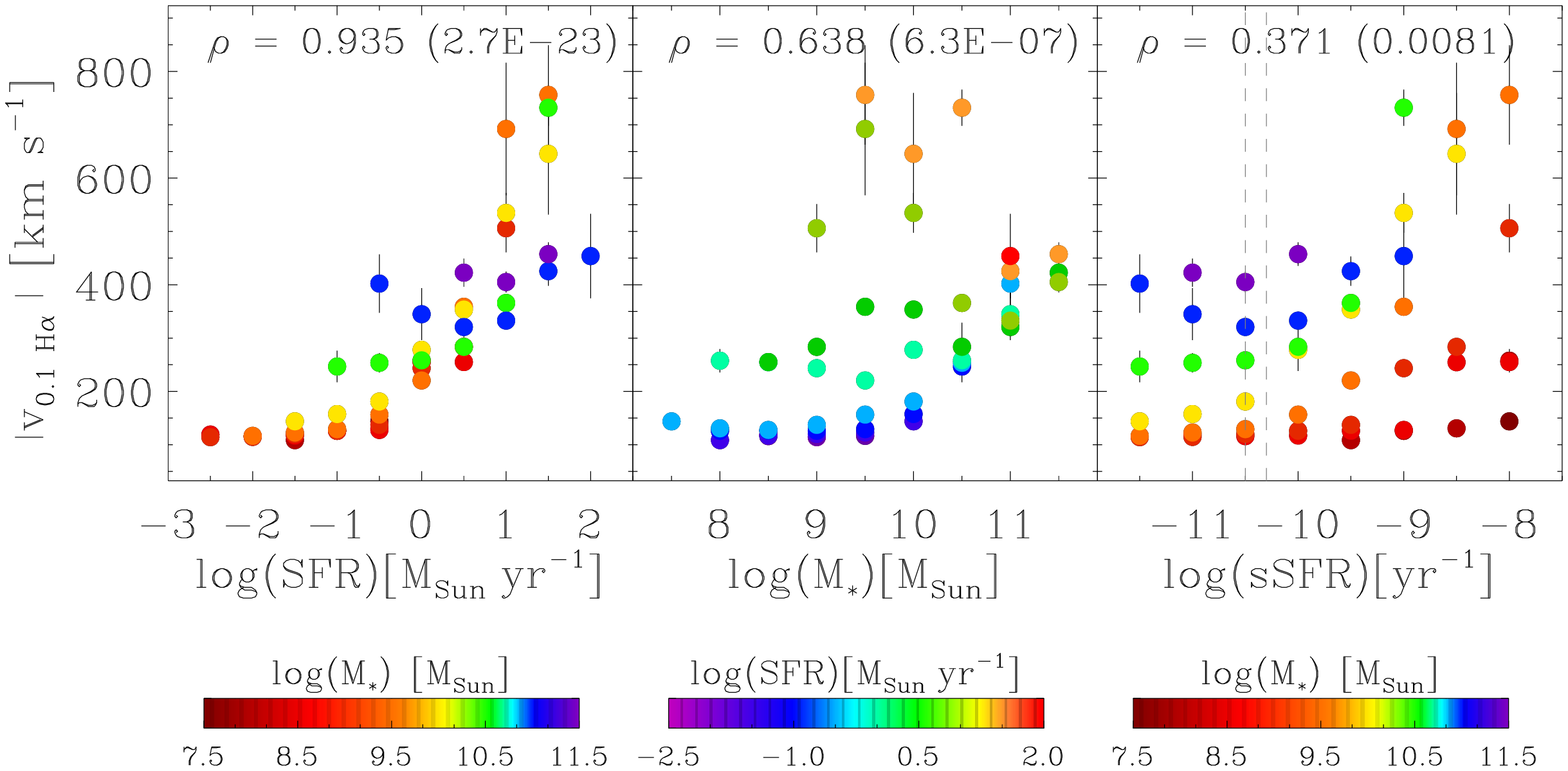}\\
   \includegraphics[clip=true, trim=1.8cm 1.cm 2.2cm 1.8cm,angle=90,width=1.\columnwidth]{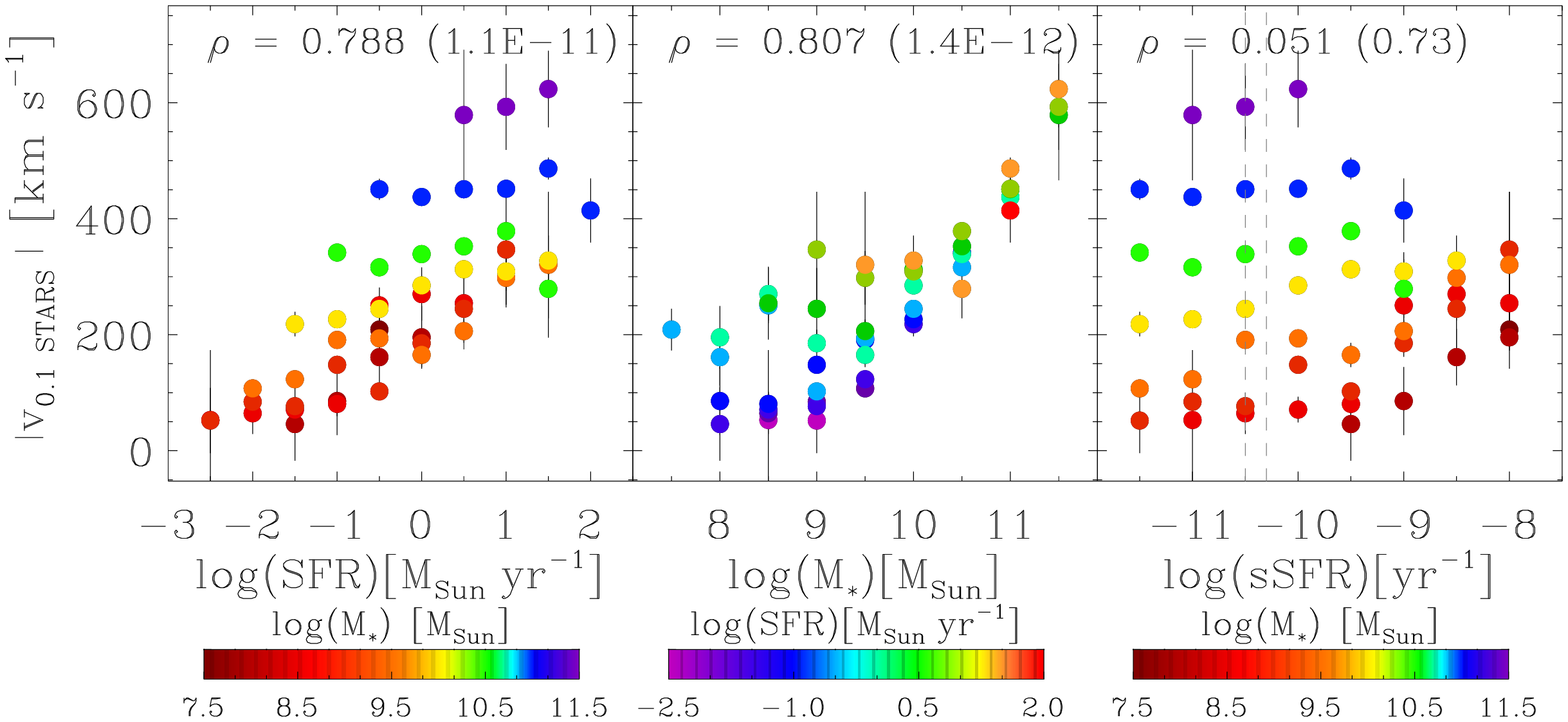}
   \caption{(Modulus of the) 0.1th percentile velocity of the LoSVD of the ionised gas as traced by the 
   		[OIII] (\emph{top panel}) and H$\alpha$ (\emph{middle panel}) emission lines, and of the
		stars (\emph{bottom panel}). For a Gaussian velocity distribution, the 0.1th percentile
		velocity corresponds to -3 standard deviations ($\sigma$) from the mean velocity, and therefore it probes the
		high velocity blueshifted tail of the LoSVD. Similarly to Fig.~\ref{fig:sigma}, we report for each plot the Spearman
		rank correlation coefficient $\rho$ along with its associated two-sided p-value.}
   \label{fig:v01}
\end{figure}

To this purpose, we calculate the 0.1th, 2.3th,15.9th, 84.1th, 97.7th and 99.9th percentile velocities 
of the LoSVDs of gas and stars. We chose these values because,
in a Gaussian distribution, they correspond respectively to $-3\sigma$, $-2\sigma$, $-1\sigma$,
$1\sigma$, $2\sigma$ and $3\sigma$, where $\sigma$ is the standard deviation of the Gaussian distribution.
The $N$-th percentile velocity, $v_N$, is defined as: 
\begin{equation}
P(v < v_N) = \int_{-\infty}^{v_N} \mathcal{L}(v) dv = N,
\end{equation}
where $P(v < v_N)$ is the probability of observing, with the given line-of-sight velocity distribution $\mathcal{L}(v)$, velocities
lower than $v_N$, and ${\int_{-\infty}^{+\infty} \mathcal{L}(v) dv = 1}$.

Figure~\ref{fig:v01} shows the 0.1th percentile velocities of the [OIII], H$\alpha$ and stellar LoSVDs as a function of SFR,
M$_*$ and sSFR. The corresponding plots obtained for the other 
percentile velocities are shown in the Appendix (Figures~\ref{fig:v16} to ~\ref{fig:v99}).
The results of the Spearman rank test performed on the relationships between the various LoSVD parameters 
($\sigma$ and percentile velocities) and galaxy parameters (SFR, M$_*$ and sSFR), for both gas and stars, are 
reported in Table~\ref{table:rho_values} to facilitate the comparison between the different relationships shown in 
Figs.~\ref{fig:sigma}, \ref{fig:v01} and \ref{fig:v16}--\ref{fig:v99}.

\begin{table}[tb]
\small
 \begin{minipage}{80mm}
  \caption{Spearman rank correlation parameters$^\dag$}
  \label{table:rho_values}
\begin{tabular}{@{}rccc@{}}
\hline
\hline
  	   			&  [OIII] 						& H$\alpha$  			&   stars   	\\
				& $\rho$ (p-value)	         	& 	 $\rho$ (p-value)      	& $\rho$ (p-value) 	\\
\hline
$\sigma$ vs SFR	&	0.906 (1.5E-19)		& 0.918 (7.5E-21)		&	0.795 (5.4E-12)	\\
$v_{15.9}$ vs SFR	&	0.826 (1.6E-13)		& 0.852 (4.4E-15)		&	0.825 (1.7E-13) 	\\
$v_{2.3}$ vs SFR    	&	0.891 (4.7E-18)		& 0.925 (8.5E-22)		&	0.792 (7.3E-12)	\\
$v_{0.1}$ vs SFR	&	0.938 (8.9E-24)		& 0.935 (2.7E-23)		&	0.788 (1.1E-11)	\\
$v_{84.1}$ vs SFR	&	0.760 (1.5E-10)		& 0.735 (1.2E-09)		&	0.787 (1.3E-11)	\\
$v_{97.7}$ vs SFR    &	0.889 (6.5E-18)		& 0.885 (1.6E-17)		&	0.756 (2.3E-10)	\\
$v_{99.9}$ vs SFR	&	0.921 (2.5E-21)		& 0.894 (2.2E-18)		&	0.731 (1.7E-09)	\\
\hline
$\sigma$ vs M$_*$	&	0.810 (9.9E-13)		& 0.752 (3.0E-10)		&	0.810 (1.1E-12)	\\
$v_{15.9}$ vs M$_*$ &	0.832 (7.0E-14)		& 0.790	(9.4E-12)		&	0.802 (2.7E-12)	\\
$v_{2.3}$ vs M$_*$   &  	0.808 (1.3E-12)		& 0.730 	(1.8E-09)		&	0.808 (1.3E-12)	\\
$v_{0.1}$ vs M$_*$	 &	0.678 (6.2E-08)		& 0.638	(6.3E-07)		&	0.807 (1.4E-12)	\\
$v_{84.1}$ vs M$_*$	 &	0.886 (1.3E-17)		& 0.820 (3.4E-13)		&	0.792 (7.3E-12)	\\
$v_{97.7}$ vs M$_*$  &	0.810 (1.1E-12)		& 0.754	(2.5E-10)		&	0.815 (6.2E-13)	\\
$v_{99.9}$ vs M$_*$	 &	0.566 (1.9E-05)		& 0.657	(2.2E-07)		&	0.811 (8.8E-13)	\\
\hline
$\sigma$ vs sSFR	&	0.175 (0.22)			& 	0.246 (0.085)		&	0.059 (0.69)		\\
$v_{15.9}$ vs sSFR	&	0.069 (0.64)			& 	0.142 (0.33)			&	0.094 (0.52)		\\
$v_{2.3}$ vs sSFR    &	0.159 (0.27)			& 	0.268 (0.059)		&	0.054 (0.71)		\\
$v_{0.1}$ vs sSFR	&	0.332 (0.019)		& 	0.371 (0.0081)		&	0.051 (0.73)  	\\
$v_{84.1}$ vs sSFR	&	-0.044 (0.76)		& 	-0.002 (0.99)		&	0.070 (0.63)		\\
$v_{97.7}$ vs sSFR  &	0.160 (0.27)			& 	0.207 (0.15)			&	0.019 (0.90)		\\
$v_{99.9}$ vs sSFR	&	0.435 (0.0016)		& 	0.314 (0.026)		&	-0.005 (0.97)	\\
\hline
\end{tabular}
\end{minipage}

\begin{flushleft}
\small
\textbf{Notes:} $^\dag$ This table lists the Spearman rank correlation
coefficients ($\rho\in(0.1)$) and corresponding two-sided p-values calculated for the relationships between LoSVD 
parameters (velocity dispersion and percentile velocities) and galaxy properties (SFR, M$_*$, sSFR)
that are shown in Figs.~\ref{fig:sigma}, \ref{fig:v01} and Figs. ~\ref{fig:v16}--\ref{fig:v99}.
A high value of $\rho$ (e.g. $\rho\gtrsim 0.5$) combined with a low p-value ($\leq 0.05$) indicates
the presence of a statistically significant correlation.
\end{flushleft}
\end{table}

It is clear from Table~\ref{table:rho_values} that the velocity dispersions and the
percentile velocities show all significant correlations 
with SFR and M$_*$, for both the gas and the stars. There are however important differences
among the various LoSVD parameters, in particular between the 15.9th
and the 0.1th percentile velocities.
More specifically, by moving towards the high velocity blueshifted tail of the LoSVD
(i.e. from the 15.9th to the 0.1th percentile velocity), the correlation between gas velocity and SFR 
tightens (the Spearman correlation coefficient $\rho$ increases), whereas the correlation between
stellar velocity and SFR weakens slightly ($\rho$ decreases). A similar trend is evidenced also for the 
redshifted tail of the LoSVD (i.e. from the 84.1th to the 99.9th percentile velocity). 
These results strengthen the hypothesis that the high velocity tail of the ionised gas LoSVD traces
mainly star formation-feedback related mechanisms and is little affected by other mechanisms, which instead dominate the relationships
between stellar velocity and SFR.
The correlation between gas velocity and SFR is slightly tighter at blueshifted than at redshifted velocities
(especially if using H$\alpha$ as a tracer), probably because of dust extinction affecting
mostly the receding (i.e. redshifted) side of the outflow. 
Furthermore, Table~\ref{table:rho_values}
shows that the correlation between ionised gas velocity and M$_*$  
weakens (and almost breaks down, as clearly shown in the corresponding plots in Fig.~\ref{fig:v01} and Figs.~\ref{fig:v84}--\ref{fig:v99}) 
towards the high velocity tail of the LoSVD ($\rho$ decreases), suggesting that gravity (as traced by M$_*$)
does not play an important role in determining the dynamics of the high-velocity gas.

In summary, by exploring the relationships between the different percentile velocities
of the LoSVDs of gas and stars and the galaxy parameters, we 
 infer that the correlation with the SFR is tighter for the ionised gas at higher percentile velocities and, in particular,
that the high velocity (blueshifted or redshifted) tails of the gaseous LoSVDs depend tightly on the SFR (correlation coefficient
$\rho\gtrsim0.9$) and more weakly ($\rho<0.7$) on the stellar mass. This result supports the hypothesis that, 
in the high velocity regime, the
gas kinematics progressively ceases to probe the dynamical mass of galaxies (which is in first approximation
traced by the stellar mass), but it is instead intrinsically related to the rate at which stars are formed, likely because of the
presence of stellar feedback mechanisms. 

However, regardless of the correlations observed between the various LoSVD parameters ($\sigma$ and percentile
velocities) and galaxy properties (SFR, M$_*$, sSFR) that have been discussed in this Section (Table~\ref{table:rho_values}), 
in order to isolate the effects of outflows from other mechanisms we need to investigate the {\it differences} between the 
LoSVDs of gas and stars. Indeed, by looking at the galaxy stacks where the difference 
$(v_{\rm gas}-v_{\rm stars})$ is significantly greater than zero, we can identify galaxies where the 
motions of gas and stars are clearly decoupled, and so galaxies where outflows are most likely taking place. Differences between
the gaseous and stellar kinematics and their dependency on galaxy properties 
will be investigated in $\S$~\ref{sec:outflows}.

\subsection{Asymmetries in the LoSVDs of gas and stars}\label{sec:asym}

Asymmetric wings of nebular emission lines usually trace perturbed gas, 
whose dynamics is not consistent with purely
virial motions but that can be instead explained with radial (outward or inward) motions in conjunction with dust
extinction effects. The presence of a blue asymmetry is commonly interpreted in terms of 
outflows, because obscuration by dust in the galaxy disk affects primarily the backside receding gas
(e.g. \citealt{Lehnert+Heckman96,Villar-Martin+11,Soto+12}).

\begin{figure}[tb]
	 \centering
   \includegraphics[clip=true, trim=5.75cm 1cm .5cm 3.1cm,angle=90,width=1.\columnwidth]{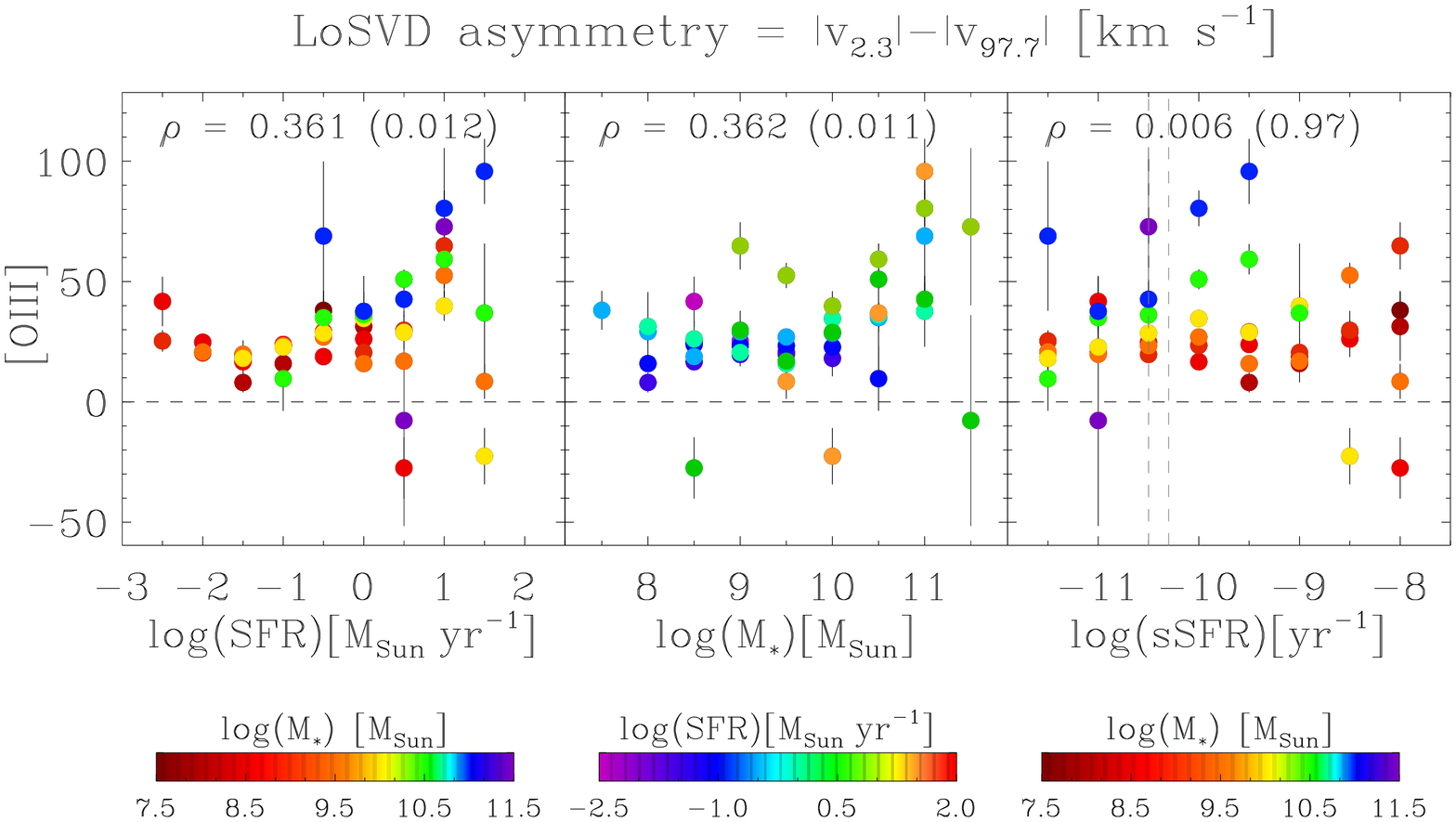}\\
   \includegraphics[clip=true, trim=5.75cm 1cm 2.2cm 3.1cm,angle=90,width=1.\columnwidth]{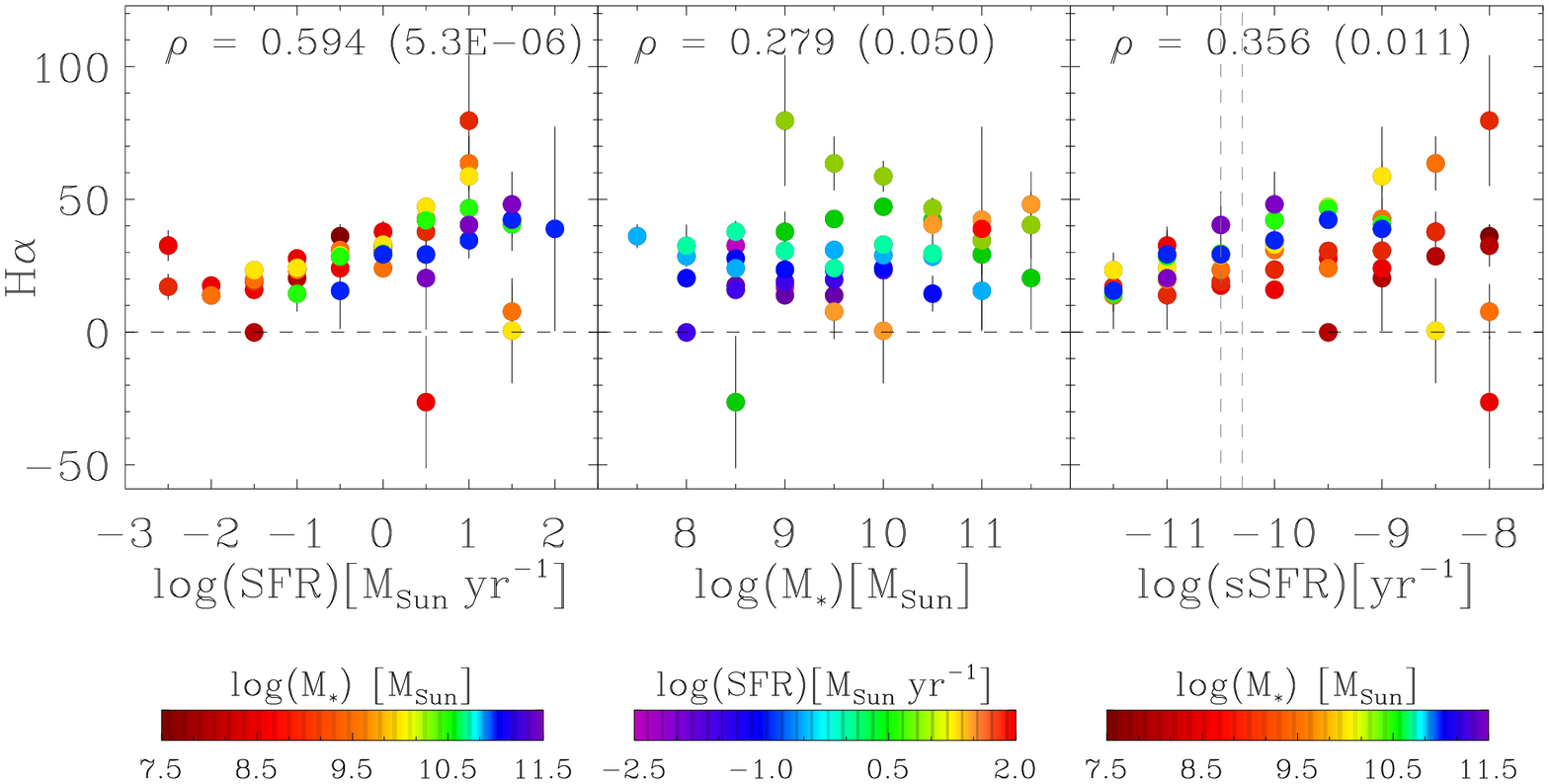}\\
   \includegraphics[clip=true, trim=1.8cm 1cm  2.2cm 3.1cm,angle=90,width=1.\columnwidth]{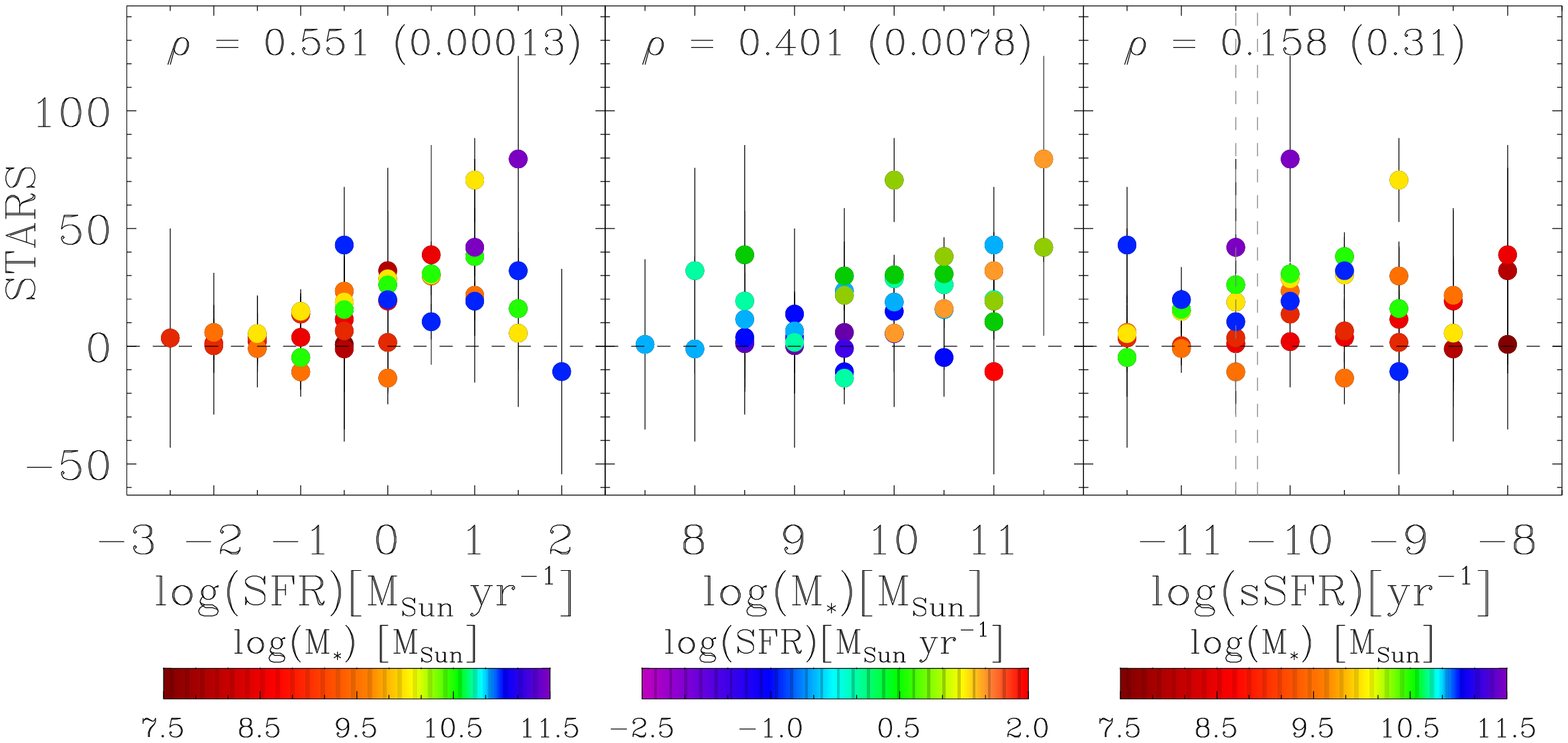}
   \caption{Blue asymmetry of the [OIII] (\emph{top panel}),  H$\alpha$ (\emph{middle panel}) 
   	       and stellar (\emph{bottom panel}) LoSVDs, as measured by the 
   	       difference between the (moduli of the) 2.3th and 97.7th percentile velocities
	       (i.e. $y = |v_{2.3}|-|v_{97.7}|$). For display purposes, only the less noisy bins are plotted, i.e.
	       bins with either $\sigma_{y} < 50~{\rm km~s^{-1}}$ or $|y| \geq 2\sigma_{y}$.
	       The Spearman rank correlation coefficient ($0<\rho<1$, higher values of $\rho$ indicate stronger correlation) 
	       is reported for each plot, along with the corresponding 
	       two-sided p-value (if the p-value is $\leq \alpha$, where $\alpha=0.05$ is the level of significance, the observed correlation is 
	       statistically significant). }
   \label{fig:v2_v98_diff}
\end{figure}

Visual inspection of the LoSVD profiles of gas and stars shown in Figs.~\ref{fig:losvd_oiii_2} 
to \ref{fig:losvd_stars_1} suggests that the line-of-sight
velocity distributions of both gas and stars are not symmetrical. We further investigate
this possibility in Fig.~\ref{fig:v2_v98_diff}, where we plot, as a function of SFR, M$_*$ and sSFR, the
 difference between the (moduli of the) 
2.3th and 97.7th percentile velocities, i.e. ${|v_{2.3}|-|v_{97.7}|}$, evaluated for 
[OIII], H$\alpha$ and stars.
Figure~\ref{fig:v2_v98_diff} shows that 
almost all galaxy bins (a part form a few, noisier stacks) exhibit
a clear blue asymmetry in the LoSVDs of [OIII] and H$\alpha$, with ${|v_{2.3}|-|v_{97.7}|}$ values
ranging between $\sim0$~km~s$^{-1}$ and $\sim100$~km~s$^{-1}$. 
Very surprisingly, also the stellar LoSVDs are affected by a blueward asymmetry, although less pronounced
than in the gas. The velocity difference ${|v_{2.3}|-|v_{97.7}|}$ measured for the stars is
consistent with zero at low SFRs. However, in galaxy bins with
${\rm SFR \gtrsim 0.1~M_{\odot}~yr^{-1}}$, ${|v_{2.3}|-|v_{97.7}|}$ increases up to
values of a few tens of km~s$^{-1}$, with
large uncertainties, but still significantly larger than zero and therefore inconsistent 
with the absence of any asymmetry.

The presence of a blue asymmetry in the LoSVDs of [OIII] and H$\alpha$ 
is suggestive of galactic outflows.
The hypothesis of star formation-driven outflows
is corroborated by the correlation observed between
the H$\alpha$ blue asymmetry (as traced by ${|v_{2.3}|-|v_{97.7}|}$)
and the SFR ($\rho$=0.594), and by the hint of a 
correlation observed for [OIII] ($\rho$=0.361).
We note however that the dust properties of galaxies may also affect the trends observed
for [OIII] and H$\alpha$ in Fig.~\ref{fig:v2_v98_diff}.
Indeed \cite{Santini+14}, by using {\it Herschel} observations of local and intermediate-redshift ($z\lesssim2.5$)
galaxies, found a tight correlation between dust mass
and star formation rate, which they interpreted as a consequence of the Schmidt--Kennicutt law
(since dust and gas mass are related). Therefore, the positive correlation between 
${|v_{2.3}|-|v_{97.7}|}$ and SFR observed for the ionised gas is 
likely due to the combined effect of star formation-driven outflows and 
${\rm M_{dust}}$--SFR proportionality, since dustier galaxy disks result in more pronounced line
asymmetries. The evidence for a relationship between LoSVD blue
asymmetry and stellar mass is instead more marginal ($\rho=0.362$ and $\rho=0.279$, respectively for [OIII] and H$\alpha$,
second column of Fig.~\ref{fig:v2_v98_diff}). The absence of a clear correlation is consistent
with the recent results by \cite{Santini+14}, who showed that the positive correlation 
between dust and stellar mass found by previous studies significantly flattens when separating galaxies 
according to their SFR.

Figure~\ref{fig:v2_v98_diff} shows that, in the stellar LoSVDs, blue asymmetries are also present but, due to
the large uncertainties, it is very difficult to investigate possible relationships with galaxy properties. However,
we note that there is some evidence for a weak correlation between stellar blue asymmetry and SFR ($\rho=0.551$),
and an even more marginal one with stellar mass ($\rho=0.401$), vaguely recalling the corresponding trends
observed in ionised gas.

In Figure~\ref{fig:appendixB} ($\S$~\ref{sec:losvd_stars}) we showed a check on the stellar continuum fitting for two galaxy stacks: one exhibiting 
a blue asymmetry in the stellar LoSVD, and one with no asymmetry.  These plots demonstrate that the detection of such a faint
feature in the stellar kinematics of external galaxies is made possible thanks to the high signal-to-noise reached in the composite
spectra, which allows us to fit the stellar continuum (and so extract the stellar kinematics) with unprecedented accuracy.
To our knowledge, blue asymmetries in the stellar LoSVDs of external galaxies have never been observed before.
Indeed, while an intrinsic kinematic asymmetry or ``lopsidedness'' in the stellar distribution is certainly possible for individual galaxies 
or at specific locations within galaxies, for example
due to stellar bars, galaxy interactions or even 
counter-rotating disks (e.g. \citealt{Krajnovic+15}), such effects should average out when 
combining the integrated spectra of thousands of different galaxies. The observation
of an asymmetry of the stellar LoSVD that is {\it consistently blue} (and never red) could be explained by 
the presence of stars in radial outward motions in galaxies, combined with obscuration by dust lanes in the galactic disk.
Dust in the disk would hamper mainly the detection of stars moving away from the galaxy with a redshifted (line-of-sight) velocity component, 
hence skewing the average stellar LoSVD towards blueshifted velocities.

The blue asymmetries in the stellar LoSVDs suggest 
that high-velocity runaway stars, hypervelocity stars (HVS, \cite{Brown+05})
and possibly high-velocity stellar clusters \citep{Caldwell+14}
in radial motions may be a rather common phenomenon in star forming galaxies.
High velocity stars in radial trajectories are usually very difficult to detect even in our own galaxy.
\cite{Palladino+14} discovered a sample of 20 candidate hypervelocity stars in the Milky Way, 
the bulk of which have velocities between 600 and 800~km~s$^{-1}$, and in at least
half of the cases exceed the escape velocity from the Galaxy.
These are G- and K-type dwarf stars, hence less massive and older than the
typical massive B-type HVSs \citep{Brown+12a}.
Interestingly, the trajectories of the stars detected by \cite{Palladino+14} are not consistent with ejection
 via three-body interactions between a binary system and the SMBH of our Galaxy or of M31, which
 is currently the most popular explanation for young B-type hypervelocity stars \citep{Hills88,Yu+Tremaine03}
 and predicts that one component of the binary system remains in a bound orbit around the SMBH.
On the contrary, the HVS candidates detected by \cite{Palladino+14} appear to 
originate from all directions, suggesting 
that another mechanism must be capable of accelerating stars to high speeds, such
as a supernova explosion in a close binary system \citep{Blaauw61, Zubovas+13b}, which is
currently the most likely origin of lower speed runaways ejected from the galactic disk \citep{Brown+15}.
Notably, the fastest HVS ever detected, US~708, with a Galactic rest-frame velocity of 1200~km~s$^{-1}$, has been recently
confirmed to be a solid candidate for an ejected Type Ia supernova donor remnant \citep{Geier+15}.
The supernova binary ejection scenario may be a viable explanation for our observations, supported by the tentative
correlation observed in Fig.~\ref{fig:v2_v98_diff} between the stellar blue asymmetry and the SFR.

 However, although HVSs may provide a plausible explanation for the blue asymmetries observed
in the stellar LoSVDs of local star forming galaxies, the properties and origin of these features
need further and in-depth investigation with high signal-to-noise data such as will be delivered by ongoing 
surveys (Gaia, MaNGA) and new facilities (MUSE, JWST, 30m class telescopes).
  
\subsection{Outflow properties}\label{sec:outflows}

Overall, the presence and properties of the line-of-sight velocity distributions of ionised gas
as derived from composite spectra of nearby non-active galaxies suggest ubiquity of 
star formation-driven galactic outflows. However, our analysis has also highlighted the presence of
additional factors, other than outflows, that are likely affecting the kinematics of
both the ionised gas and the stars, namely
virial motions, projection effects, gas content, and dust extinction. 
The next step is to try to isolate in the data the effects of gas outflows,
disentangling them from other mechanisms, in order to get a completely unbiased picture of their
occurrence and properties in local galaxies.

\begin{figure}[tp]
	 \centering
   \includegraphics[clip=true, trim=5.75cm 1.cm .8cm 2.3cm,angle=90,width=1.\columnwidth]{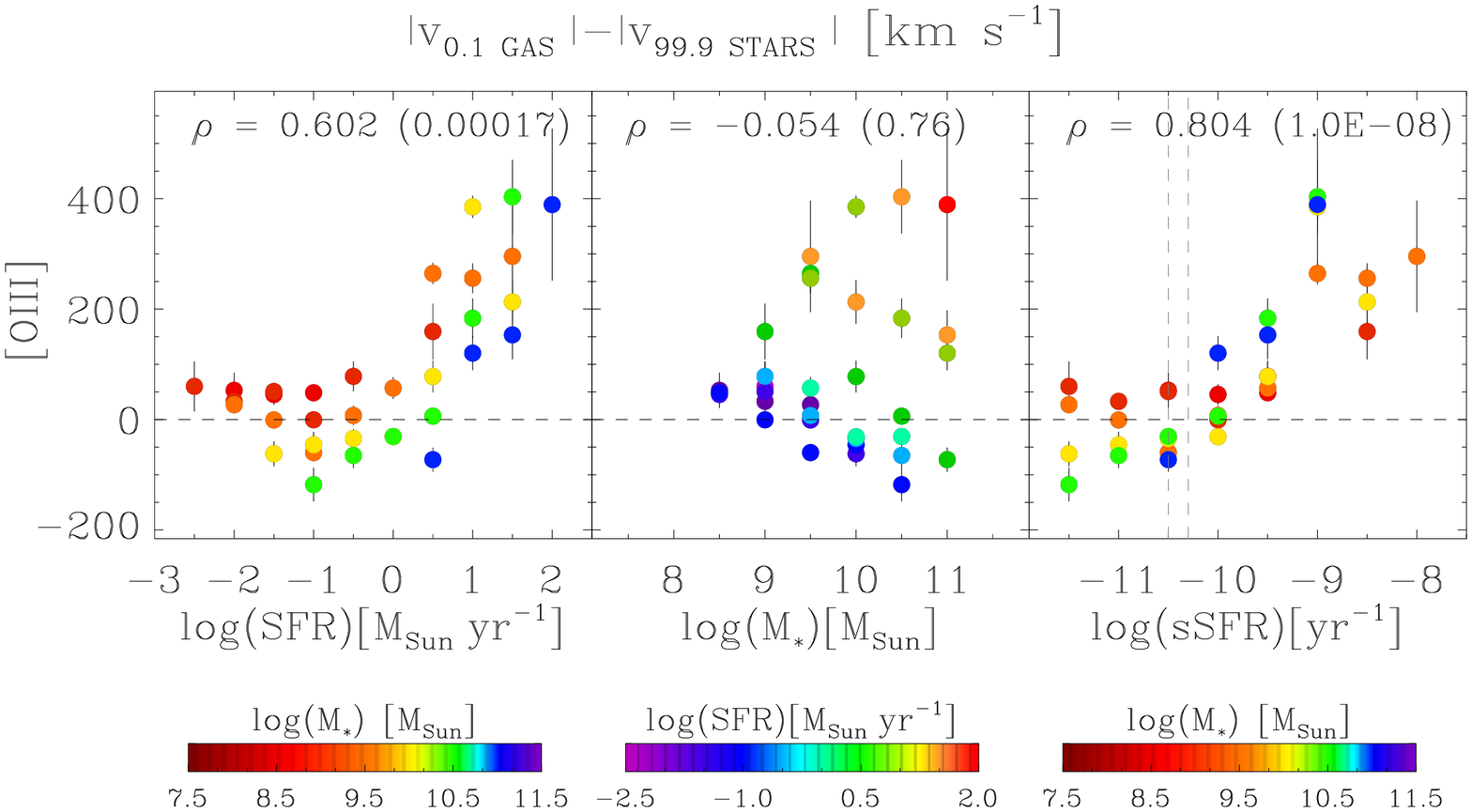}\\
   \includegraphics[clip=true, trim=1.8cm 1.cm 2.2cm 2.3cm,angle=90,width=1.\columnwidth]{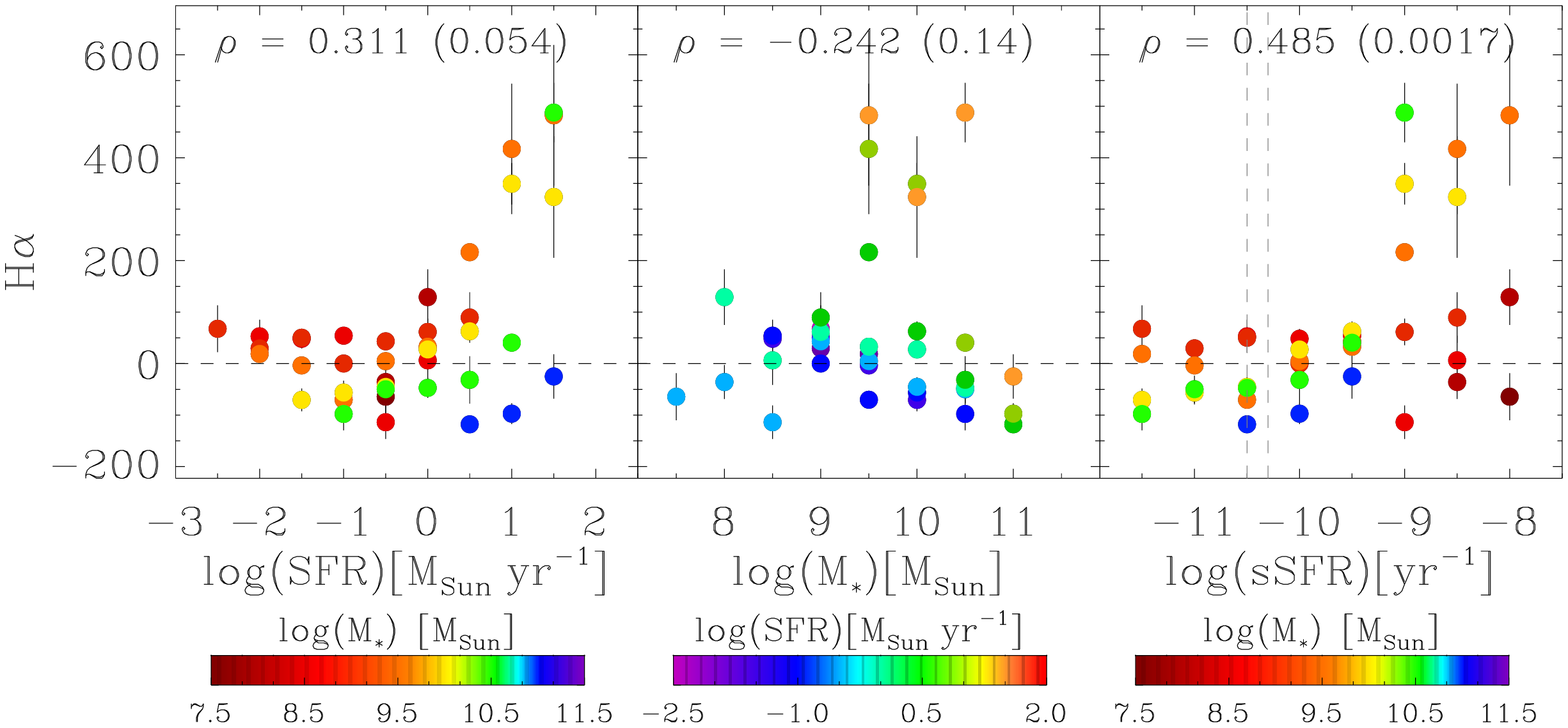}
   \caption{Excess of line-of-sight velocity of the ionised gas with respect to the stars ($v_{\rm gas}-v_{\rm stars}$) as given 
   		by $y = |v_{\rm 0.1~gas}|-|v_{\rm 99.9~stars}|$
   	        for [OIII] (\emph{upper panel}) and for H$\alpha$ (\emph{bottom panel}).
	       For display purposes, only the less noisy bins are plotted, i.e. only bins 
	       with either $\sigma_{y} < {\rm 50~km~s^{-1}}$ or $|y| \geq 2\sigma_{y}$.
	       As explained in $\S$~\ref{sec:outflows}, the quantity $y$ is a reliable tracer of 
	       ionised outflows only when it is
	       positive. 
	       The Spearman rank correlation coefficient ($0<|\rho|<1$, higher values of $|\rho|$ indicate stronger correlation, 
	       $\rho<0$ if there is an anti-correlation) 
	       is reported in each plot, along with the corresponding 
	       two-sided p-value (if the p-value is $\leq \alpha$, where $\alpha=0.05$ is the level of significance, the observed correlation is 
	       statistically significant). }
   \label{fig:plot_outflow_3sigma}
\end{figure}

\begin{figure}[tp]
\centering
   \includegraphics[clip=true, trim=-2cm 0cm 0cm -2.1cm,angle=90,width=\columnwidth]{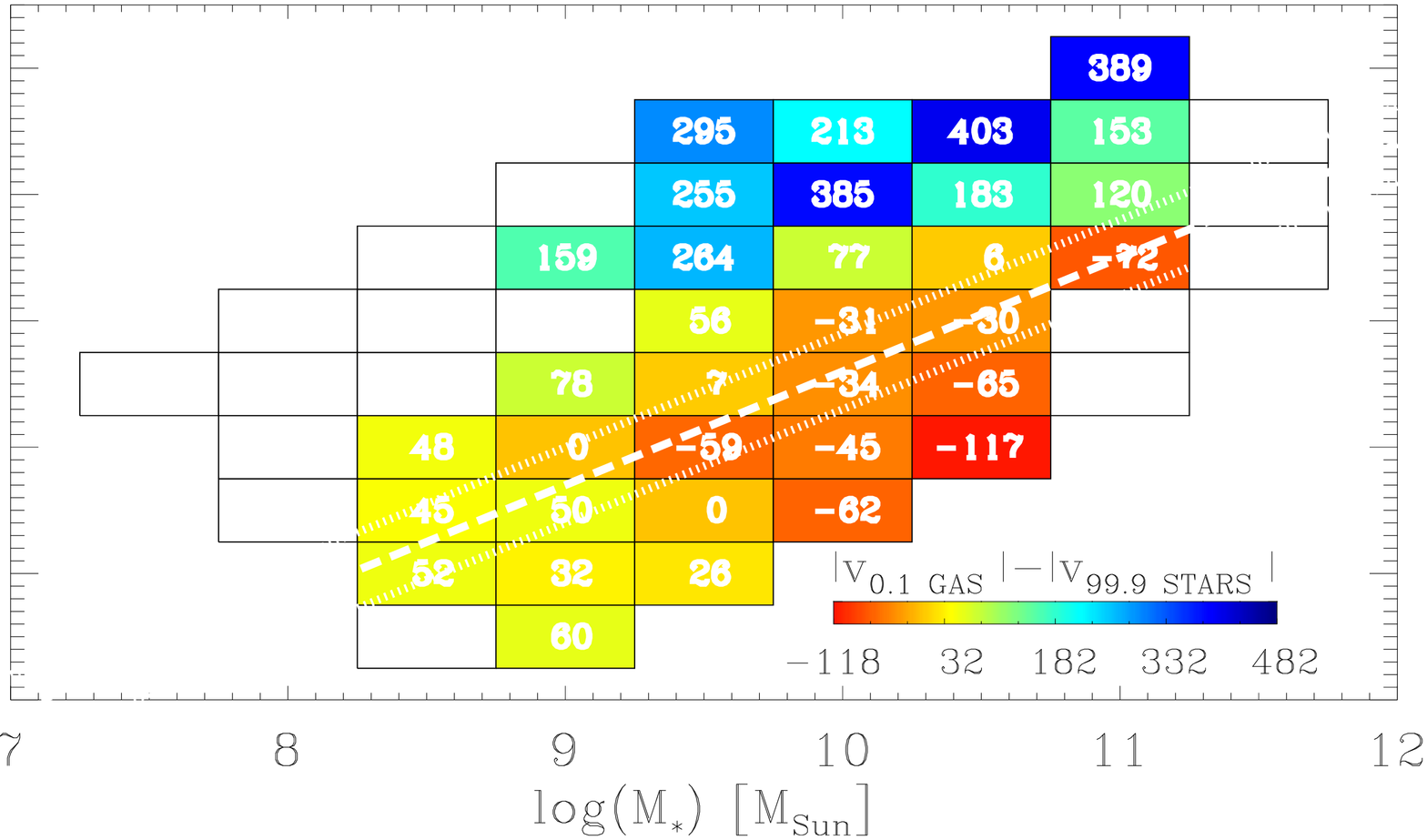}\\
   \includegraphics[clip=true, trim=-2cm 0cm 0cm -2.1cm,angle=90,width=\columnwidth]{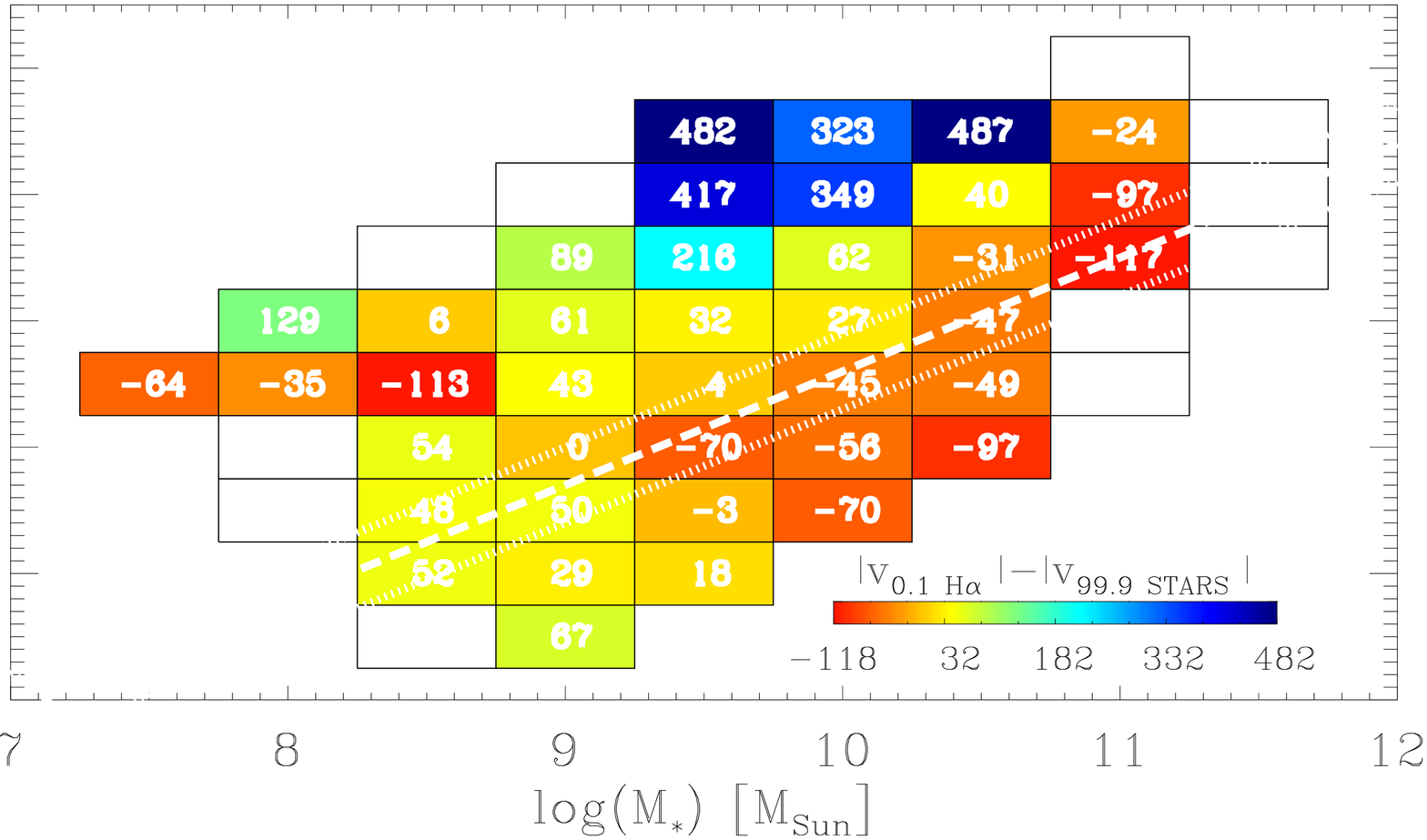}
   \caption{Grid of galaxy bins in the M$_*$--SFR parameter space used for the spectral stacking
   	(see also Fig.~\ref{fig:grid_sf}). Each bin is colour-coded according to the line-of-sight
      velocity difference $v_{\rm gas}-v_{\rm stars}$ as given by $|v_{\rm 0.1~gas}|-|v_{\rm 99.9~stars}|$ for [OIII]
     ({\it upper panel}) and for H$\alpha$ ({\it lower panel}). The $|v_{\rm 0.1~gas}|-|v_{\rm 99.9~stars}|$ values are also
     reported for each bin.
     The same bins as in Fig.~\ref{fig:plot_outflow_3sigma} are plotted.}
   \label{fig:grid_outflow_3sigma}
\end{figure}

The plots in Figs.~\ref{fig:sigma}, ~\ref{fig:v01} and \ref{fig:v2_v98_diff}
provided important clues on the concomitant mechanisms that have an impact on the 
LoSVDs of gas and stars. In the following, we briefly summarise
some logical deductions that will help us understand outflow properties:
\begin{enumerate}
\item The effects of virial motions on gas kinematics can be reduced considerably by looking at
the high velocity tail of the LoSVD of the ionised gas. In particular,
 the correlation between the line-of-sight velocity of the ionised gas and M$_*$ (which
is related to the dynamical mass) is weaker for the 0.1th and 99.9th
($\rho\sim0.6$) than for the 15.9th and 84.1th ($\rho\sim0.8$) percentile velocities (Table~\ref{table:rho_values}). Moreover, the highest 
velocities show the tightest correlation with the SFR ($\rho\gtrsim0.9$), indicating that the effects of star formation-driven 
feedback is indeed predominant.
\item The stellar LoSVDs appear quite broad, with stellar 
velocity dispersions as high as $\sim170$~km~s$^{-1}$, 
i.e. as high as the highest $\sigma$ observed in the ionised gas and, for large
stellar masses, even higher than the gas, probably because of the presence of stellar bulges
in massive galaxies combined with projection effects, as explained in $\S$~\ref{sec:kinematics}. 
As a consequence, only the {\it excess of gas
velocity with respect to the stars}, i.e. $v_{\rm gas}-v_{\rm stars}$, can reliably trace
gas motions due to gaseous outflows.
\item Because of dust attenuation, the blueshifted side of the LoSVD of the gas
is a better probe of outflowing gas than the redshifted side. Therefore, following
points (i) and (ii), gaseous outflows will be investigated through the 0.1th percentile velocity of the 
LoSVD of [OIII]
and H$\alpha$, from which we will subtract the corresponding stellar velocity.
\item Another complication for the study of ionised outflows is given by the presence of (modest) blue 
asymmetries in the stellar LoSVDs, 
whose origin is not clear and may be related to
non virial motions of stars in galaxies. In the following analysis we
will assume that this effect can be reduced (if not completely removed) by 
simply using the redshifted stellar velocities, i.e. by subtracting
from the 0.1th percentile LoS velocity of the gas, the 99.9th (rather than the 0.1th) percentile LoS 
velocity of the stars.
\end{enumerate}
By considering points 1 to 4, the most appropriate tracer of ionised outflows
is then given by the velocity difference $|v_{\rm 0.1~gas}|-|v_{\rm 99.9~stars}|$\footnote{We  
note that the velocity differences 
$|v_{\rm 0.1~gas}|-|v_{\rm 99.9~stars}|$ and $|v_{\rm 0.1~gas}|-|v_{\rm 0.1~stars}|$ 
used as outflow tracers produce results that are overall consistent. However, $|v_{\rm 0.1~gas}|-|v_{\rm 99.9~stars}|$
is probably a better choice because it allows us to minimise the effects due the 
presence of a blue asymmetry in the stellar LoSVDs, whose main consequence is lowering our
estimate of $v_{\rm gas}-v_{\rm stars}$, especially when using H$\alpha$ emission to trace the gas.}:
this quantity will be used to track ``secure'' outflows in
galaxy stacks. Indeed, if $(v_{\rm gas}-v_{\rm stars})>0$, then the
kinematics of gas and stars are decoupled and we are reliably tracing gaseous outflows. 
The only other mechanism
that would explain such decoupling would be galaxy mergers, but the merger fraction is
negligible in the local Universe \citep{Patton+Atfield08}, so gaseous 
outflows are the only viable explanation.

Figure~\ref{fig:plot_outflow_3sigma} shows the
relationships between the quantity $|v_{\rm 0.1~gas}|-|v_{\rm 99.9~stars}|$, evaluated for both
[OIII]- and H$\alpha$-emitting gas, and the 
galaxy physical parameters (i.e. SFR, ${\rm M_{*}}$ and sSFR). In Fig.~\ref{fig:grid_outflow_3sigma} we show
the stacking grid in the M$_*$--SFR parameter space defined by the sample of local star forming
galaxies used in this study (i.e. the same grid as Fig.~\ref{fig:grid_sf}), with each bin colour-coded according to the
value of $|v_{\rm 0.1~gas}|-|v_{\rm 99.9~stars}|$ (measured for [OIII] and H$\alpha$).

It is worth emphasising that the negative values of $|v_{\rm 0.1~gas}|-|v_{\rm 99.9~stars}|$
estimated in some galaxy bins (especially at ${\rm log(M_{*} [M_{\odot}]) \geq 10.0}$), 
are probably a consequence of
the conjunction of projection effects and of the absence of detectable galactic outflows.
More specifically, ignoring galactic outflows (which usually expand out perpendicular to the galactic disk, 
along its minor axis, as shown by observations, e.g.
\citealt{Lehnert+Heckman96, Heckman+00, Chen+10, Rubin+14, Cazzoli+14}), 
the ionised gas is confined to the plane of the galactic disk, and the projection effects described in $\S$~\ref{sec:kinematics},
affecting mostly observations of face-on galaxies with a prominent stellar bulge (which dominate
the population of high-M$_*$ galaxies, e.g. \citealt{Gavazzi+15}), can be even more accentuated.
For this reason, negative values of $|v_{\rm 0.1~gas}|-|v_{\rm 99.9~stars}|$ simply indicate galaxy
bins in which outflows are not present, or in which outflows are not prominent enough to be detected:
as a consequence, this quantity can reliably trace outflows
{\it only when it is positive}.

Although many attempts have been made to establish fundamental
``scaling'' relationships linking outflows and galaxy properties 
\citep{Martin05, Rupke+05,Chen+10, Westmoquette+12, Martin+12, Rubin+14, Arribas+14, Heckman+15},
the general picture remains unclear and contradictory. The main reason
for this impasse may be the absence of a study on galactic outflows of large statistical significance: 
most of previous studies relied on limited samples of objects,
showing extreme characteristics and spanning only a narrow range
of galaxy properties (e.g. SFR, M$_*$). As a consequence, previous results do not adequately 
represent the whole local population of galaxies. Our analysis instead capitalises on a large statistical
and almost unbiased sample of local star forming galaxies, particularly suitable for determining scaling
relationships between galactic outflows and galaxy properties.
Moreover, our method allows, to accurately identify and separate of the effects of star-formation driven 
outflows from the other mechanisms affecting the LoSVD of the gas. 

The general picture emerging from Figs.~\ref{fig:plot_outflow_3sigma} 
and \ref{fig:grid_outflow_3sigma} is the following\footnote{We stress that all results discussed in this section and in the subsequent
analyses do not depend on the choice of using $v_{\rm 99.9~stars}$ instead of $v_{\rm 0.1~stars}$ to measure $v_{\rm stars}$.}: the excess of gas velocity
 with respect to the stars (i.e. $v_{\rm gas}-v_{\rm stars}$, as traced by $|v_{\rm 0.1~gas}|-|v_{\rm 99.9~stars}|$), which we use in this study to trace galactic outflows,
depends weakly on the SFR ($\rho=0.602$ for [OIII], but for H$\alpha$ the correlation with SFR is much weaker, $\rho=0.311$, and not
statistically significant as indicated by the high p-value) and more tightly on the sSFR ($\rho=0.804$ and $\rho=0.485$ for [OIII] and H$\alpha$ respectively).
However, we note that for galaxy bins identified by higher stellar masses
the correlation between $|v_{\rm 0.1~gas}|-|v_{\rm 99.9~stars}|$ and SFR {\it at a fixed M$_*$} 
becomes significantly tighter for both [OIII] and H$\alpha$, i.e. $\rho>0.8$ (with $\rm p-value\ll0.05$) 
at M$_*\geq10^{9.5}$~M$_{\odot}$ and even more so at M$_*\geq10^{10}$~M$_{\odot}$, for which we measure 
$\rho>0.95$.

Globally, there is instead no relation between 
$|v_{\rm 0.1~gas}|-|v_{\rm 99.9~stars}|$ and M$_*$ (second column of Fig.~\ref{fig:plot_outflow_3sigma}): 
$\rho=-0.054$ for [OIII] and $\rho=-0.242$ for H$\alpha$, but with high p-values, implying 
that such weak anti-correlation is not significant.
For some galaxy bins, an anti-correlation appears between $|v_{\rm 0.1~gas}|-|v_{\rm 99.9~stars}|$ and M$_*$ {\it at a fixed SFR}.
This effect is however highly significant ($\rho<-0.9$) for both [OIII] and H$\alpha$ only at
$\rm log(SFR [M_{\odot}~yr^{-1}])=-1.0$ and $\rm log(SFR [M_{\odot}~yr^{-1}])=0.5$.
Since the stellar mass traces the depth of the gravitational potential in galaxies, these observations
would suggest that, on equal SFRs, it becomes increasingly 
difficult to launch gaseous outflows in larger gravitational potentials 
by means of star formation feedback only.
This qualitatively agrees with theoretical predictions (further discussed
in $\S$~\ref{sec:discussion_outflows}) as well as with 
indirect observational evidence for a mass dependance of negative feedback 
provided by studies of the ``mass-metallicity'' relationship in galaxies
\citep{Tremonti+04, Maiolino+08}.
These studies showed that 
the gas-phase metallicity increases steadily with stellar mass, 
at least up to ${\rm log(M_{*} [M_{\odot}])\simeq 10.5}$,
suggesting that low mass galaxies are less able than massive galaxies to retain in their gas 
phase the metals provided by supernova explosions, probably because they are more vulnerable to galactic outflows.

In addition, Fig.~\ref{fig:plot_outflow_3sigma} reveals some interesting differences
between [OIII] and H$\alpha$ as tracers of ionised outflows. In particular, H$\alpha$ emission
seems to be a worse tracer of galactic outflows than [OIII] emission:
(i) The correlation between $v_{\rm gas}-v_{\rm stars}$ and SFR is considerably weaker and less
significant for H$\alpha$ than for [OIII], and the same holds for the correlation with sSFR; 
(ii) for bins with $\rm log(M_{*} [M_{\odot}])= 11.0$, high values
of $|v_{\rm 0.1~gas}|-|v_{\rm 99.9~stars}|$ are inferred by using [OIII], whereas no velocity excess is detected
when using H$\alpha$ emission as a gas tracer.
Qualitatively, an higher outflow ionisation in more massive galaxies
could explain the differences that we evidence between [OIII] and H$\alpha$ as outflow tracers.
We note that it is unlikely that such a difference between the H$\alpha$ and [OIII] LoSVDs is a consequence
of residual stellar H$\alpha$ absorption. Indeed, the H$\alpha$ LoSVDs are obtained by fitting the H$\alpha$ and
[NII]$\lambda\lambda$6548,6583 lines simultaneously (see $\S$~\ref{sec:losvd_fit}), by constraining the kinematics 
of all components employed in the fit to be the same for the three lines. Therefore, the resulting LoSVDs reflect also
the kinematics traced by the [NII] lines, which are much less affected by possible residual stellar absorption.

Finally, we note that there is an intriguing property of galactic outflows that stands out immediately 
from Fig.~\ref{fig:grid_outflow_3sigma}:
local non-active galaxies hosting powerful outflows are located {\it above} the main sequence of star
forming galaxies. 
The absence (or insignificance) of galactic outflows along and/or below the MS, which
we show here for the first time in a clear way, probably constitutes a striking feature of the local galaxy population, and
it may persist to higher redshifts, as marginally shown in outflow studies at $z\sim1$ \citep{Martin+12}.
The relationship between galactic outflows and the MS of star forming galaxies will be further discussed in
$\S$~\ref{sec:discussion_MS}.


\section[]{Discussion}

\subsection{Implications of outflow properties for feedback models}\label{sec:discussion_outflows}

Intense star formation activity
conveys energy and momentum to its surroundings, which can then be transmitted by various
physical mechanisms to larger scales, thus eventually impacting the entire host galaxy
and affecting its capability of forming stars (negative feedback).
Although negative feedback from star formation can manifest under various forms (such as galactic outflows
of ionised, neutral and molecular gas, shocks, hot bubbles and cavities, metal-enrichment of the IGM), its direct and unambiguous 
signature can be very difficult to detect in most galaxies.
As a consequence, the detailed physics underlying this phenomenon is not yet completely understood, as
well as its relevance for quenching star formation in galaxies.
However, significant advances have been accomplished by theorists in this field, who,
by also exploiting the valuable information provided by observational studies,
propose two main scenarios for star formation-driven feedback to occur:

\begin{enumerate}
\item According to the first,``canonical'' scenario, the kinetic energy released by supernova explosions and,
especially in the very first stages of the starburst ($\Delta t \leq 10^7 \rm yrs$) and at high metallicities ($\mathcal{Z} \geq \mathcal{Z}_{\odot}$), 
by stellar winds from OB and Wolf-Rayet stars, plays the most important role in counteracting and self-regulating star formation
( ``energy-driven'' scenario, e.g. \citealt{Chevalier77, Leitherer+92, Chevalier+Clegg85, Springel+Hernquist03, Veilleux+05}). 
This feedback mechanism is subject to the thermalisation efficiency of the energy deposited into the ISM:
such efficiency may vary, but it is believed to be around 1-10\%; therefore, up to $\sim$99\% of 
the total kinetic energy injected by star formation may be dissipated radiatively in dense gas \citep{Chevalier77, 
Murray+05, Thornton+98, MacLow+Ferrara99}.
If the transfer from kinetic energy from SNe and stellar winds to cloud motions is efficient enough to unbind the gas in the galaxy, it can 
drive strong and large-scale outflows \citep{Chevalier+Clegg85}. 
A natural outcome of this model is that energy-driven outflows are more important, at a given
SFR (which is directly related to the kinetic energy injected by supernova ejecta and by stellar winds into the ISM),
in low mass galaxies residing in less massive dark matter halos, because it is easier 
for gas to escape from their shallower gravitational wells \citep{Dekel+Silk86,Springel+Hernquist03,Dave+11a}.
\item In the second scenario, the momentum transferred by the UV radiation from young and massive
stars to dust and the momentum injected by supernova
explosions\footnote{The latter is thought to be important especially at later stages and in regions where the infrared optical depth is low,
such as at large radii from the galactic centre.} (``momentum-driven'' scenario) dominate 
the feedback mechanism \citep{Murray+05, Dave+11a}.
The advantage of momentum-driven outflows is that they can be effective even when radiative losses are high, because 
momentum cannot be radiated away \citep{Murray+05}. However, this feedback mechanism, in order to be effective, requires
ISM conditions that may not be satisfied in normal local star-forming galaxies, such as 
large optical depths to infrared radiation \citep{Dave+11a, Hopkins+11}, as well as collisional coupling of gas and dust in the outflow
\citep{Murray+05, Dave+11a}. For these reasons, momentum-driven outflows may be most effective at launching
cold and dense gas (which is likely hydrodynamically coupled to dust) in dusty star-forming
galaxies and AGNs \citep{Murray+05, Fabian12}. 
We note that in this model, according to the original formulation by \cite{Murray+05}, the condition
that must be satisfied in order to launch gas outward via momentum injection
is that the total momentum flux exceeds the gravitational force, which
translates into a condition on the luminosity of the starburst $L \geq L_{M}$, where $L_M$ is an
``Eddington-like'' luminosity threshold. If $L\geq L_{M}$, a momentum-driven wind
develops, with velocity $v_{\rm out} \gtrsim 2\sigma_{*}$ (where $\sigma_*$ is the stellar velocity dispersion;
\citealt{Murray+05}).
\end{enumerate}

In $\S$~\ref{sec:outflows} we have identified the galaxy stacks displaying the signature of
galactic outflows as those for which the excess of gas velocity with respect to the stellar velocity, as estimated
from the high-velocity tail of the LoSVD, is significant. More specifically,
we have considered values of ${|v_{\rm 0.1~gas}|-|v_{\rm 99.9~stars}|}$ significantly greater than zero
as a clear signature of outflows.
The next step would be comparing the scaling relationships obtained for the observed
outflows with the expectations of theoretical models.
However, the quantity ${|v_{\rm 0.1~gas}|-|v_{\rm 99.9~stars}|}$, while being suitable
for {\it identifying} outflows, is not appropriate for the comparison with models, because 
it has no clear interpretation in terms of the physical quantities usually employed by theorists.
Unfortunately, because of the information lost with the stacking, it is not
possible to estimate the mass entrained in the observed outflows,
and therefore to calculate some important parameters such as mass-loss rate, 
kinetic power and momentum rate of the outflows, which would help to discriminate
between different feedback models. Most importantly, spatially resolved spectroscopic information
are needed to infer these quantities \citep{Rupke+Veilleux13, Cazzoli+14, Arribas+14}. 
However, we can obtain a proxy of the outflow velocity, and this can be exploited to shed light on the
physical mechanism powering the ionised outflows.

The LoSVD of ionised gas traces a combination of dynamical
motions of gas clouds within the gravitational potential of the galaxy and of perturbed
motions due to star formation feeedback-related processes such as outflows and turbulence 
in the disk. In the following analysis, since we consider only the high-velocity tail of the LoSVD, we
assume that outflows dominate over turbulence, and so, simplistically, that any non-gravitational motions of the gas
at such high velocities are due to outflows and not to turbulence. Therefore, the LoSVD of the gas may be regarded as the 
convolution of two line-of-sight velocity distributions, one tracing the virial motions,
and the other one tracing the outflow:

\begin{equation}\label{eq:outflow_conv}
\mathcal{L}_{\rm gas}(v) = \mathcal{L}_{\rm virial}(v)  \ast \mathcal{L}_{\rm outflow}(v).
\end{equation}

We note that such approximation reflects the fact that 
the typical outflows detected in our galaxy sample are rather modest, and most likely
consist of localised bubbles and chimneys, emanating from star forming regions spread across
the stellar disk (see also the case of NGC~1569 discussed by \citealt{Martin+02}). In this scenario, the outflowing
material participates, at least to some extent, also to the virial (rotational) motions in the galaxy.
However, we stress that Eq.~\ref{eq:outflow_conv} is not universally applicable. In particular,
Eq.~\ref{eq:outflow_conv} does not hold if the kinematics of the high velocity gas 
is largely {\it dominated} by a large-scale ``bulk outflow'' expanding perpendicular to the disk. This is the case
of powerful starburst-driven superwinds (e.g. M~82, NGC~253) if they are probed on scales of at least a few kpc, i.e.
sufficiently larger than the typical vertical scale height of the stellar disk, where the motions of the gas in outflow is clearly 
distinct from the disk motions. Indeed, even the strong ``bulk outflows'' that develop in powerful starbursts such as M~82 
appear to originate as a collection of individual chimneys expanding out from the disk that subsequently merge to form the classical
biconical outflows observed on kpc scales \citep{Wills+99,Lehnert+Heckman96}.

Following from Eq.~\ref{eq:outflow_conv}, by (simplistically) approximating both $\mathcal{L}_{\rm virial}(v)$ and $\mathcal{L}_{\rm outflow}(v)$
with Gaussian distributions, the line-of-sight velocity dispersion of the gas corresponds to:
\begin{equation}
\sigma_{\rm gas}^2 =  \sigma_{\rm virial}^2 + \sigma_{\rm outflow}^2,
\end{equation}
where $\sigma_{\rm virial}$ can be approximated by the stellar line-of-sight velocity dispersion,
i.e. $\sigma_{\rm virial} \approx \sigma_{\rm stars}$.
As a result, the ``outflow'' line-of-sight velocity dispersion would be:
\begin{equation}\label{eq:outflow_sigma}
\sigma_{\rm outflow} =  (\sigma_{\rm gas}^2 - \sigma_{\rm stars}^2)^{1/2}.
\end{equation}
We note that Eq.~\ref{eq:outflow_sigma} can be extended also to the percentile
velocities as, in the approximation of Gaussian distributions, they correspond to multiples
of the velocity dispersion. Therefore, according to this line of reasoning, the 
outflow velocity can be approximated by the quantity $(v_{N,{\rm gas}}^2-v_{N,{\rm stars}}^2)^{1/2}$,
where $v_{N}$ indicates the $N$-th percentile velocity. 

As we commented earlier, because of the observed blue asymmetries observed in the {\it stellar} LoSVDs 
of some galaxy stacks ($\S$~\ref{sec:asym}), 
the percentile velocities on the blue side of the stellar LoSVD, i.e. the 15.9th, 2.3th and 0.1th, may be
not appropriate to trace virial motions in the galaxy. Therefore, in the following, we 
adopt the same strategy as in $\S$~\ref{sec:outflows} and define the outflow velocity as:
\begin{equation}\label{eq:v_out}
v_{\rm out} = (v_{N,{\rm gas}}^2-v_{100-N,{\rm stars}}^2)^{1/2},
\end{equation}
where the percentile velocity of the gas is chosen on the blue side of the LoSVD (because it
is less affected by dust), i.e. $N$= 0.1, 2.3 or 15.9, and, for the stars,
we use $100-N$, which gives the corresponding percentile velocity on the red side of the LoSVD. 
We note that $v_{\rm out}$ can be inferred from Eq.~\ref{eq:v_out} only in those cases for
which $v_{N,{\rm gas}} > v_{100-N,{\rm stars}}$: this means that $v_{\rm out}$ cannot be calculated
for quite a few bins below the MS, as illustrated by Fig.~\ref{fig:grid_outflow_3sigma}.

\begin{figure}[tbp]
	 \centering
   \includegraphics[clip=true, trim=5.75cm 1.cm .8cm 2.7cm,angle=90,width=1.\columnwidth]{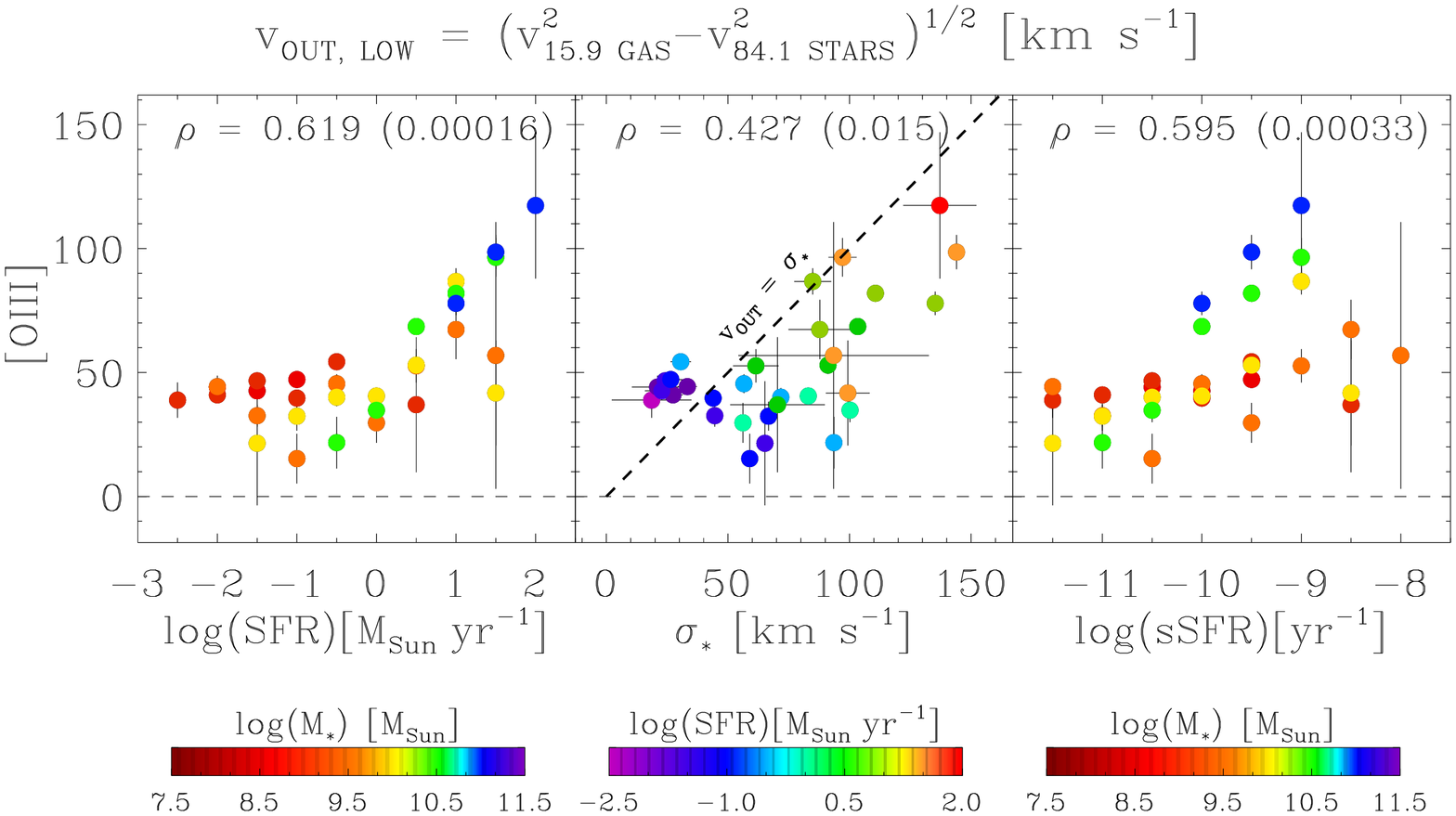}\\
   \includegraphics[clip=true, trim=1.8cm 1.cm 2.2cm 2.7cm,angle=90,width=1.\columnwidth]{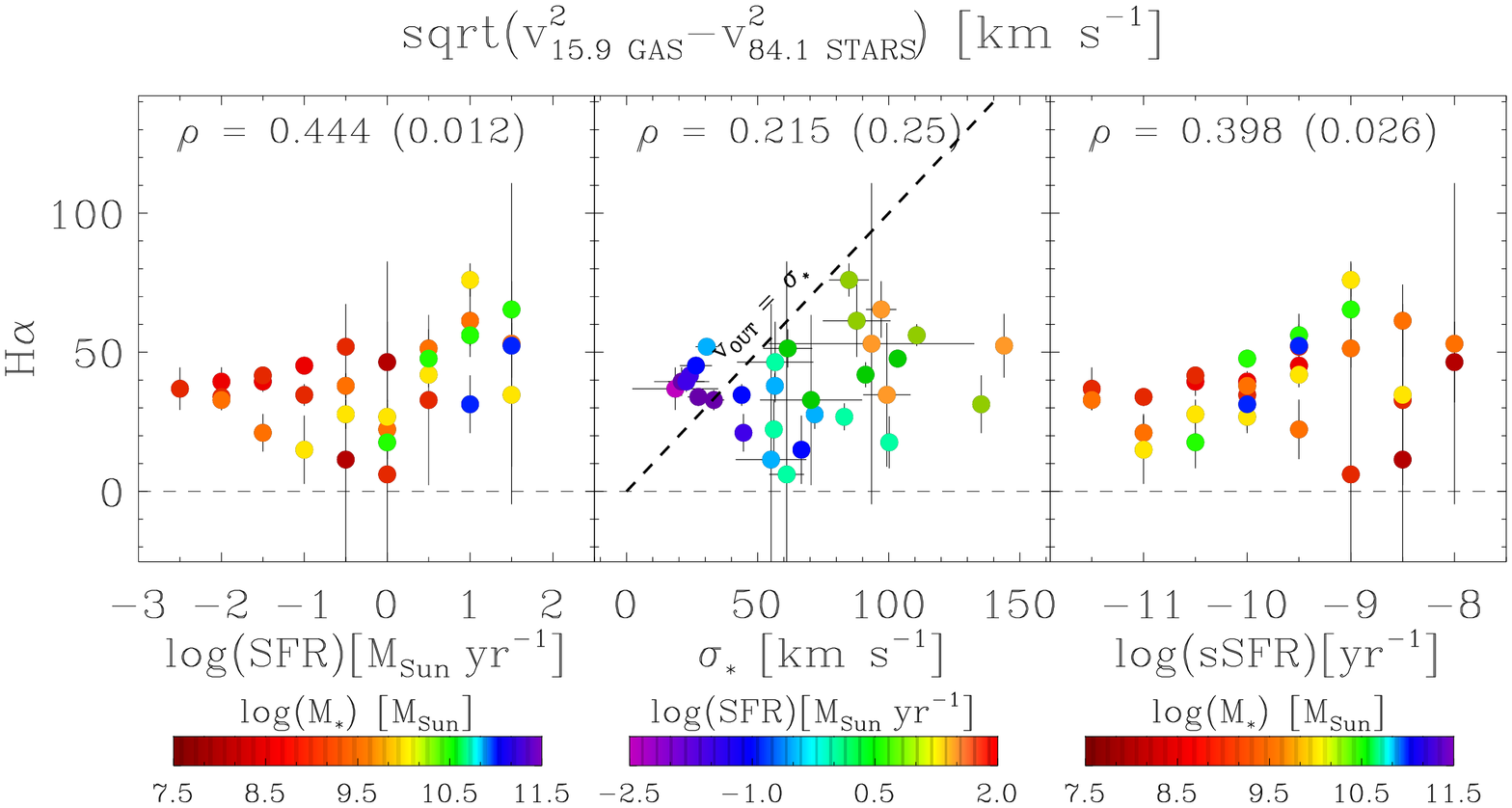}
   \caption{Outflow velocity as given 
   		by $(v_{\rm 15.9~gas}^2-v_{\rm 84.1~stars}^2)^{1/2}$ (see explanation in $\S$~\ref{sec:discussion_outflows})
		inferred from [OIII] ({\it upper panel}) and H$\alpha$ emission ({\it bottom panel}), plotted as a function
	       of SFR, stellar velocity dispersion (as obtained from the fit to the stellar continuum), and sSFR.
	       The same bins as Fig. \ref{fig:plot_outflow_3sigma} are plotted, excluding those for which
	       $(v_{\rm 15.9~gas}^2-v_{\rm 84.1~stars}^2)<0$. In the middle panels, the dashed lines indicate the 
	       locus of points where $v_{\rm out} = \sigma_{\rm stars}$.
	       The Spearman rank correlation coefficient ($0<\rho<1$, higher values of $\rho$ indicate stronger correlation) 
	       is reported for each plot, along with the corresponding 
	       two-sided p-value (if the p-value is $\leq \alpha$, where $\alpha=0.05$ is the level of significance, the observed correlation is 
	       statistically significant).}
   \label{fig:plot_outflow_vel_1sigma}
\end{figure}

\begin{figure}[tbp]
	 \centering
   \includegraphics[clip=true, trim=5.75cm 1cm .8cm 2.5cm,angle=90,width=1.\columnwidth]{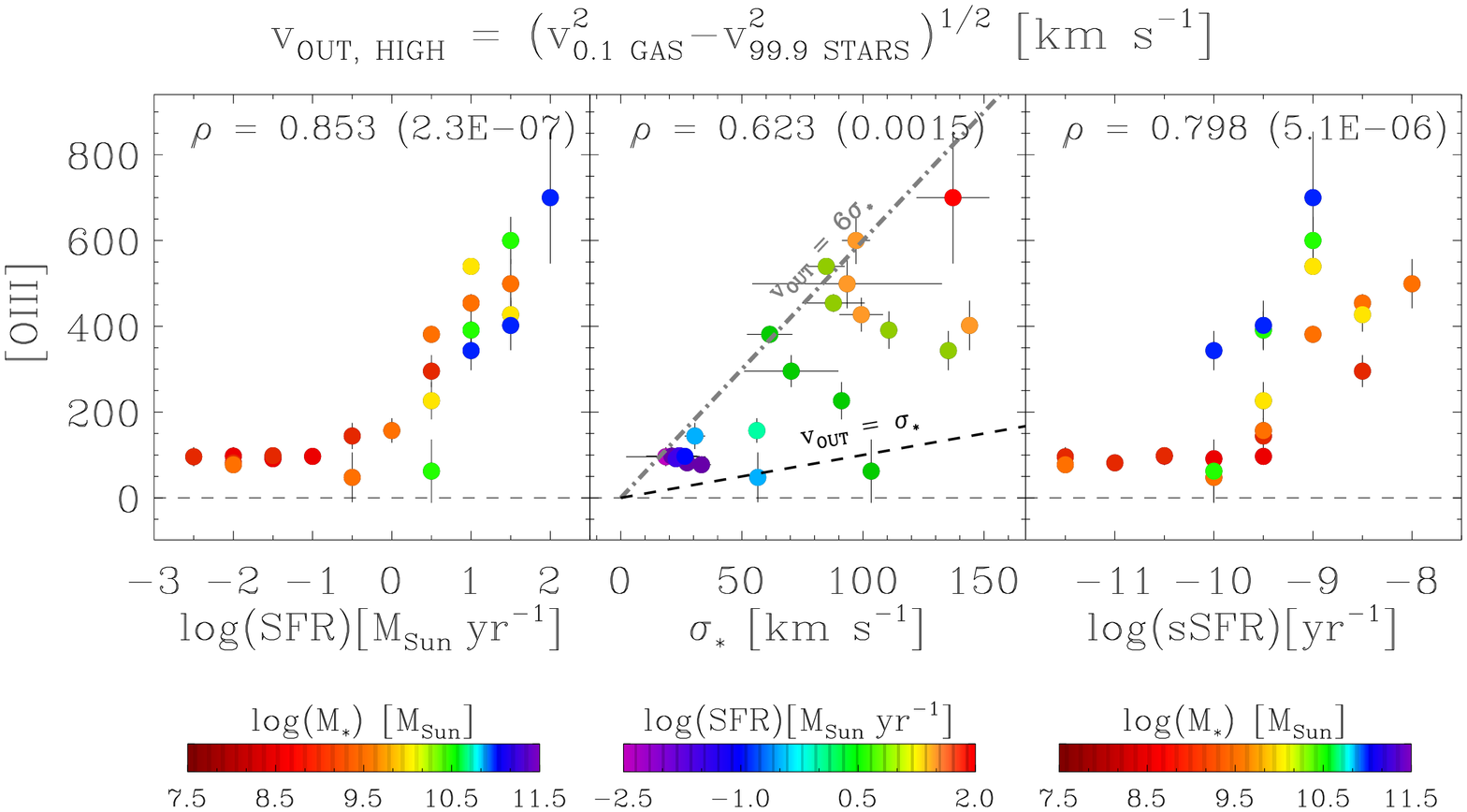}\\
   \includegraphics[clip=true, trim=1.8cm 1cm 2.2cm 2.5cm,angle=90,width=1.\columnwidth]{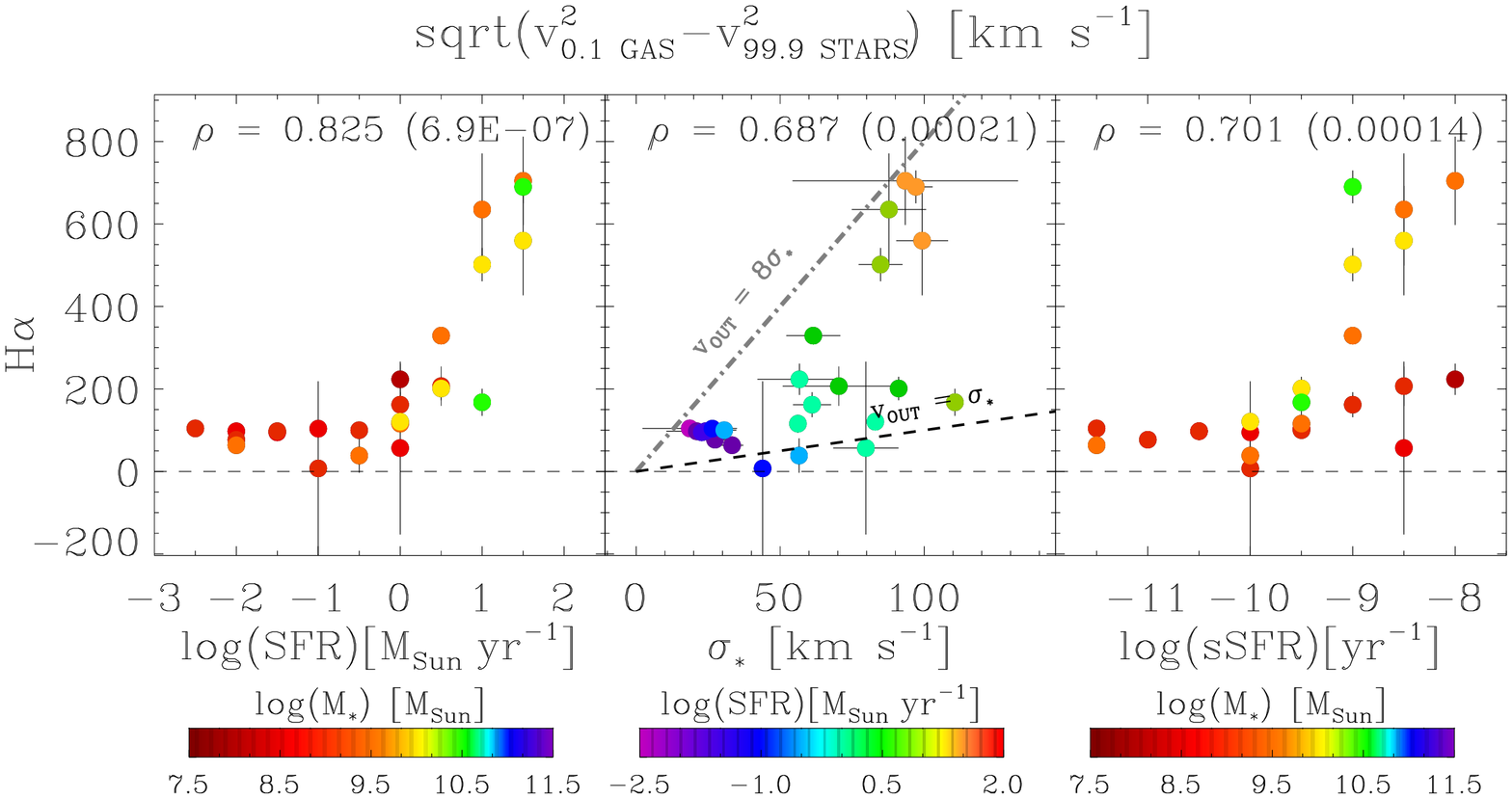}
   \caption{Outflow velocity as given 
   		by $(v_{\rm 0.1~gas}^2-v_{\rm 99.9~stars}^2)^{1/2}$ (see explanation in $\S$~\ref{sec:discussion_outflows})
   	       inferred from [OIII] (\emph{upper panel}) and H$\alpha$ emission (\emph{bottom panel}), plotted as a function
	       of SFR, stellar velocity dispersion (as obtained from the fit to the stellar continuum), and sSFR.
	       The same bins as Fig.~\ref{fig:plot_outflow_3sigma} are plotted, excluding those for which
	       $(v_{\rm 0.1~gas}^2-v_{\rm 99.9~stars}^2)<0$. In the middle panels, the black dashed lines indicate the 
	        locus where $v_{\rm out} = \sigma_{\rm stars}$, and the grey dot-dashed
	        lines indicate the approximate upper envelopes of the distributions of values, given by the relationships:
	        $v_{\rm out} = 6\sigma_{\rm stars}$ for [OIII], and
	        $v_{\rm out} = 6\sigma_{\rm stars}$ for H$\alpha$.
	        The Spearman rank correlation coefficient ($0<\rho<1$, higher values of $\rho$ indicate stronger correlation) 
	       is reported for each plot, along with the corresponding 
	       two-sided p-value (if the p-value is $\leq \alpha$, where $\alpha=0.05$ is the level of significance, the observed correlation is 
	       statistically significant). 
	    }
   \label{fig:plot_outflow_vel_3sigma}
\end{figure} 

Figures~\ref{fig:plot_outflow_vel_1sigma} and \ref{fig:plot_outflow_vel_3sigma} show
the outflow velocity ($v_{\rm out}$) as a function of galaxy properties (e.g. SFR, $\sigma_*$ and sSFR), for two
different estimates of the outflow velocity, indicated as $v_{\rm out, low}$ and $v_{\rm out, high}$,
 probing respectively the low and the high velocity gas within the ionised outflows, 
 and defined as $v_{\rm out, low}\equiv(v_{\rm 15.9~gas}^2-v_{\rm 84.1~stars}^2)^{1/2}$
and $v_{\rm out, high}\equiv(v_{\rm 0.1~gas}^2-v_{\rm 99.9~stars}^2)^{1/2}$.
We note that in the middle panels of 
Fig.~\ref{fig:plot_outflow_vel_1sigma} and \ref{fig:plot_outflow_vel_3sigma}
the outflow velocity is shown as a function of the stellar velocity dispersion 
instead of the stellar mass, so that these trends can be readily compared to the predictions of 
models of star formation-driven feedback.
Figures~\ref{fig:plot_outflow_vel_1sigma} and \ref{fig:plot_outflow_vel_3sigma} indicate
that the outflow velocity scales with SFR and sSFR. In particular, $v_{\rm out, high}$, shows a very tight 
($\rho>0.8$) correlation with the SFR. Previous studies found a similar increase of outflow
speed with SFR (e.g.  \citealt{Martin05} using a sample of IR luminous galaxies at $z\sim0$\footnote{ Although the
correlation in their Fig.~6 is mainly driven by the
three dwarf starbursts with detected NaID outflows from \cite{Schwartz+Martin04}.}; \citealt{Arribas+14} for outflows
detected in $z\sim0$ (U)LIRGs; \citealt{Chisholm+14} in local star forming and starburst galaxies; 
\citealt{Heckman+11} and \citealt{Heckman+15} using a a sample of UV-luminous starbursts and Lyman break
analogs at $z\sim0$, \citealt{Bradshaw+13} and \citealt{Weiner+09} by means of a stacking analysis of star forming galaxies at the typical redshifts of $z\sim1.2$
and $z\sim1.4$, respectively) and with sSFR \citep{Bradshaw+13,Heckman+15}. However, the correlations found by previous studies are in most cases
less significant than in our study, and inevitably based on smaller number statistics. 
Furthermore, the literature also reports many cases in which observations failed
at finding a clear correlation between outflow velocity and SFR 
(e.g. \citealt{Heckman+00}, \citealt{Rupke+05}, \citealt{Chen+10}, and \citealt{Westmoquette+12} at $z\sim0$ \footnote{By using their data only,
neither \cite{Rupke+05} nor \cite{Westmoquette+12} found a significant correlation between outflow velocity and SFR, the correlation
appears only when including the dwarf starbursts from \cite{Schwartz+Martin04}.}; \citealt{Sato+09} at $z\sim0.4$; \citealt{Rubin+14} at 
$z\sim0.5$; \citealt{Kornei+12} and \citealt{Martin+12} at $z\sim1$; 
\citealt{Erb+12} at $z\sim1.5$). In most cases, the main issue appears to be the narrow dynamical range of host galaxy properties
probed by such outflow studies, which is crucial to recover scaling relations \citep{Rupke+05,Westmoquette+12,Martin+12}.
Inspection of Figs.~\ref{fig:plot_outflow_vel_1sigma} and \ref{fig:plot_outflow_vel_3sigma} suggests that a strong
correlation between outflow velocity and SFR is characteristic of galaxies with 
 $\rm log(SFR [M_{\odot}~yr^{-1}]) \gtrsim 0$; at $\rm log(SFR[M_{\odot}~yr^{-1}]) < 0$ 
the dependence of $v_{\rm out}$ on SFR is nearly flat.

Overall we find that the outflow velocity increases with the stellar velocity dispersion,
as shown by Fig.~\ref{fig:plot_outflow_vel_3sigma} 
(a positive correlation between $v_{\rm out, high}$ and $\sigma_*$ is observed with
$\rho>0.6$ and p-value$\ll 0.05$ 
for both [OIII] and H$\alpha$ outflows)
and, to some extent, also 
by Fig.~\ref{fig:plot_outflow_vel_1sigma} 
(a weak correlation between $v_{\rm out, low}$ and $\sigma_*$ is found for [OIII], although
not for H$\alpha$). 
 While the existence of a correlation between gas velocity and SFR is a more or
less established property of galactic outflows, the relationship with stellar
mass (or with the dynamical mass, as probed by the galaxy rotation speed or by the 
stellar velocity dispersion) is more controversial. 
\cite{Lehnert+Heckman96}, in their sample of local, 
edge-on IR luminous galaxies did not find any relationship between the kinematical properties
(emission line widths, velocity shear) of the ionised gas, as measured 
along the minor axis
of the disks (where outflows are found to dominate the gas kinematics), 
and galaxy rotation speeds. Similarly, \cite{Heckman+00} showed that the NaI
outflow velocity is unrelated to the galaxy rotational velocity. On the contrary, \cite{Martin05} found an almost linear
relation with galaxy rotation speed (although they could only
test this relationship for a small fraction of the galaxy sample with available rotation curves).
The positive correlation between maximum outflow velocity and  
galaxy circular velocity found by \cite{Rupke+05} is completely driven
by the dwarf starburst sample of \cite{Schwartz+Martin04}, highlighting once again the importance
of probing a wide dynamical range in galaxy properties.  
In line with this hypothesis, \cite{Arribas+14} found that the distribution of ionised 
outflow velocity as a function of galaxy dynamical mass
is rather scattered in the (U)LIRG regime, i.e. for massive and highly star forming galaxies. 
The results by \cite{Heckman+11}, whose sample spans a wider range of galaxy properties, suggest
that outflow velocity and stellar mass are instead related.
Studies at higher redshifts seem to find more consistently a correlation between outflow
velocity and M$_*$ in star forming galaxies \citep{Weiner+09,Bradshaw+13,Rubin+14}.

Higher stellar velocity dispersions trace deeper gravitational potentials.
The positive correlation that we observe in Fig.~\ref{fig:plot_outflow_vel_3sigma} 
between $v_{\rm out, high}$ and $\sigma_*$ suggests that ionised
outflows may still gain a velocity sufficiently high for gas to escape also from massive galaxies, and not
only from lower-mass galaxies (as it would be the case if the outflow velocity did not depend on M$_*$, in which
case the escape of the gas from low-mass galaxies would be favoured).
However, we note that our analysis may be biased towards higher velocity outflows.
Indeed, our method probably selects only outflows with
velocity close to or above the escape velocity from the galaxy, because we are selecting
gas velocities in excess of the stellar velocity in order
to avoid contamination by gas in virial motion.
Figure~\ref{fig:plot_outflow_vel_3sigma} also shows that, {\it for a fixed SFR}, 
the outflow velocity (in the high velocity regime) decreases with $\sigma_*$. A possible explanation
is that for a given SFR, i.e. for a given energy and momentum budget available
to accelerate the gas, outflows are slowed down and re-captured 
by the gravitational potential in more massive galaxies,
and therefore they become increasingly ineffective at higher masses.

The relationships between outflow velocity and stellar velocity dispersion
(central panels of Figs.~\ref{fig:plot_outflow_vel_1sigma} and \ref{fig:plot_outflow_vel_3sigma}) suggest
that the physical mechanisms powering the lower and the higher velocity clouds in the observed ionised
outflows may be slightly different.
Indeed, at lower velocities ($v_{\rm out, low}$), 
which trace the bulk of the gas in outflow
\footnote{By definition, $v_{\rm out, high}$ probes only 0.1\% of the LoSVD of ionised gas,
i.e. the very high velocity tail of the outflow, which however only represents
a small fraction of the total LoSVD of the gas.}, the data points are placed
just below the $v_{\rm out} = \sigma_{*}$ locus, indicating that 
radiation pressure-driven (or ``momentum-driven'') winds, which are characterised
by $v_{\rm out}\gtrsim2\sigma_*$ \citep{Murray+05,Dave+11a}, 
are unlikely to dominate in this regime. 
Instead, these outflows may be 
generated by the kinetic energy injected into the ISM by
supernovae and stellar winds, i.e. ``energy-driven''.
Similarly, \cite{Bouche+12}, by using background quasars to trace MgII blueshifted absorption
in $z\sim0$ galaxies, estimated outflow speeds that are on average half the escape velocity, although
these authors did not find any correlation with the SFR.
On the other hand, the situation changes significantly in the high velocity regime, 
as shown by Fig.~\ref{fig:plot_outflow_vel_3sigma}, where the data
lie mostly above the $v_{\rm out} = \sigma_*$ relationship. In this
case, momentum-driven winds may also be at play (see also
findings by \citealt{Martin05}, \citealt{Heckman+15} for what they define as ``strong outflows'' at $z\sim0$, 
and by \citealt{Weiner+09}, at $z\sim1.4$).
Figure ~\ref{fig:plot_outflow_vel_3sigma} also suggests that the asymptotic velocity of
the observed ionised outflows is as high as $6-8~\sigma_{\rm stars}$, i.e. comparable to or 
higher than the escape velocity ($v_{\rm esc} \simeq (5-6)\sigma_*$, where $\sigma_*$ is the line-of-sight
stellar velocity dispersion, gives a conservative
estimate of the escape velocity, e.g. \citealt{Weiner+09}).

\subsection{The connection between galactic outflows and the main sequence of star-forming galaxies}\label{sec:discussion_MS}

\begin{figure*}[tbp]
\centering
   \includegraphics[clip=true, trim=1.3cm 5.5cm 0.9cm 2cm,angle=90,width=1.5\columnwidth]{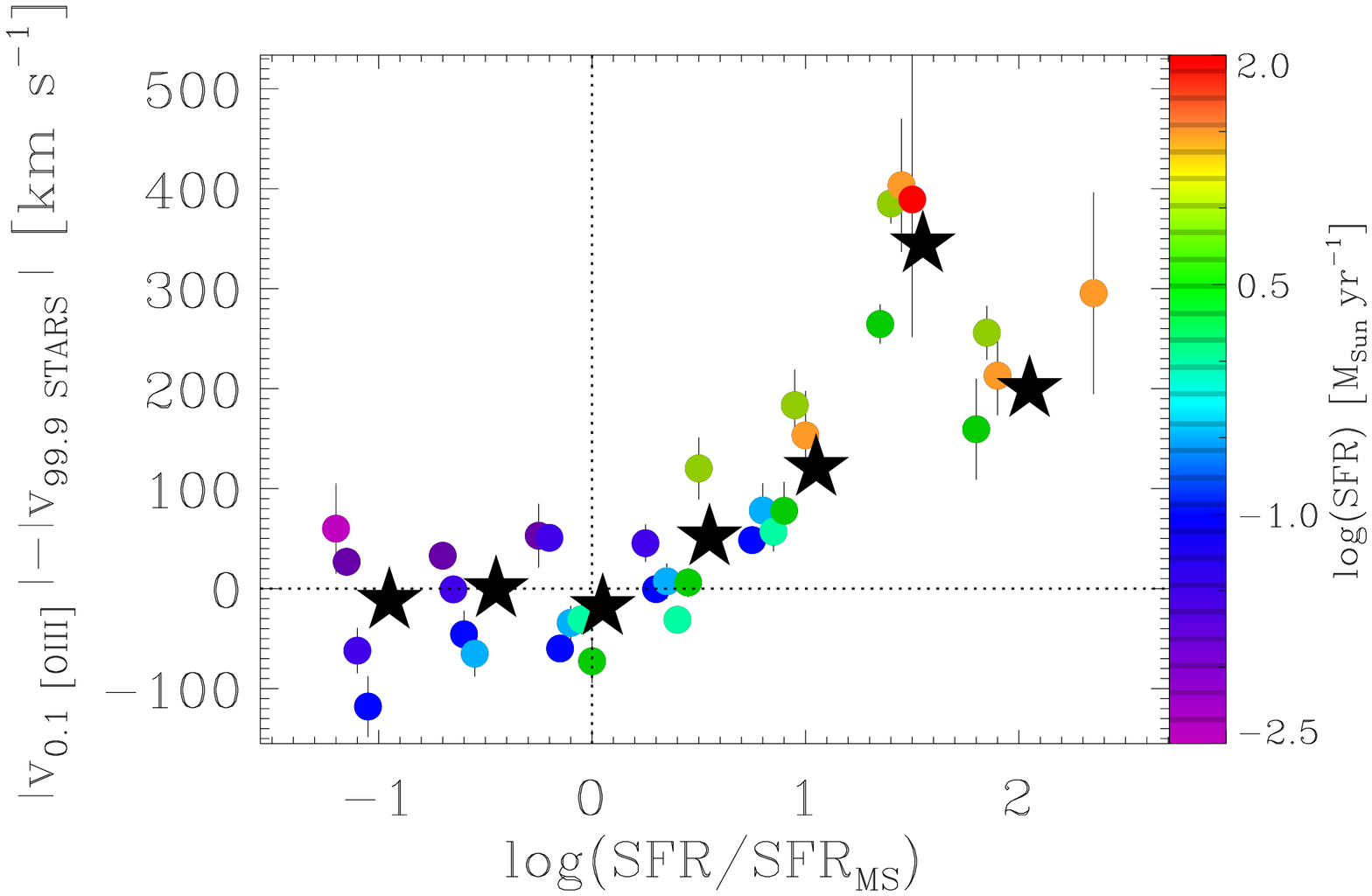}\\
  \includegraphics[clip=true, trim=1.3cm 5.5cm 1.2cm 2cm,angle=90,width=1.5\columnwidth]{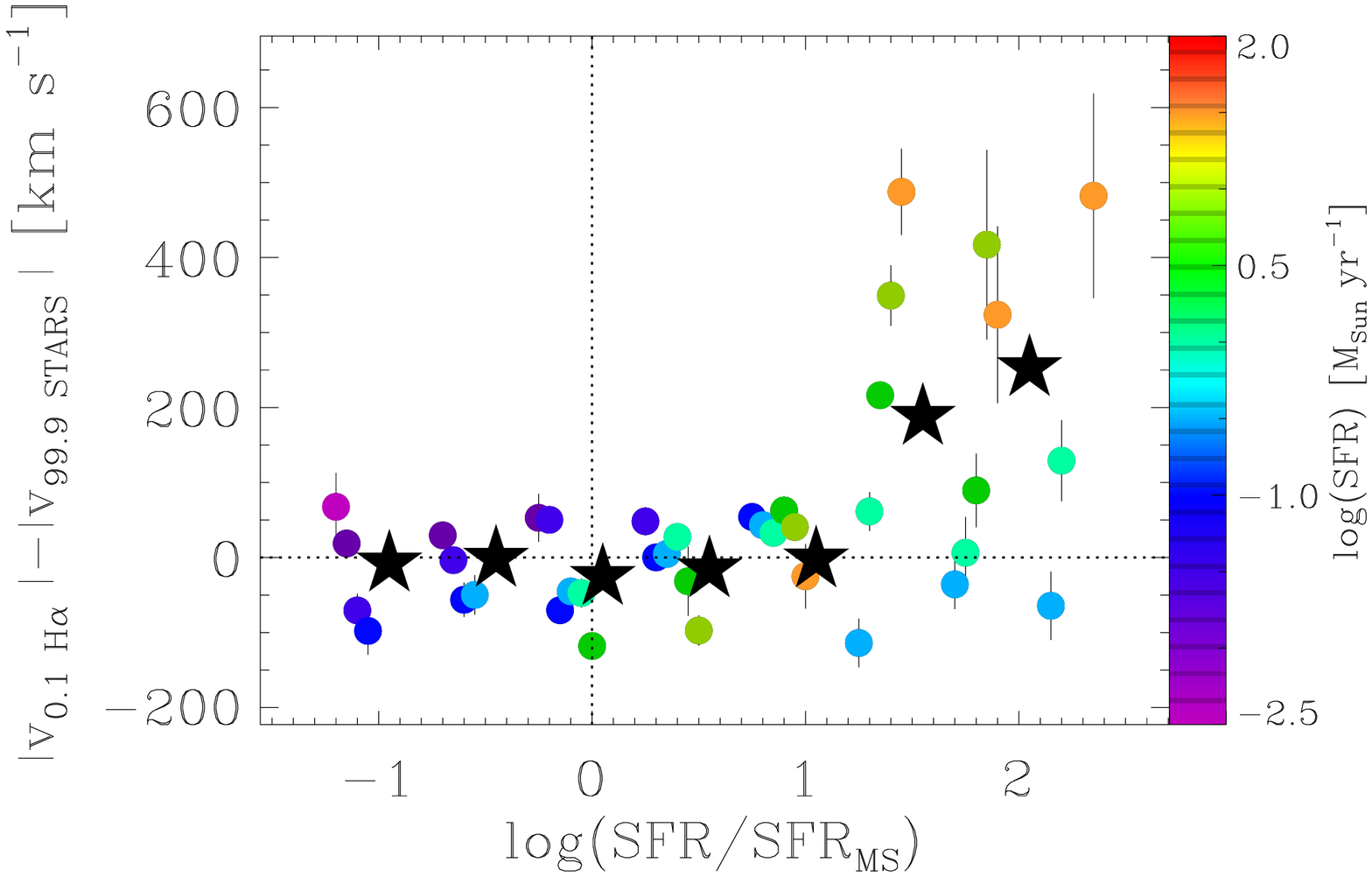}\\
   \caption{Excess of line-of-sight velocity of the ionised gas with respect to the stars (${v_{\rm gas}-v_{\rm stars}}$) as 
   given by ${y = |v_{\rm 0.1~gas}|-|v_{\rm 99.9~stars}|}$, plotted as a function
   of the offset from the local MS of star-forming galaxies. In the {\it upper panel} the outflow velocity is measured
   from [OIII] and in the {\it lower panel} from H$\alpha$. 
   As in Fig.~\ref{fig:plot_outflow_3sigma}, only bins with either $\sigma_{y} < {\rm 50~km~s^{-1}}$ or ${|y| \geq 2\sigma_{y}}$
   are shown. The black stars indicate the mean trend, obtained by averaging the data points (weighted with their errors) over 
   bins of  ${\rm \Delta log(SFR/SFR_{MS}) =0.5}.$}
   \label{fig:plot_of3_deltams}
\end{figure*}

Galactic outflows and cosmic inflows are invoked as key moderators of the SFR 
(and sSFR) in galaxies through cosmic times. 
Specifically, powerful galactic-scale outflows driven by AGNs and/or by 
extreme starburst episodes (possibly merger-driven) are thought to constitute the observational evidence
for the so called ``quenching'' mechanism acting on galaxies at all cosmic times
(or ``mass-quenching'', distinct from the ``environment-quenching'',
\citealt{Peng+10, Peng+12}).
Such mechanism,
through different physical processes yet poorly understood, 
should eventually transforms disky star-forming galaxies into passive spheroids,
thus generating the observed dichotomy between blue star forming and 
red passive galaxies \citep{Baldry1+04}.
Cosmic inflows are believed to 
feed galaxies through a smooth and continuous process, hence 
directly regulating the rate at which star formation occurs \citep{Dekel+09,Bouche+10,Lilly+13}.
The tight correlation between stellar mass and SFR observed 
from $z\sim0$ up to $z\sim4$ \citep{Schreiber+15} 
provides the major observational support to the hypothesis
of a ``cosmic feeding of star formation'', but direct observational evidence is still lacking.

In this scenario there is an effect that has been little explored from an observational point of view, that is
the possible role of galactic outflows in shaping the MS of star forming galaxies and,
in particular, its {\it upper envelope}.
Figure~\ref{fig:grid_outflow_3sigma} shows that ionised outflows are detected only in galaxies
located above the MS on the ${\rm M_{*}-SFR}$ diagram.
This effect can be visualised even better in Fig.~\ref{fig:plot_of3_deltams}, where
we plotted the excess of line-of-sight velocity of the ionised gas with respect to the stars, i.e. 
${|v_{\rm 0.1~gas}|-|v_{\rm 99.9~stars}|}$ (for both [OIII] and H$\alpha$), as a function of
the offset from the local MS (as derived in $\S$~\ref{sec:stacking}).
Figure~\ref{fig:plot_of3_deltams} shows a sharp increase of
outflows at $\rm SFR > SFR_{MS}$, where, for a given stellar mass, $\rm SFR_{MS}$ represents
the SFR of a galaxy located on the MS.
In other words, we empirically observe that for a given stellar mass, when the SFR
exceeds SFR$_{\rm MS}$, the gas velocity increases significantly, with the
effect of producing observable outflows. These outflows may eventually contribute to the suppression
of the SFR by reducing gas available for star formation, therefore bringing the galaxy
back onto the MS. Our observational result may reveal
a self-adjusting mechanism that explains the tightness of the local MS (i.e. $\pm0.3$ dex). We note
that this hypothesis is consistent with the scenario recently proposed by \cite{Tacchella+15}, who used 
simulations of high-redshift star forming galaxies to explore the mechanisms that confine galaxies into a
narrow sequence in the ${\rm M_{*}-SFR}$ plane.

The picture emerging from Figs.~\ref{fig:grid_outflow_3sigma} and \ref{fig:plot_of3_deltams} is
consistent with the abundant observational evidence for outflows in local starbursts and (U)LIRGs.
Outflows of ionised, neutral and molecular gas are
a common feature of galaxies with intense star formation activity, as traced by their
high infrared luminosities $L_{\rm IR}(8-1000~{\rm\mu m}) \geq 10^{11} L_{\odot}$
\citep{Chung+11,Sturm+11,Westmoquette+12,Rupke+Veilleux13,Rodriguez-Zaurin+13,Veilleux+13,
Spoon+13,Bellocchi+13, Cazzoli+14,Arribas+14, Cicone+14}.
These galaxies have larger SFR and sSFR than local MS galaxies (i.e. ${\rm SFR > 10~M_{\odot} yr^{-1}}$ and
sSFR typically higher than ${\rm 10^{-9}yr^{-1}}$, e.g. \citealt{Combes+13}) 
and populate the upper-right region of the M$_{*}$--SFR diagram. 
Previous observations at $z\sim0$ are therefore consistent with our findings, i.e. that star formation-driven 
galactic outflows are preferentially hosted by galaxies located above the local MS in the ${\rm M_{*}-SFR}$ diagram.

On the contrary, except from \cite{Martin+12}, outflows searches at higher redshifts revealed in general 
no dependency of galactic outflows on the galaxy position on the ${\rm M_{*}-SFR}$ diagram with respect to 
the MS \citep{Rubin+14,Genzel+14, Forster-Schreiber+14}, although in some of these cases the observed outflows
are plausibly driven by AGNs \citep{Genzel+14,Forster-Schreiber+14}.
The majority of NaID outflows detected by \cite{Sato+09} are hosted by
galaxies lying on the red sequence in the rest-frame $(U-B)$ color vs $M_B$ diagram and showing early-type morphologies. 
These authors however note that most of the red sequence galaxies with detectable outflows in their sample may have
some residual star formation, as suggested by their still high IR luminosities, and/or show signs of recent
star formation, as indicated by their $NUV-R_{AB}$ color.  Furthermore, since AGNs
are not excluded from their sample, an accreting super massive black hole may be
responsible for driving the observed outflows, especially in those galaxies which do not 
show significant star formation activity.

We note that there are secondary effects emerging from Fig.~\ref{fig:plot_of3_deltams}.
On the one hand, outflows
seem to ``saturate'' at $\rm log(SFR/SFR_{MS}) > 1.5$ when using [OIII] as a tracer, while
the scatter on the corresponding plot derived using H$\alpha$ does not allow us to assess
whether a similar saturation is present also for this tracer or not.
On the other hand, the minimum ${\rm SFR/SFR_{MS}}$ required to launch observable outflows
seems to be lower (i.e. closer to 1) for [OIII], while it is about a factor of 10 larger
for H$\alpha$. 

An indirect implication of Fig.~\ref{fig:plot_of3_deltams} is that galaxies located on or just below the local MS
do not host significant outflows, although we cannot exclude the presence of low-velocity outflows (``weak outflows'', e.g. 
\citealt{Heckman+15}), or of fewer galaxies with important outflow activity but where the outflow is seen edge-on.
This implies that star formation-driven outflows are likely not responsible for the migration of blue star forming galaxies
into the ``red sequence'', and so they are not responsible for what we generally refer to as ``quenching'', consistent with
what is found by simulations \citep{Hopkins+12} and also consistent with recent observational evidence that, for most galaxies, 
quenching results from the cutoff of gas inflow (a process refereed to as ``strangulation'' or ``starvation'', \citealt{Peng+Maiolino15}).
However, our results clearly highlight a link between
high sSFRs (i.e. sSFR in excess with respect to MS galaxies) and fast star formation-driven outflows.

\section{Summary and conclusions}

Galactic outflows, ejecting gas out of the galaxy, are a manifestation of feedback.
Powerful outflows have been observed in galaxies for decades, however it is still not clear
what is their role in galaxy evolution and whether they are actually responsible for reducing or, in the most
extreme cases, shutting off star formation in galaxies. Some key questions that we have addressed in this
work are: how can we reliably trace galactic outflows and estimate their properties? How do outflow properties
relate to the properties of the host galaxy? How do outflow properties compare to the predictions
of galaxy evolutionary models?

The most obvious source of feedback in galaxies is star formation itself, which, by conveying 
energy and momentum to the interstellar medium over prolonged timescales, can potentially affect future star formation
and hence galaxy evolution.
In this paper we have investigated the presence and properties of star formation-driven outflows
of ionised gas in normal galaxies by using a large spectroscopic sample of
$\sim160,000$ local non-active galaxies drawn from the SDSS.
The galaxy sample was divided into a fine grid of bins in the $\rm M_{*} - SFR$ 
parameter space, for each of which we produced a composite spectrum by stacking together the SDSS
spectra of the galaxies contained in that bin.
We exploited the high signal-to-noise of the stacked spectra to study
the emergence of faint features of
optical emission lines that trace galactic outflows, which otherwise would be too faint to detect in individual
galaxy spectra. Not only has the stacking technique allowed us to explore the presence of
galactic outflows in a large and representative sample of normal galaxies, spanning a wide range
 of galaxy properties (${\rm M_{*} \in [2 \times 10^7, 6 \times 10^{11}]~M_{\odot}}$ and
${\rm SFR \in [2 \times 10^{-3}, 2 \times 10^2 ]~M_{\odot}~yr^{-1}}$), but also to
break the degeneracy between SFR and M$_*$ that affects the bulk of the local star forming galaxy population
(i.e. the MS galaxies).

Based on the assumption that feedback-related processes (outflows, turbulence) should affect only
the gas kinematics, leaving the motions of stellar populations unperturbed, we traced the kinematical signature of 
feedback by adopting a novel approach that relies on the comparison between the LoSVD of the ionised gas 
(as traced by the [OIII] and H$\alpha$+[NII] emission lines) and the LoSVD of the stars, which were used as a reference tracing virial motions.
More specifically, significant deviations of the ionised gas kinematics from the stellar kinematics
in the high velocity tails of the LoSVDs (to minimise the effects of virial motions and turbulence) were interpreted as a signature
of outflows.

Our results suggest that the incidence of ionised outflows in local star forming galaxies 
increases with SFR and sSFR. Moreover, at a given SFR, more
massive galaxies are increasingly less efficient at launching powerful outflows.
In galaxy stacks displaying the clearest signature of galactic outflows, the outflow velocity
($v_{\rm out}$) was found to correlate tightly with the SFR for 
${\rm SFR > 1~M_{\odot}~yr^{-1}}$, whereas at lower SFRs the dependance
of $v_{\rm out}$ on SFR appears nearly flat. 
The outflow velocity, although with a much larger scatter, increases also 
with the stellar velocity dispersion ($\sigma_*$), and in a few galaxy stacks 
reaches values as high as $v_{\rm out} \sim (6-8) \sigma_*$, indicating that 
both energy-driven and momentum-driven winds may be at play.

Strikingly, the kinematical signature of ionised outflows was detected {\it only} in
galaxies located 
{\it above the main sequence} of star forming galaxies in the ${\rm M_{*}-SFR}$ diagram, 
and the incidence of high velocity outflows increases sharply with the offset from the MS.
On the one hand, this result clearly highlights a link between (specific) star formation rates in excess with respect to MS galaxies and
fast star formation-driven outflows, suggesting that such outflows may be responsible for shaping
the upper envelope of the MS by providing a self-regulating mechanism for star formation.
On the other hand, our findings suggest that the observed outflows may have little or no role
in the ``quenching'' of star formation in galaxies, 
i.e. the migration of blue star forming galaxies into the ``red sequence''.

Finally, although our study focussed mainly on galactic outflows, a
complementary analysis of the stellar kinematics revealed
blue asymmetries (of a few 10 km s$^{-1}$) in the stellar LoSVDs of galaxies with higher SFRs and M$_*$. 
The large uncertainties did not allow us to draw any firm conclusion
about the origin of such asymmetries, but a possibility is that these trace 
 the presence of a large number of runaway stars and hypervelocity stars in radial
trajectories within local galaxies.

In this paper, besides demonstrating the advantages and limitations of applying the spectral
stacking technique to optical emission lines for outflow studies, we proposed a new strategy 
to study galactic outflows, which relies on the direct comparison between the LoSVD of the gas and the LoSVD of the stars.
Such a new approach was tested with the SDSS spectroscopic sample that, although certainly valuable 
for its large statistics, is obviously limited in terms of spatial information. However, this methodology can
reveal its full potential if applied to resolved (IFS) observations of large galaxy samples,
which are becoming available thanks to the ongoing IFS surveys (MaNGA, CALIFA and SAMI) and new powerful
instruments such as MUSE.

\begin{acknowledgements}
We thank the anonymous referee for his/her thorough review of the manuscript that significantly helped 
us improving both the analysis and the presentation of the results.
CC gratefully acknowledges support from the Isaac Newton Studentship (University of Cambridge) and from
the Swiss National Science Foundation Professorship grant PP00P2\_138979/1 (ETH Zurich).
This work is based on SDSS-DR7 data and makes use of the 
MPA-JHU release of spectral measurements.
We thank David Schlegel and Christy Tremonti for providing crucial information on the SDSS arc lamp calibration spectra,
which allowed us to estimate the SDSS spectral instrumental profile shown in Fig.~\ref{fig:res}.
CC thanks Kevin Schawinski, Kurt Soto, Adriano Agnello and George Privon for stimulating discussions.
\end{acknowledgements}

\bibliography{ref}
\bibliographystyle{aa}

\appendix

\section{The percentile velocities of the LoSVDs}

\begin{figure}[tbp]
   \includegraphics[clip=true, trim=5.75cm 1.cm 2.2cm 1.8cm,angle=90,width=.9\columnwidth]{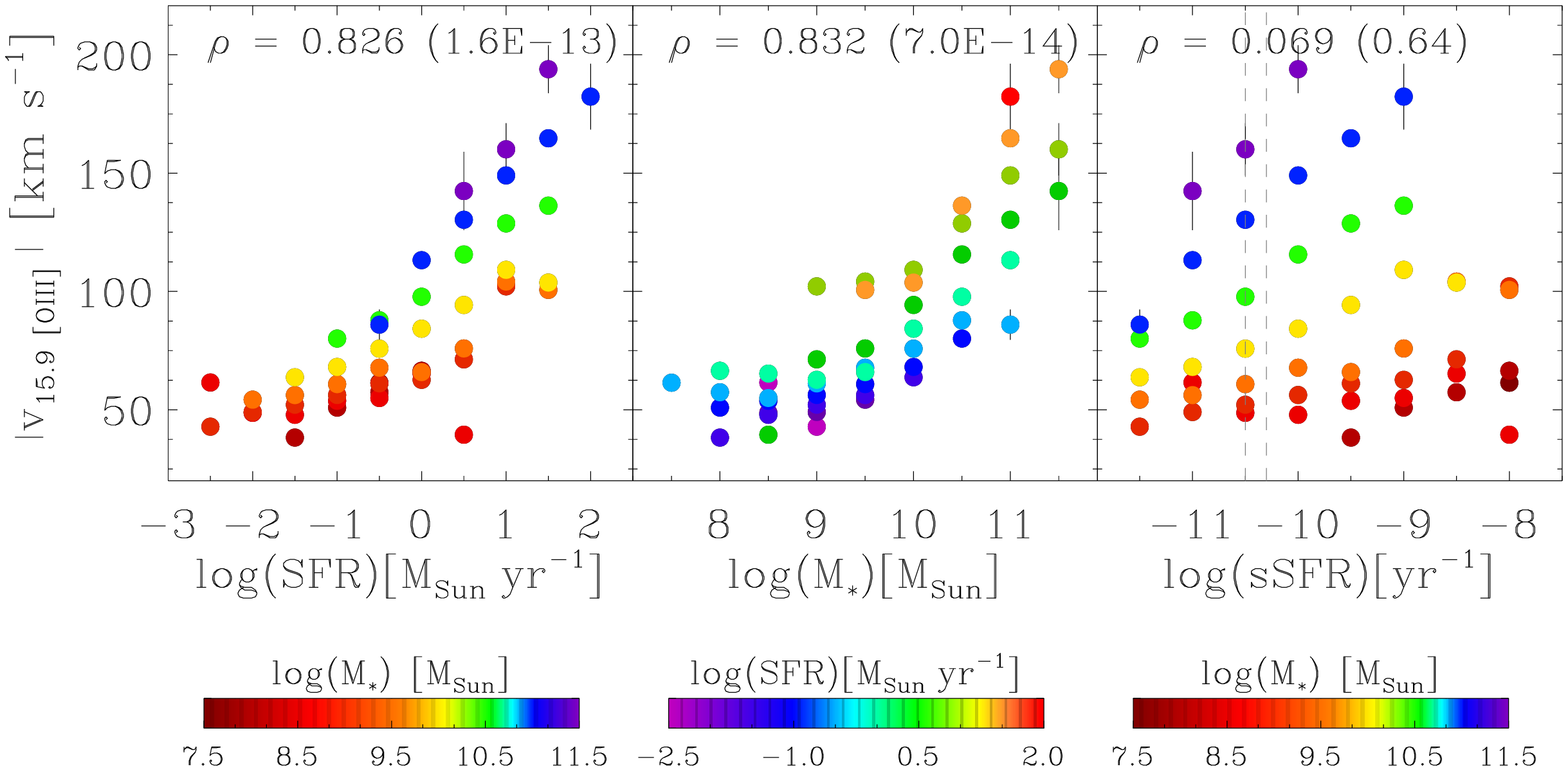}\\
   \includegraphics[clip=true, trim=5.75cm 1.cm 2.2cm 1.8cm,angle=90,width=.9\columnwidth]{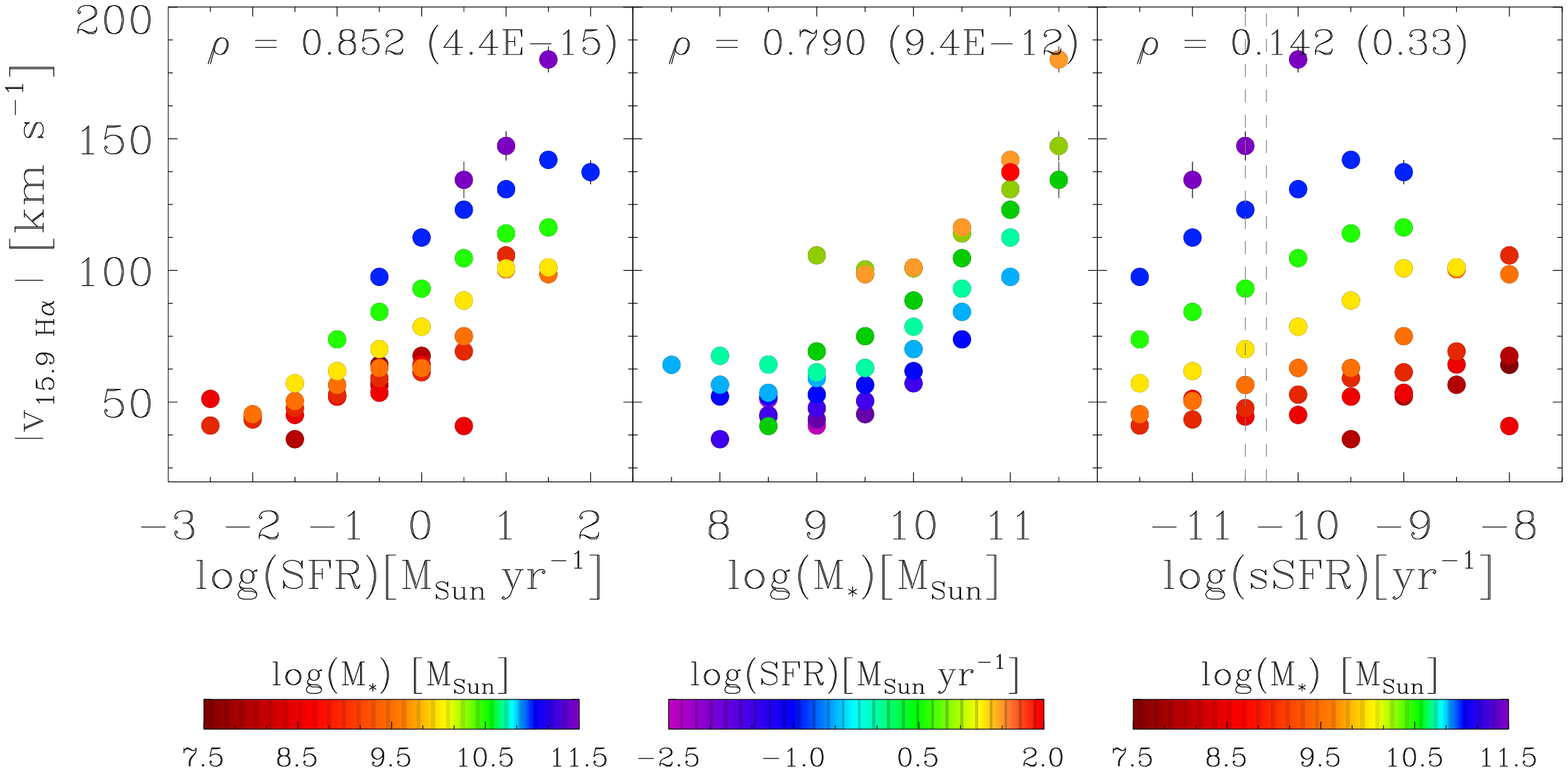}\\
   \includegraphics[clip=true, trim=2.cm 1.cm 2.2cm 1.8cm,angle=90,width=.9\columnwidth]{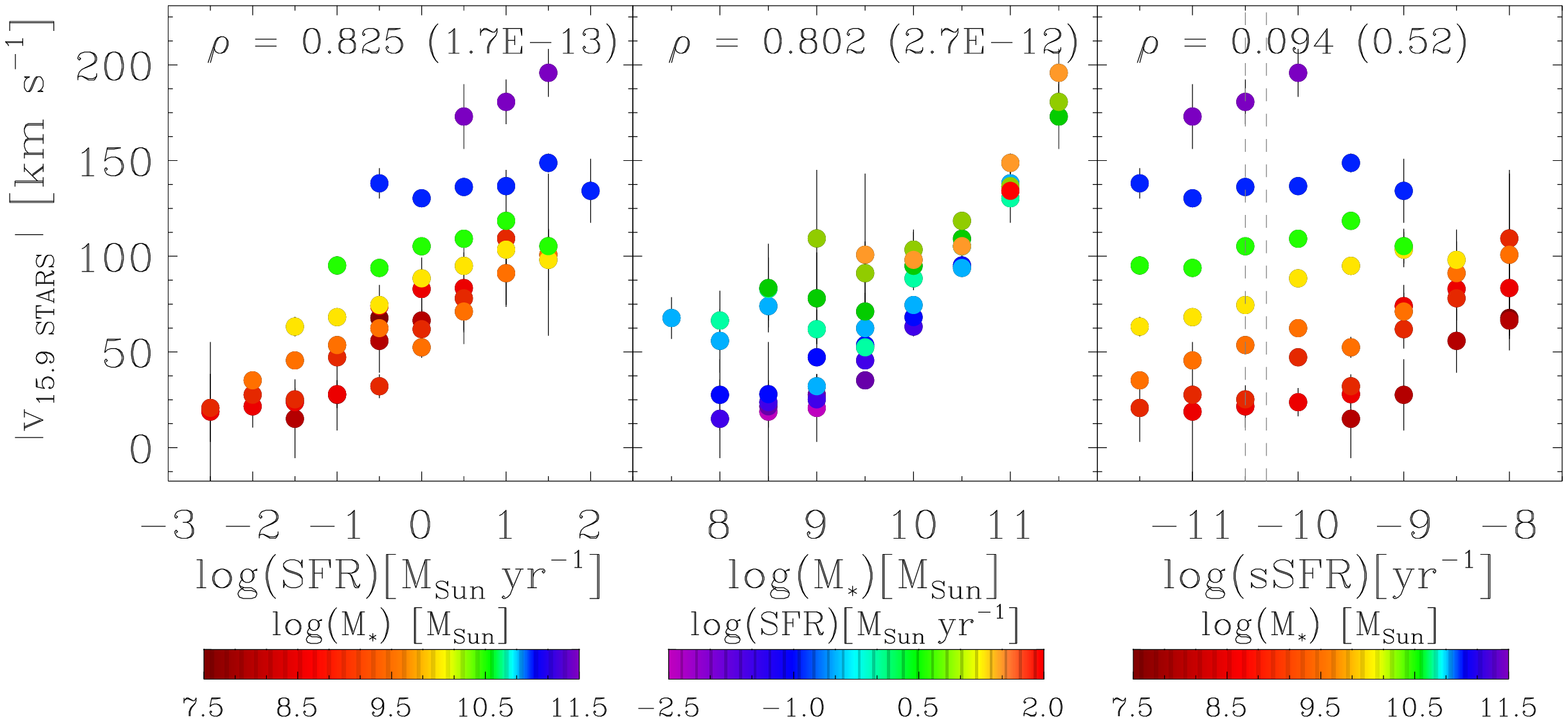}
   \caption{(Modulus of the) 15.9th percentile velocity of LoSVD of the ionised gas as traced by the 
   		[OIII] (\emph{top panel}) and H$\alpha$ (\emph{middle panel}) emission lines, and of the
		stars (\emph{bottom panel}). For a Gaussian velocity distribution, the 15.9th percentile
		velocity corresponds to -1 standard deviation ($-\sigma$) from the mean velocity. Similarly to Fig.~\ref{fig:sigma}, we report for
		each plot the Spearman rank correlation coefficient $\rho$ along with its associated two-sided p-value.}
   \label{fig:v16}
\end{figure}

\begin{figure}[tbp]
   \includegraphics[clip=true, trim=5.75cm 1.cm 2.2cm 1.8cm,angle=90,width=.9\columnwidth]{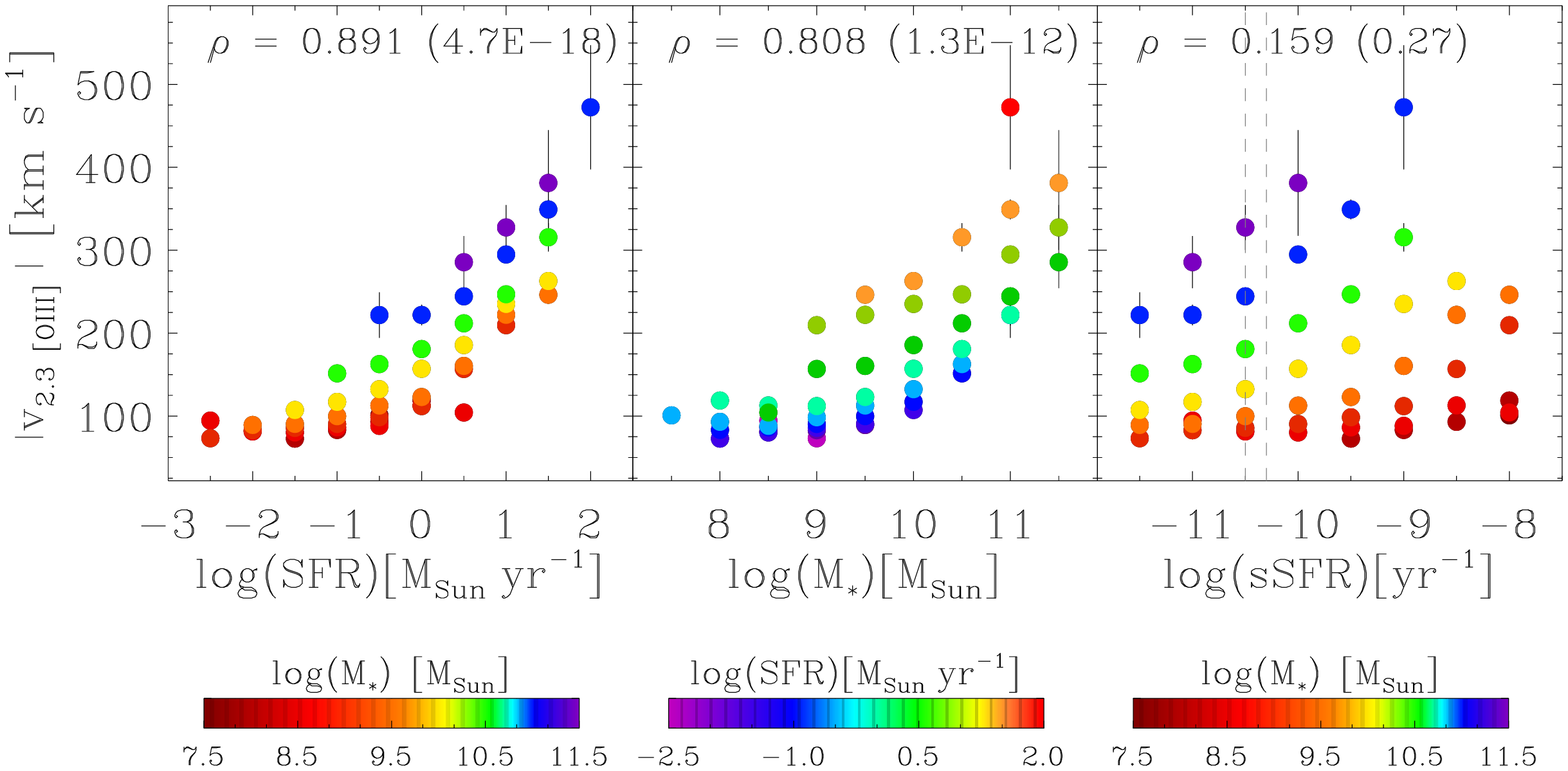}\\
   \includegraphics[clip=true, trim=5.75cm 1.cm 2.2cm 1.8cm,angle=90,width=.9\columnwidth]{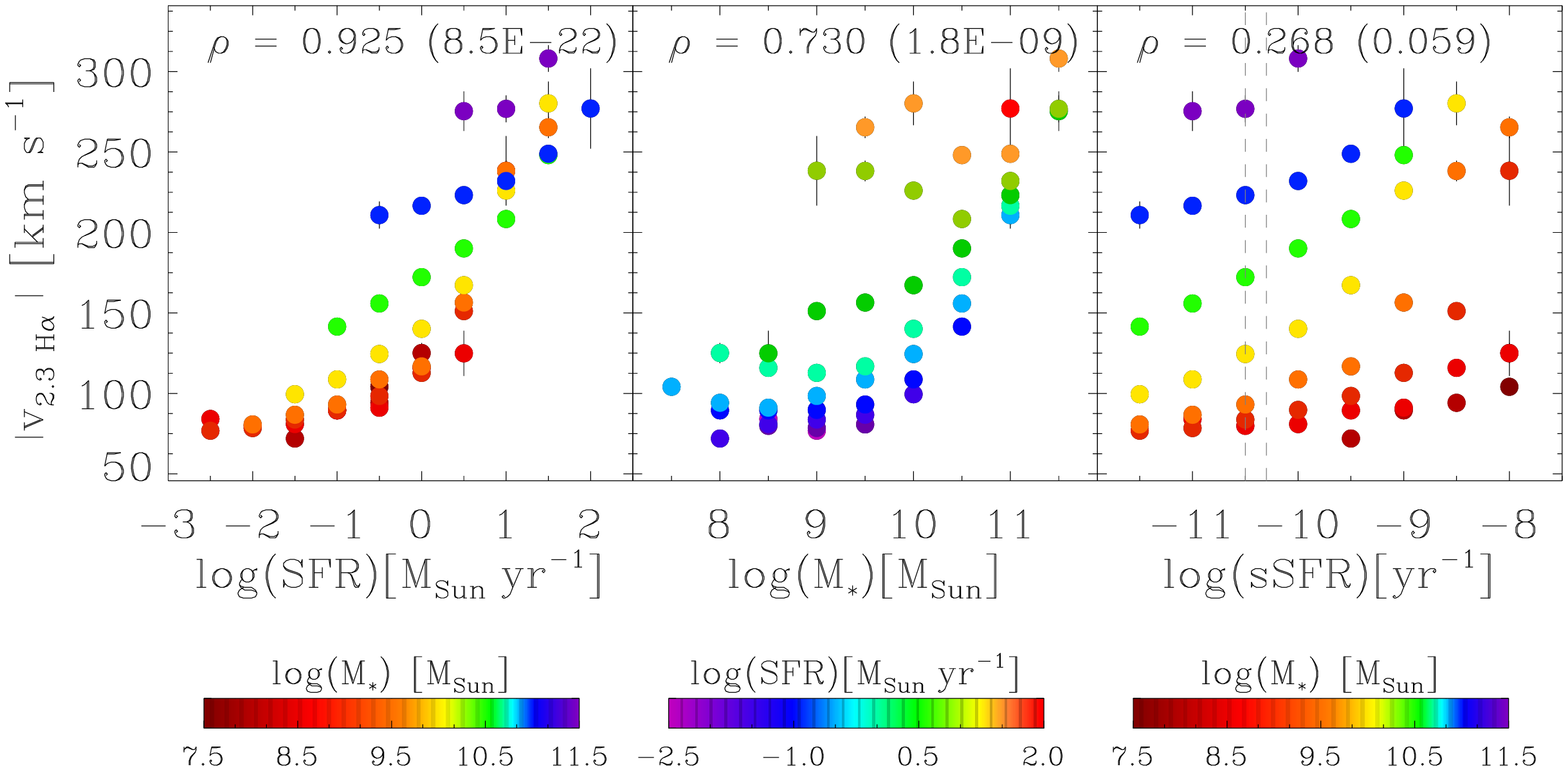}\\
   \includegraphics[clip=true, trim=2.cm 1.cm 2.2cm 1.8cm,angle=90,width=.9\columnwidth]{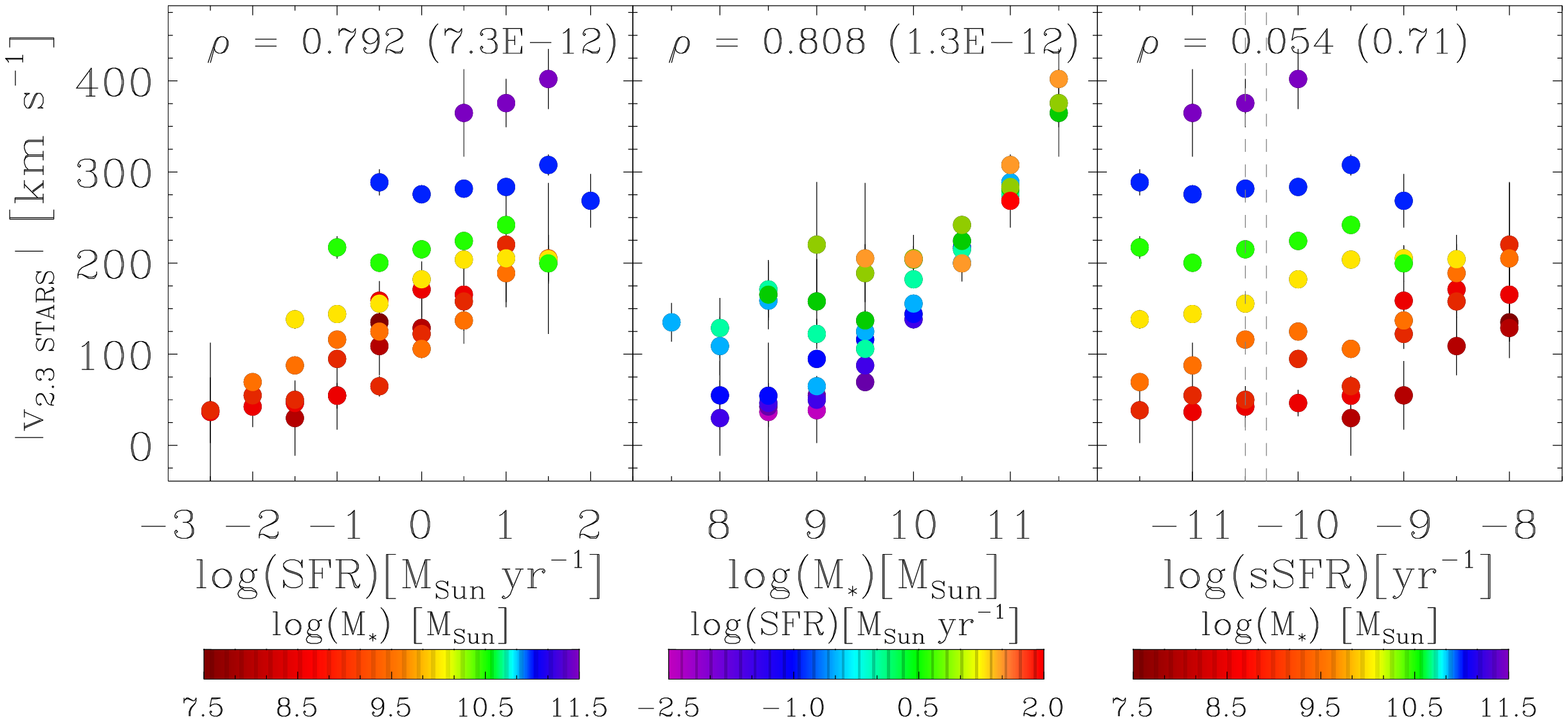}
   \caption{(Modulus of the) 2.3th percentile velocity of the LoSVD of the ionised gas as traced by the 
   		[OIII] (\emph{top panel}) and H$\alpha$ (\emph{middle panel}) emission lines, and of the
		stars (\emph{bottom panel}). For a Gaussian velocity distribution, the 2.3th percentile
		velocity corresponds to -2 standard deviations ($-2\sigma$) from the mean velocity.
		Similarly to Fig.~\ref{fig:sigma}, we report for
		each plot the Spearman rank correlation coefficient $\rho$ along with its associated two-sided p-value.}
   \label{fig:v2}
\end{figure}

\begin{figure}[tbp]
   \includegraphics[clip=true, trim=5.75cm 1.cm 2.2cm 1.8cm,angle=90,width=.9\columnwidth]{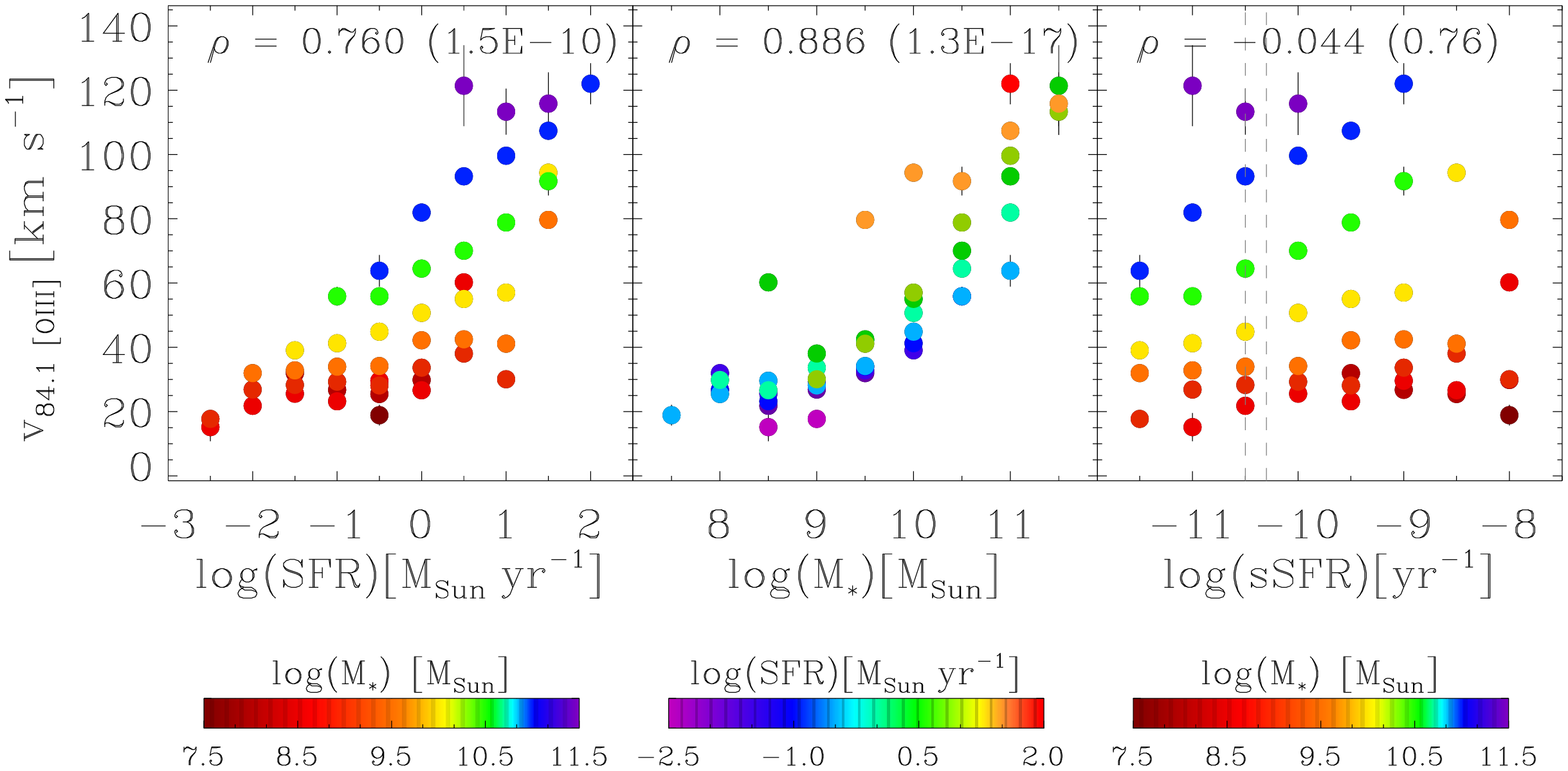}\\
   \includegraphics[clip=true, trim=5.75cm 1.cm 2.2cm 1.8cm,angle=90,width=.9\columnwidth]{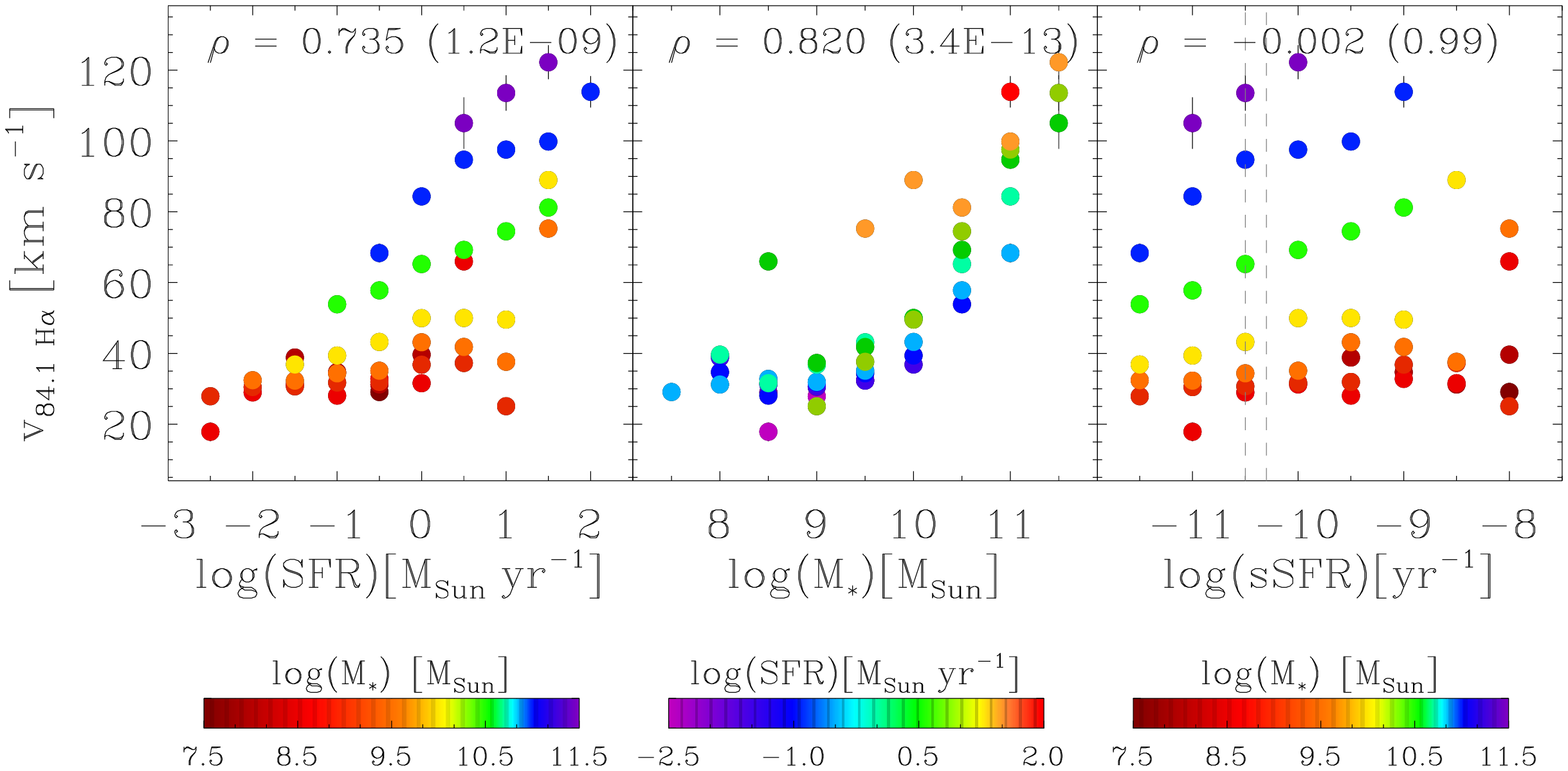}\\
   \includegraphics[clip=true, trim=2.cm 1.cm 2.2cm 1.8cm,angle=90,width=.9\columnwidth]{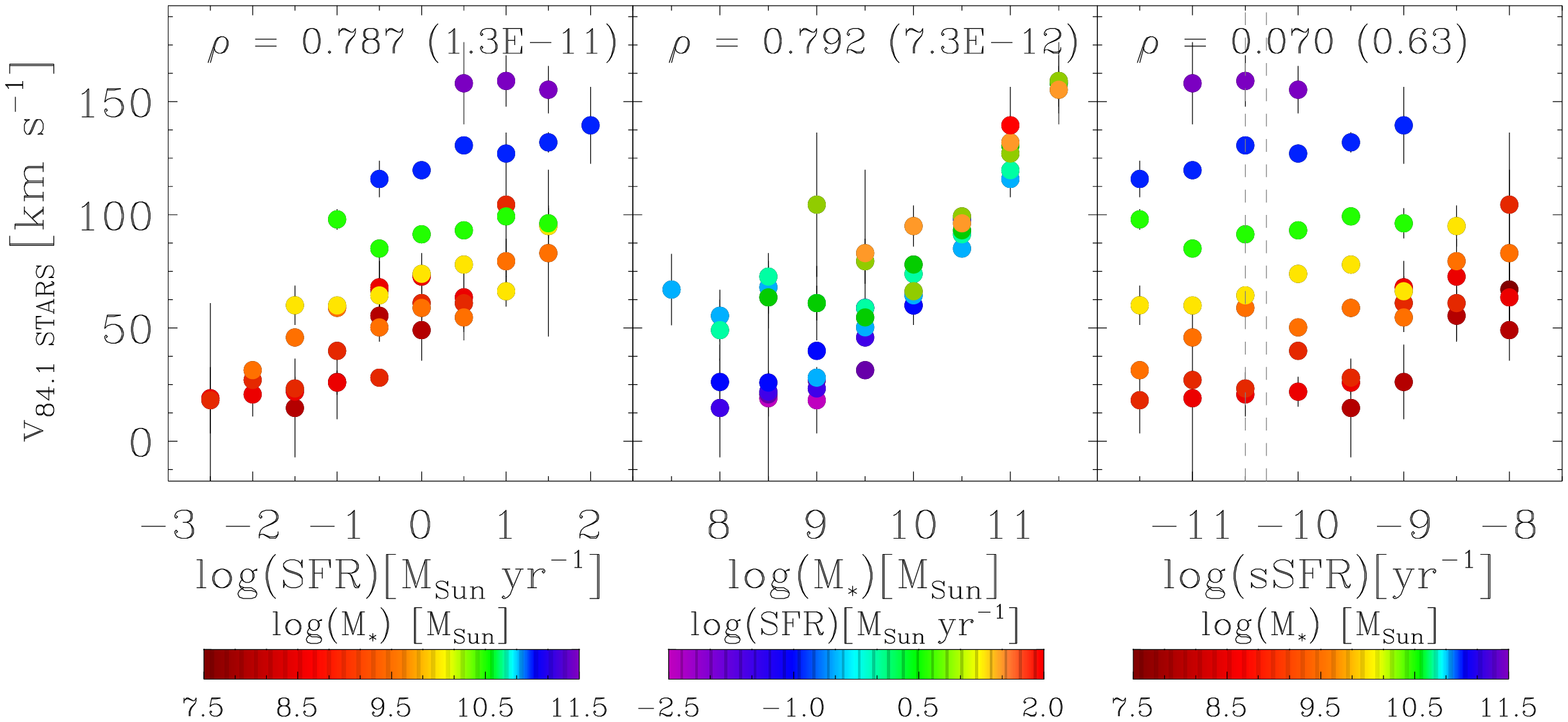}
   \caption{84.1th percentile velocity of the LoSVD of the ionised gas as traced by the 
   		[OIII] (\emph{top panel}) and H$\alpha$ (\emph{middle panel}) emission lines, and of the
		stars (\emph{bottom panel}). For a Gaussian velocity distribution, the 84.1th percentile
		velocity corresponds to +1 standard deviation ($\sigma$) from the mean velocity.
		Similarly to Fig.~\ref{fig:sigma}, we report for
		each plot the Spearman rank correlation coefficient $\rho$ along with its associated two-sided p-value.}
   \label{fig:v84}
\end{figure}

\begin{figure}[tbp]
   \includegraphics[clip=true, trim=5.75cm 1.cm 2.2cm 1.8cm,angle=90,width=.9\columnwidth]{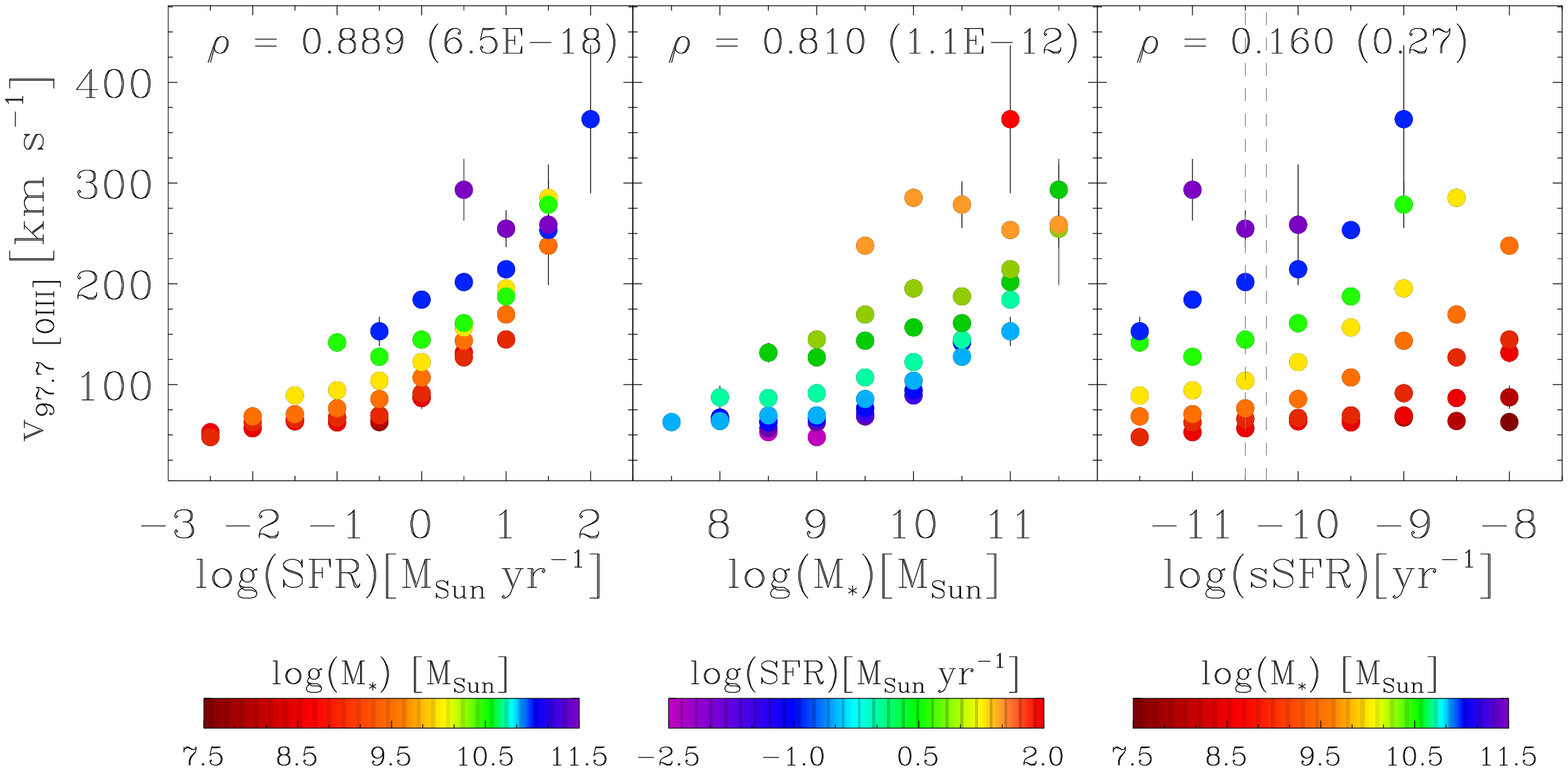}\\
   \includegraphics[clip=true, trim=5.75cm 1.cm 2.2cm 1.8cm,angle=90,width=.9\columnwidth]{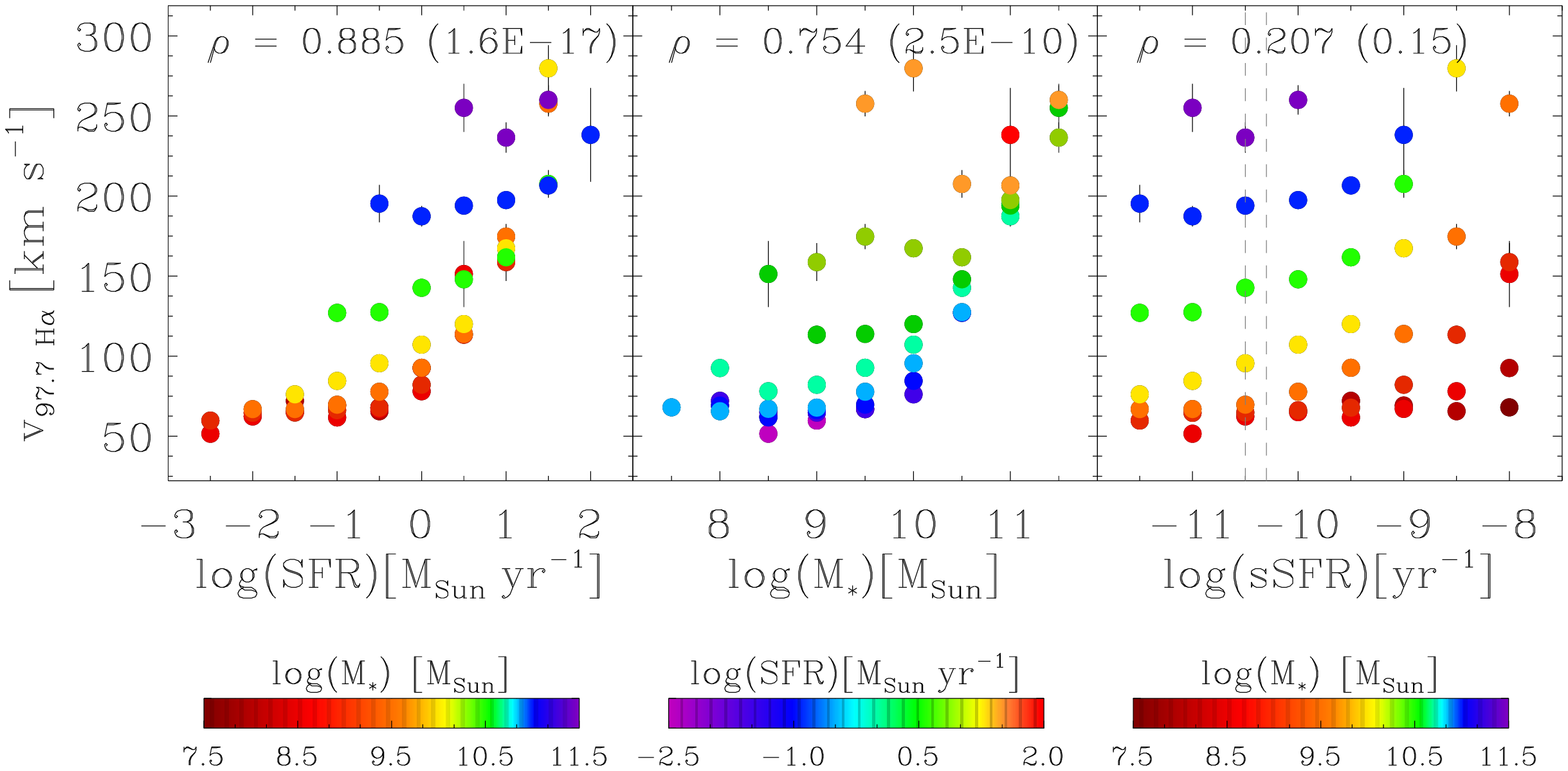}\\
   \includegraphics[clip=true, trim=2cm 1.cm 2.2cm 1.8cm,angle=90,width=.9\columnwidth]{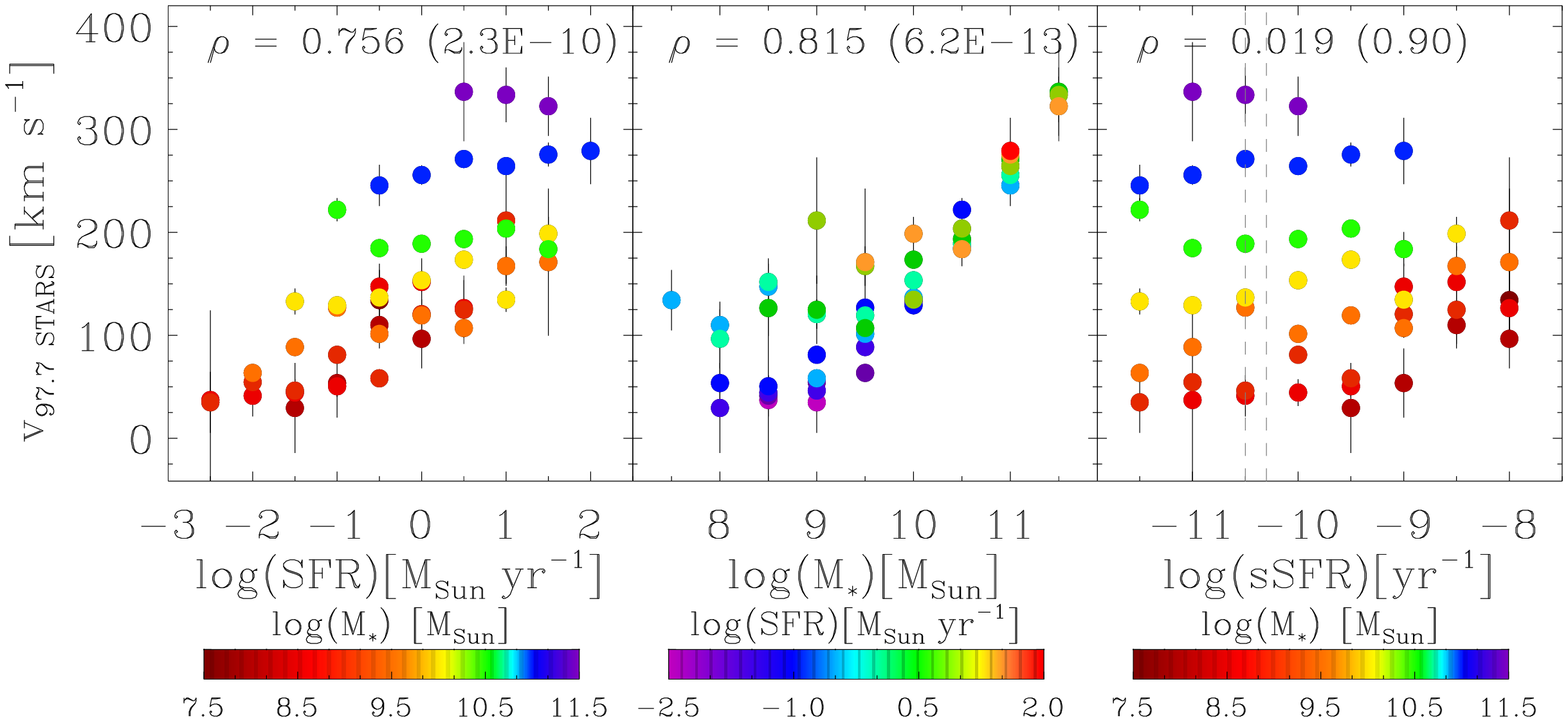}
   \caption{97.7th percentile velocity of the LoSVD of the ionised gas as traced by the 
   		[OIII] (\emph{top panel}) and H$\alpha$ (\emph{middle panel}) emission lines, and of the
		stars (\emph{bottom panel}). For a Gaussian velocity distribution, the 97.7th percentile
		velocity corresponds to +2 standard deviations ($2\sigma$) from the mean velocity.
		Similarly to Fig.~\ref{fig:sigma}, we report for
		each plot the Spearman rank correlation coefficient $\rho$ along with its associated two-sided p-value.}
   \label{fig:v98}
\end{figure}

\begin{figure}[tbp]
   \includegraphics[clip=true, trim=5.75cm 1.cm 2.2cm 1.8cm,angle=90,width=.9\columnwidth]{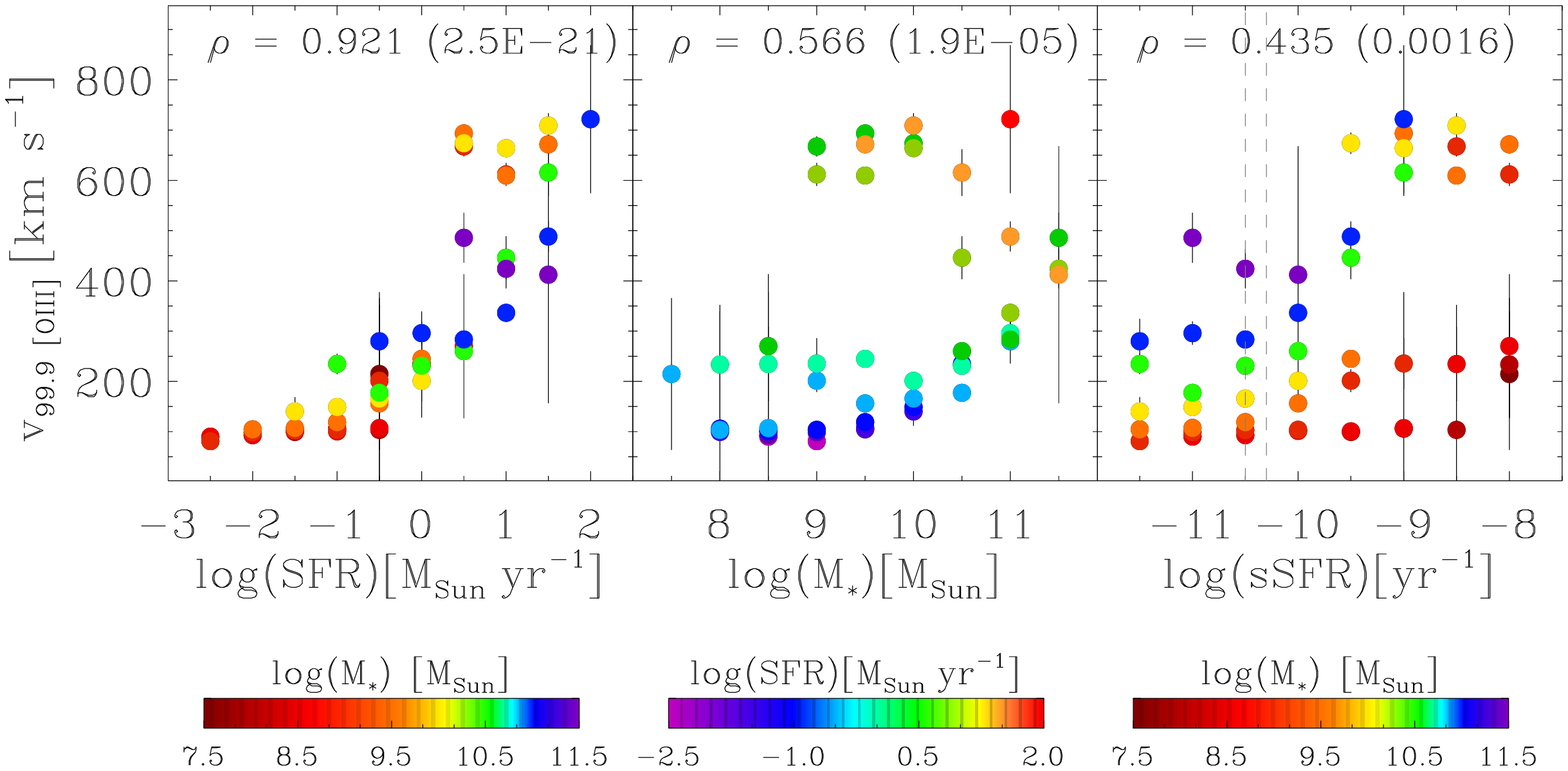}\\
   \includegraphics[clip=true, trim=5.75cm 1.cm 2.2cm 1.8cm,angle=90,width=.9\columnwidth]{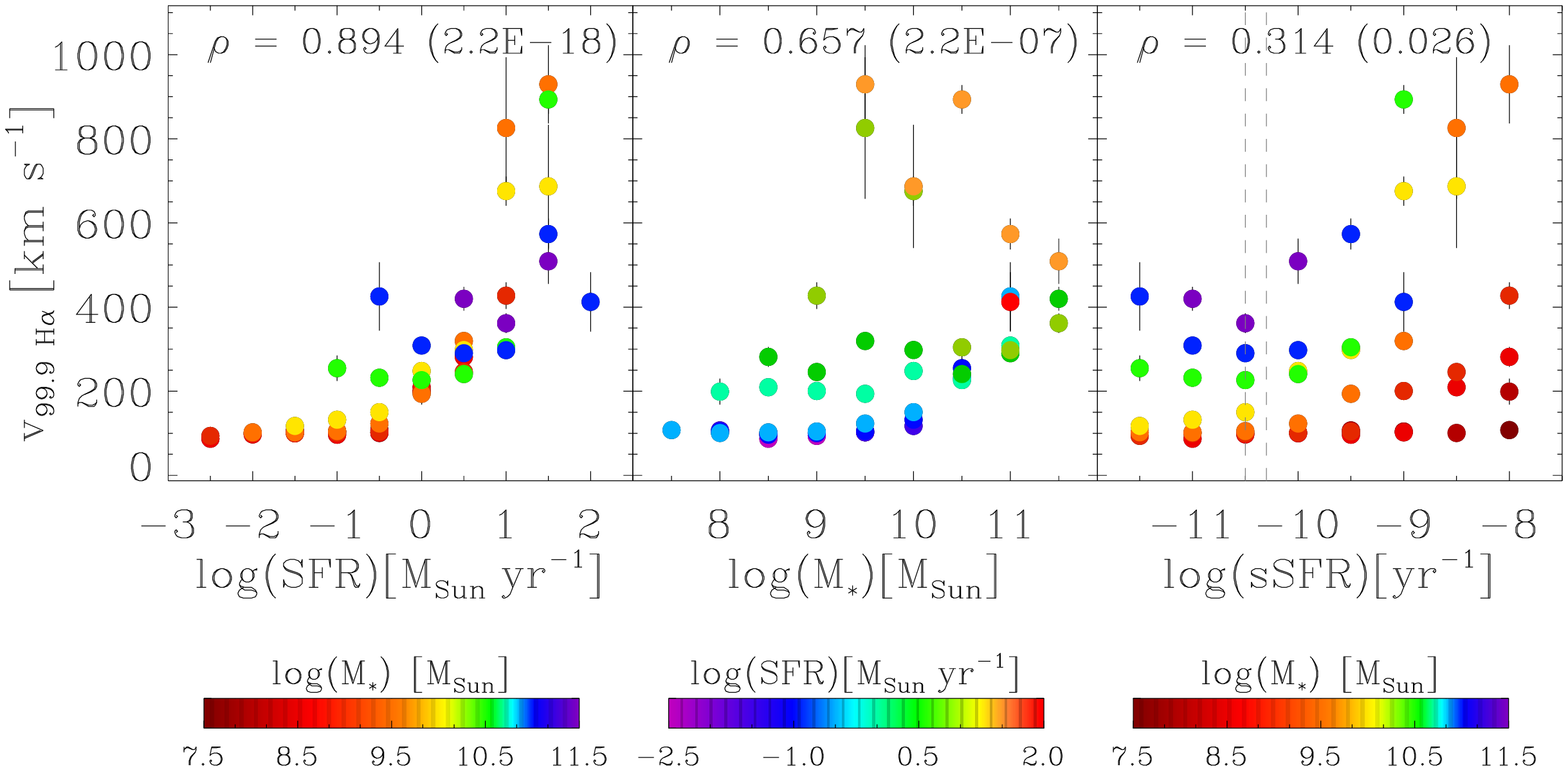}\\
   \includegraphics[clip=true, trim=2cm 1.cm 2.2cm 1.8cm,angle=90,width=.9\columnwidth]{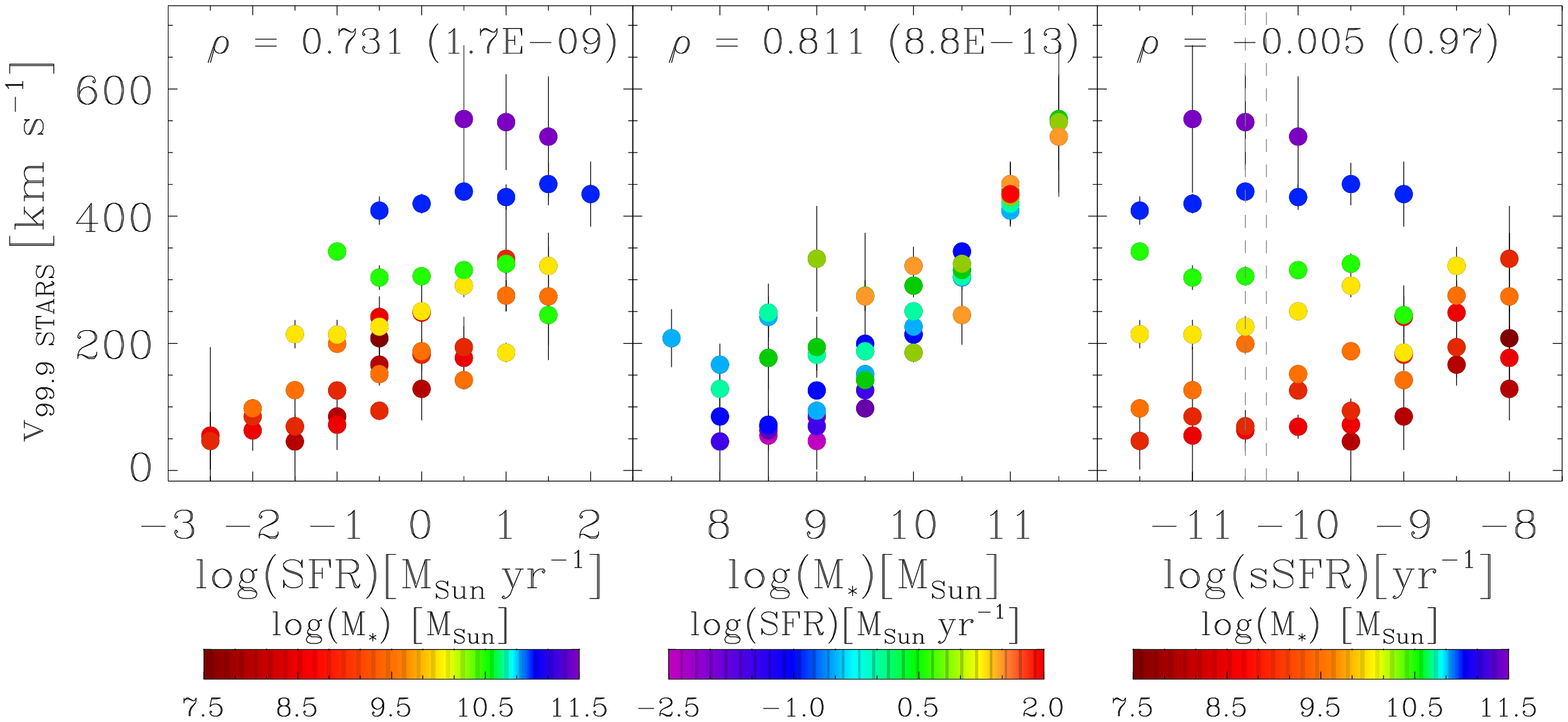}
   \caption{99.9th percentile velocity of the LoSVD of the ionised gas as traced by the 
   		[OIII] (\emph{top panel}) and H$\alpha$ (\emph{middle panel}) emission lines, and of the
		stars (\emph{bottom panel}). For a Gaussian velocity distribution, the 99.9th percentile
		velocity corresponds to +3 standard deviations ($3\sigma$) from the mean velocity.
	        Similarly to Fig.~\ref{fig:sigma}, we report for
		each plot the Spearman rank correlation coefficient $\rho$ along with its associated two-sided p-value.}
   \label{fig:v99}
\end{figure}

Figures~\ref{fig:v16}, \ref{fig:v2}, \ref{fig:v84}, \ref{fig:v98} and \ref{fig:v99} show, respectively, 
the 15.9th (-$\sigma$), 2.3th (-2$\sigma$), 84.1th ($\sigma$),
97.7th (2$\sigma$) and 99.9th (3$\sigma$) percentile velocities of the [OIII], H$\alpha$ and stellar LoSVDs
as a function of SFR, M$_*$ and sSFR. 
The results of the Spearman rank test performed on these relationships are 
reported in Table~\ref{table:rho_values}.
See further discussion in $\S$~\ref{sec:kinematics}. 

\end{document}